\documentclass[%
 aip,
 amsmath,amssymb,
 reprint,%
]{revtex4-1}

\usepackage{graphicx}% Include figure files
\usepackage{dcolumn}% Align table columns on decimal point
\usepackage{bm}% bold math
\usepackage{epstopdf}
\usepackage[utf8]{inputenc}
\usepackage[T1]{fontenc}
\usepackage{mathptmx}
\usepackage{etoolbox}
\usepackage{chemformula} % Formula subscripts using \ch{}
\usepackage{listings}
\usepackage[T1]{fontenc} % Use modern font encodings
\usepackage{mathptmx}
\usepackage{mathrsfs}
\usepackage{etoolbox}
\usepackage{amsmath}
\usepackage{amssymb}

\usepackage{braket}
\usepackage{color}
\usepackage{ulem}
\usepackage{accents}
\usepackage{chemscheme}
\usepackage{graphicx}
\usepackage{multirow}
\usepackage{notes2bib}
\usepackage{pgfplotstable}
\usepackage{rotating}
\usepackage{tabularx,booktabs}
\usepackage{adjustbox}
\usepackage{sidecap}
\usepackage{gensymb}
\usepackage{float}
\renewcommand{\thetable}{\arabic{table}}
\renewcommand{\tablename}{Table}
\renewcommand{\figurename}{\textbf{Figure}}
\renewcommand{\thefigure}{\textbf{\arabic{figure}}}
\DeclareUnicodeCharacter{2212}{-}
\newcommand{\bfepsilon}{\mathbf{\epsilon}\hspace{-4.5pt}\mathbf{\epsilon}}
\newcommand{\bfmu}{\mathbf{\mu}\hspace{-7.4pt}\mathbf{\mu}}
\usepackage{hyperref}
\hypersetup{
colorlinks=true,
    linkcolor=.,
    filecolor=.,      
    urlcolor=.,
    citecolor =.,
}

\makeatletter
\def\@email#1#2{%
 \endgroup
 \patchcmd{\titleblock@produce}
  {\frontmatter@RRAPformat}
  {\frontmatter@RRAPformat{\produce@RRAP{*#1\href{mailto:#2}{#2}}}\frontmatter@RRAPformat}
  {}{}
}%
\makeatother
\begin{document}

\preprint{AIP/123-QED}

\title[MesoHOPS]{MesoHOPS: Size-invariant scaling calculations of multi-excitation open quantum systems}

% Force line breaks with \\
\author{Brian Citty}
\altaffiliation{Co-first author: These authors contributed equally. All authors agree that these authors may list themselves in either order for their CV/Resume and other purposes.}

\author{Jacob K. Lynd}%
\altaffiliation{Co-first author: These authors contributed equally. All authors agree that these authors may list themselves in either order for their CV/Resume and other purposes.}

\author{Tarun Gera}
\affiliation{ 
Department of Chemistry, University of Texas at Austin, Austin, TX, 78712, USA
}%
 
\author{Leonel Varvelo}
\affiliation{Department of Chemistry, Southern Methodist University, PO Box 750314, Dallas, TX, USA}

\author{Doran I.G.B. Raccah}
\affiliation{ 
Department of Chemistry, University of Texas at Austin, Austin, TX, 78712, USA
}%
 \email{doran.raccah@utexas.edu}
\date{\today}% It is always \today, today,

\begin{abstract}
The photoexcitation dynamics of molecular materials on the 10-100 nm length scale depend on complex interactions between the electronic and vibrational degrees of freedom, rendering exact calculations difficult or intractable. The adaptive Hierarchy of Pure States (adHOPS) is a formally exact method that leverages the locality imposed by interactions between thermal environments and electronic excitations to achieve size-invariant scaling calculations for single-excitation processes in systems described by a Frenkel-Holstein Hamiltonian. Here, we extend adHOPS to account for arbitrary couplings between the thermal environments and vertical excitation energies, enabling formally exact, size-invariant calculations that involve multiple excitations or states with shared thermal environments. In addition, we introduce a low-temperature correction and effective integration of the noise to reduce the computational expense of including ultrafast vibrational relaxation in HOPS simulations. We present these advances in the latest version of the open-source MesoHOPS library and use MesoHOPS to characterize charge separation at a one-dimensional organic heterojunction when both the electron and hole are mobile.

\end{abstract}

\maketitle

\section{Introduction}
 
The photophysics of excited electronic states in molecular materials depend on their interaction with the vibrational degrees of freedom. The importance of the coupling between electronic and nuclear degrees of freedom has long been recognized in individual molecules and small aggregates for a variety of non-adiabatic phenomena, ranging from internal conversion via conical intersections\cite{domcke_conical_intersections} to excitation energy transfer between molecular chromophores.\cite{may_kuhn} There is a growing recognition that the complicated interplay between vibrational motion and electronic dynamics can also control processes that occur over tens-to-hundreds of nanometers length scales (i.e., the mesoscale), including excitation energy transfer,\cite{akselrod_subdiffusive_2014,sneyd2021efficient} singlet exciton-exciton annihilation,\cite{caram2016room,kumar2023exciton} charge separation and recombination,\cite{jailaubekov2013hot,Gelinas_2014_OPV,Kato_OPV} and triplet fusion.\cite{zhu2018exciton,wan2018transport} While the rapid development of spatially-resolved nonlinear spectroscopy\cite{delorImagingMaterialFunctionality2020, ginsbergSpatiallyResolvedExciton2020}  provides new experimental insights into mesoscale excited-state processes, the corresponding simulations often remain intractable.

Simulations of mesoscale excited-state processes in molecular materials require theoretical methods with size-invariant scaling. In molecular materials, a cube 100 nm on a side can contain more than $10^6$ molecules. As a result, computational methods where the number of operations, and hence CPU time, scale even linearly with the number of molecules ($n^1$, notated here as $\mathcal{O}(n^1)$ scaling) are incapable of mesoscale simulations. In some contexts, periodic boundary conditions allow simulations of effectively infinite materials,\cite{liu2017energy} but even in those cases the structural disorder characteristic of molecular materials - ranging from point defects to interfaces - often necessitates implausibly large super-cells. The development of theoretical methods with the necessary size-invariant computational scaling for simulations on the mesoscale (i.e., scaling as $n^0$ or $\mathcal{O}(1)$) remains an important open problem.

When the influence of the vibrational degrees of freedom can be approximated as a Markovian correction to the electronic dynamics, there are some reduced-scaling open quantum systems methods capable of mesoscale simulations. For example, a hybrid modified Redfield generalized Förster theory has been extended to mesoscale simulations of the photosynthetic thylakoid membrane.\cite{amarnath_multiscale_2016,bennett_energy-dependent_2018}
Similarly, the small polaron transform has been integrated within the delocalized kinetic Monte Carlo method to yield size-invariant scaling for mesoscale simulations of charge diffusion.\cite{balzer_delocalised_kmc_2021, balzer_OPV_2022, balzer_mechanism_2023, willson_jumping_2023} Methods using even more stringent approximations, including Marcus theory\cite{ shuai_marcus_and_beyond_in_organic_semiconductors} and Förster theory,\cite{lunt2009exciton} have also been used to simulate mesoscale dynamics. The rapid development of ever more incisive experimental probes of mesoscale excited-state processes, however, poses a challenge to these approximate methods, which struggle to reproduce spectroscopic observables where the underlying Markovian approximation of the electronic-vibrational coupling breaks down.\cite{ginsbergSpatiallyResolvedExciton2020}

Formally exact methods, those capable of capturing the non-Markovian interactions between electronic and vibrational degrees of freedom, often scale poorly with system size. Direct applications of methods originally developed to study molecular photochemistry by propagating the combined electronic and nuclear wave function - such as multi-configuration time-dependent Hartree (MCTDH) \cite{beck2000multiconfiguration,meyer2009multidimensional} and \it {ab initio} \rm multiple spawning \cite{bennun2000ab, sisto2017atomistic, curchod2020ssaims} - typically struggle in molecular materials due to the nearly macroscopic number of vibrational degrees of freedom. In some cases these methods - particularly MCTDH -  have been applied to molecular aggregates containing more than 10 molecules by assigning each molecule a reduced number of (often harmonic) degrees of freedom, \cite{popp2021quantum} thereby capturing key aspects of vibronic dynamics while coarse-graining over molecular details. 

When the thermal environment can  be described as a continuous set of harmonic oscillators characterized by a spectral density, a variety of formally exact methods exist, but are still plagued by high-order computational scaling with the number of molecules. One of the earliest, and still most general, approaches is the Quasi-Adiabatic Path Integral (QUAPI) method that uses a transfer tensor technique to achieve linear scaling in propagation time while maintaining an $\mathcal{O}(n^{2L+2})$ scaling with the number of pigments (n), where L is the number of time steps before the bath memory decays. \cite{makri1995tensor1,makri1995tensor2,kundu2023pathsum} A similar method, the Hierarchical Equations of Motion (HEOM), uses an exponential decomposition of the bath correlation function to write a Markovian time-evolution equation for a collection of auxiliary density matrices to track the influence of bath memory with overall scaling of $\mathcal{O}(n^{2 k_{\textrm{max}}})$, where $k_{\textrm{max}}$ is a convergence parameter that scales with the memory time of the bath.\cite{Tanimura_HEOM_Original,TanimuraNumericallyExactApproach2020} The Hierarchy of Pure States (HOPS) provides a stochastic set of hierarchical equations with a reduced scaling ($\mathcal{O}(n^{ k_{\textrm{max}}})$) due to the propagation of wave functions instead of density matrices. \cite{HOPS} The Time Evolving Density operator with Orthogonal Polynomials Algorithm (TEDOPA) re-casts the harmonic oscillator environment of each site into a semi-infinite chain of coupled harmonic oscillators that are time-evolved using a time-dependent density matrix renormalization group approach (TD-DMRG).\cite{Chin_TEDOPA_Original_2010}

Recognizing the underlying sparse structure of interactions in formally exact solutions, particularly those arising from bath memory, has led to a variety of tensor-contraction based reduced-scaling methods. One of the earliest tensor-contraction approaches was the development of multi-layer multi-configuration time-dependent Hartree (ML-MCTDH) \cite{schulzeMultilayerMulticonfigurationTimedependent2016, wangMultilayerFormulationMulticonfiguration2003} where a tensor-tree was used to provide a more efficient representation of the high-dimensional vibrational wave function. More recently, the matrix-product state representation has been used to automatically select the most important basis elements for a variety of quantum dynamics techniques including path integrals, \cite{cygorek2022simulation} HEOM,\cite{MPS_HEOM_Shi,borrelli2021expanding, Mangaud2023} HOPS,\cite{Gao_HOMPS_2022,flannigan2022many}  and explicit psuedomodes. \cite{somoza2019dissipation} Matrix product state methods scale, at minimum, as $\mathcal{O}(n^2)$, though in practice established methods do not achieve this limit. The improved performance of HEOM using more sophisticated network structures, such as the tensor trees pioneered in ML-MCTDH, suggests that further improvements in performance and robustness can be achieved by selecting more flexible contraction schemes.\cite{batista_photocatalysis_2021, Ke_Tree_Tensor_2023, Mangaud2023} The conceptual limit of this is, presumably, something similar to the modular path integral (MPI) approach developed by Makri which achieves linear scaling (i.e., $\mathcal{O}(n^1)$) by tailoring the contraction scheme to the physics of the equation-of-motion.\cite{makriCommunicationModularPath2018,makriModularPathIntegral2018, Makri_Quapi_Scaling}

The adaptive hierarchy of pure states (adHOPS) is a reduced scaling implementation of HOPS that achieves size-invariant scaling (i.e., $\mathcal{O}(1)$) in large aggregates by leveraging the local structure of the HOPS equation-of-motion.\cite{varveloFormallyExactSimulations2021, varvelo2023formally, gera_simulating_2023} The HOPS equation-of-motion arises within the non-Markovian quantum state diffusion (NMQSD) formalism where excitations are localized by system-environment interactions. When system-environment interactions link individual system states to independent baths, as commonly assumed in molecular aggregates, electronic excitations localize in the basis of molecular states, a process known as dynamic localization. In NMQSD calculations the delocalization extent of the electronic wave function depends on the relative magnitude of the system-bath coupling and the coupling between electronic states. The adHOPS algorithm constructs a moving basis that captures the dynamics relevant to the localized excitation, and, in sufficiently large aggregates where the delocalization extent of the excitation is small compared to the size of the aggregate, achieves size-invariant scaling. 

While adHOPS exhibits unprecedented scaling for a formally exact method, the algorithm developed in Ref. \onlinecite{varveloFormallyExactSimulations2021} and applied in Refs. \onlinecite{varvelo2023formally} and \onlinecite{gera_simulating_2023} requires that each vibrational environment induces fluctuations in the excitation energy of a single molecule (or analogous Hamiltonian).
This is sufficient for calculations using a Frenkel-Holstein Hamiltonian in a single-excitation manifold, which can describe exciton/charge transport dynamics and linear absorption. However, this framework does not allow for multi-excitation dynamics, which are required to simulate phenomena such as charge separation/recombination in organic semiconductors and nonlinear spectroscopy.

In this paper, we extend the adHOPS algorithm to arbitrary diagonal system-bath coupling operators, allowing simulations of systems where multiple excitations are present or multiple system states share a single environment. We also derive a low-temperature correction and effective integration of the noise that simplify calculations in the presence of ultrafast vibrational relaxation. Finally, we demonstrate the power of the new adHOPS algorithm by simulating a 1-dimensional model of interfacial charge separation, characterizing the mechanism of transport when both the electron and hole are mobile. All developments described here are available in MesoHOPS, an open-source Python package.\cite{mesohops_1_4}

\section{Theoretical Formalism}

\subsection{Open Quantum Systems}

We express an open quantum system Hamiltonian as the sum of three pieces:
\begin{equation}
    \label{eq:Ham_Full}
    \hat{H} = \hat{H}_{\textrm{S}} + \hat{H}_{SB} + \hat{H}_{B}
\end{equation}
where $\hat{H}_S$ is the system Hamiltonian, 
\begin{equation}
    \hat{H}_{B} = \sum_{n,q_n} \omega_{q_n} (\hat{a}^\dagger_{q_n} \hat{a}_{q_n} + 1/2)
\end{equation}
represents the set of independent thermal environments $\{n\}$ as baths of harmonic oscillators, and
\begin{equation}
    \hat{H}_{SB} = \sum_{n,q_n} \Lambda_{q_n} \hat{L}_n (\hat{a}^\dagger_{q_n} + \hat{a}_{q_n})
\end{equation}
describes the system-bath interactions expressed in terms of the system-bath coupling operators ($\hat{L}_n$) and linear coupling coefficients ($\Lambda_{q_n}$) given by the spectral density of the bath 
\begin{equation}
    \label{eq:J_sd}
    J_n(\omega) = \pi \sum_{q_n} \vert \Lambda_{q_n}\vert^2 \delta(\omega - \omega_{q_n}).
\end{equation}
In terms of the spectral density, the time correlation function of each thermal environment is\cite{may_kuhn,Feynman_student_thesis}:
\begin{equation}
\label{eq:C_t}
C_n(t) = \frac{1}{\pi}\int_0^\infty d\omega J_n(\omega) \big(\coth(\beta \omega /2) \cos(\omega t/\hbar) - i \sin( \omega t/\hbar)\big)
\end{equation}
where $\beta=\frac{1}{k_BT}$ is the inverse temperature. Taking advantage of the overcompleteness of the basis of complex exponentials, we decompose each bath correlation function into a (non-unique) sum of exponential modes:
\begin{equation}
\label{eq:C_exp}
    C_n(t) = \sum_{j_n} C_{j_n}(t) = \sum_{j_n} g_{j_n} e^{-\gamma_{j_n} t/\hslash}.
\end{equation}
Note that, here and below, the index $j_n$ represents the $j^{\textrm{th}}$ mode of the correlation function of the $n^{\textrm{th}}$ thermal environment.

In all calculations presented in this paper, we describe each independent environment ($n$) with an overdamped Drude-Lorentz spectral density
\begin{equation}
\label{eq:drude_lorentz}
    J_n(\omega) = 2\lambda_n\gamma_{0_n}\frac{\omega}{\omega^2 + \gamma_{0_n}^2}
\end{equation}
where $\lambda_n$ is the reorganization energy and $\gamma_{0_n}$ is the reorganization timescale of the $n^{\textrm{th}}$ environment.
The associated exponential decomposition of the correlation function at temperature $T$ (Eq. \eqref{eq:C_t}) is given by a high-temperature mode, with magnitude $g_{0_n}$ and decay frequency $\gamma_{0_n}$, and  $k_{\textrm{Mats}}$ Matsubara modes, with magnitudes $g_{\nu_n}$ and decay frequencies $\gamma_{\nu_n}$:
 \begin{equation}
\label{eq:alpha_dl_coarse}
    C_n(t) = g_{0_n} e^{-\gamma_{0_n} t/\hslash} + \sum_{\nu=1}^{k_{\textrm{Mats}}} g_{\nu_n} e^{-\gamma_{\nu_n} t/\hslash}
\end{equation}
where 
\begin{equation}
    g_{0_n} = 2\lambda_n\beta^{-1}\Bigg(1+\sum_{j_n=1}^{k_{\textrm{Mats}}}\frac{\gamma_{0_n}^2}{\gamma_{0_n}^2-\gamma_{j_n}^2}\Bigg) - i\lambda_n\gamma_{0_n}
\end{equation}
and
\begin{equation}
    g_{\nu_n} = \frac{2i}{\beta}J_n(i\gamma_{\nu_n}), \quad \gamma_{\nu_n} = \frac{2\pi \nu}{\beta}
\end{equation}
where $\beta = \frac{1}{k_BT}$.

\subsection{Hierarchy of Pure States (HOPS)}

The Hierarchy of Pure States (HOPS) method provides a numerical solution to the formally exact non-Markovian quantum state diffusion (NMQSD) equation for open quantum systems.\cite{HOPS,NMQSD} In HOPS, we express the dynamics arising from the Hamiltonian (Eq. \eqref{eq:Ham_Full}) in terms of a physical (system) wave function $\ket{\psi^{(\vec{0})}_t}$, coupled to a set (hierarchy) of auxiliary wave functions $\ket{\psi^{(\vec{k})}_t}$ corresponding to the non-Markovian system-bath interactions, where $\vec{k}$ is an auxiliary vector indexed in the space of correlation function modes ($k_{j_n} = \vec{k}[j_n]$). The time-evolution of each wave function is given by the normalized nonlinear HOPS equation-of-motion:
\begin{flalign}
\begin{aligned}
\label{eq:NormNonLinearHops}
\hslash \frac{d\vert \psi^{(\Vec{k})}_t\rangle}{dt}  
=  \big(-i\hat{H}_S - \Vec{k} \cdot \Vec{\gamma} -\Gamma_t + \sum_{n} \hat{L}_{n} (z^*_{n,t}+ \sum_{j_n}\xi_{{j_n},t})\big)\vert \psi^{(\Vec{k})}_t \rangle &\\ 
+ \sum_{n,{j_n}} k_{j_n} \gamma_{j_n} \hat{L}_{n}  \vert \psi^{(\Vec{k} -\Vec{e}_{j_n})}_t \rangle &\\
- \sum_{n,{j_n}} \Big(\frac{g_{j_n}}{\gamma_{j_n}}\Big)(\hat{L}^{\dagger}_{n} - \langle\hat{L}^{\dagger}_{n}\rangle_{t}) \vert \psi^{(\Vec{k}+\Vec{e}_{j_n})}_t\rangle. &
\end{aligned}
\end{flalign}
where
\begin{equation}
    \langle\hat{L}^{\dagger}_{n}\rangle_{t} = \langle \psi^{(\vec{0})}_t \vert \hat{L}^{\dagger}_{n}\vert \psi^{(\vec{0})}_t \rangle
\end{equation}
is the expectation value of the $n^{\textrm{th}}$ system-bath coupling operator,
\begin{flalign}
\begin{aligned}
\label{eq:normcorr}
    \Gamma_t = &\sum_{n} \braket{\hat{L}_{n}}_{t} \textrm{Re}[z^*_{n,t}+ \sum_{j_n}\xi_{j_n,t}] \\
    - &\sum_{n,j_n}  \textrm{Re}\Big[\Big(\frac{g_{j_n}}{\gamma_{j_n}}\Big)\braket{\psi^{(\vec{0})}_t |\hat{L}^{\dagger}_{n}| \psi^{(\vec{e}_{j_n})}_t}\Big] \\
    + &\sum_{n,j_n}  \braket{\hat{L}^{\dagger}_n}_{t} \textrm{Re}\Big[\Big(\frac{g_{j_n}}{\gamma_{j_n}}\Big)\braket{\psi^{(\Vec{0})}_t | \psi^{(\Vec{e}_{j_n})}_t }\Big]
\end{aligned}
\end{flalign}
is a normalization correction factor, 
\begin{equation}
    \xi_{j_n,t} = \frac{1}{\hslash}\int_{0}^{t} ds C^{*}_{j_n}(t-s) \braket{\hat{L}^{\dagger}_{n}}_s
\end{equation}
is the noise memory drift term of correlation function mode $j_n$,
and $z_{n,t}$ is the stochastic fluctuation, or noise, arising from the interaction of the system and the $n^{\textrm{th}}$ environment at time $t$.

An ensemble of HOPS simulations is a stochastic unravelling of the system reduced density matrix. The noise ($z_{n,t}$) is a complex-valued Gaussian stochastic process\cite{NMQSD, HOPS, Strunz_1996} characterized by $\mathbb{E}_z[z_{n,t}] =0$,  $\mathbb{E}_z[z_{n,t} z_{n,s}] =0$, and $\mathbb{E}_z[z^*_{n,t} z_{m,s}] = \delta_{m,n}C_n(t-s)$, where $\mathbb{E}_z[\cdot]$ denotes an ensemble average over noise trajectories. For each noise trajectory there is a corresponding time-evolution of  the physical wave function ($\vert \psi^{(\vec{0})}_t \rangle$): we refer to these system dynamics as a HOPS trajectory. The ensemble of (nonlinear) HOPS trajectories is thus a Monte Carlo sampling of the dynamics of the open quantum system, such that the system reduced density operator is given by\cite{NMQSD,HOPS}
\begin{equation}
    \hat{\rho}_t = \mathbb{E}_z\left[|\psi^{(\vec{0})}_t\rangle\langle\psi^{(\vec{0})}_t|\right].
\end{equation}

\begin{figure}
    \centering
    \includegraphics{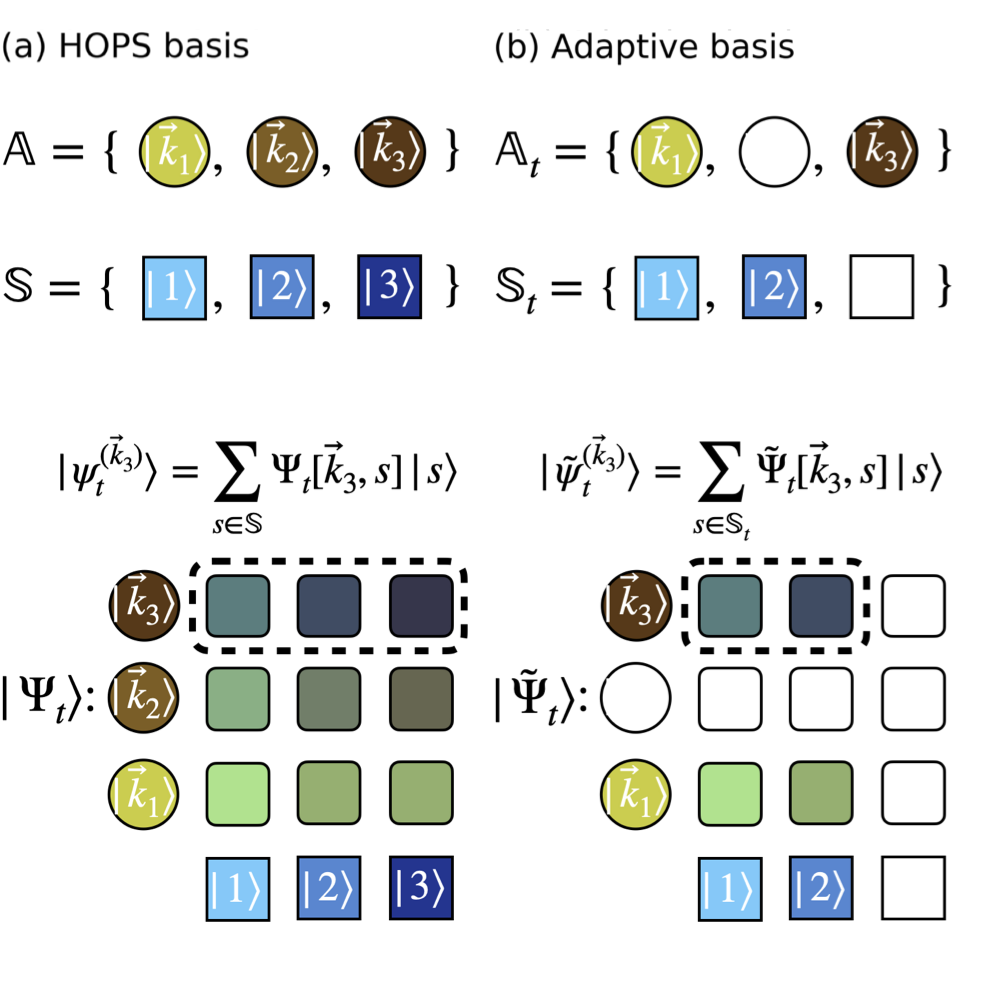}
    \caption{
    The HOPS basis. (a) The state basis $\mathbb{S}$ consists of all system states $\{|1\rangle, |2\rangle...\}$ (squares, shades of blue). The auxiliary basis $\mathbb{A}$ consists of all auxiliary vectors $\{|\vec{k}_1\rangle, |\vec{k}_2\rangle...\}$ (circles, shades of yellow). The entries of the HOPS wave function $|\Psi_t\rangle$ exist in the direct product basis $\mathbb{A}\otimes\mathbb{S}$, corresponding to $|\vec{k},s\rangle$ (rounded squares in various shades). The coefficients corresponding to an auxiliary wave function ($|\psi^{(\vec{k})}_t\rangle$) form a row inside the HOPS wave function. (b) The adaptive bases $\mathbb{S}_t$ and $\mathbb{A}_t$ are subsets of $\mathbb{S}$ and $\mathbb{A}$ formed at each time point, and $|\tilde{\Psi}_t\rangle$ is the HOPS wave function projected into the reduced basis $\mathbb{A}_t\otimes\mathbb{S}_t$.}
    \label{fig:hops_basis}
\end{figure}

We can rewrite the HOPS equation in a super-operator notation
\begin{equation}
    \label{eq:HOPS_superoperator}
    \frac{d\ket{\Psi_t}}{dt} = \mathcal{L}\ket{\Psi_t}
\end{equation}
using the HOPS wave function
\begin{equation}
    \ket{\Psi_t} = \sum_{\vec{k}\in\mathbb{A}} \sum_{s \in \mathbb{S}} \psi_t^{(\vec{k})}[s] \ket{\vec{k},s}
\end{equation}
which contains the coefficients associated with all auxiliary wave functions and exists in the direct product basis $\mathbb{A} \otimes \mathbb{S}$ (the HOPS basis), where the state basis $(\mathbb{S}=\{|s\rangle\})$ spans the Hilbert space of the system and the auxiliary basis $(\mathbb{A}=\{|\vec{k}\rangle\})$ is comprised of all auxiliary vectors (Fig. \ref{fig:hops_basis}a). For notational convenience, we express the amplitude associated with $\ket{\vec{k}, s}$ in the HOPS wave function using matrix notation as $\Psi_t[\vec{k},s]=\psi_t^{(\vec{k})}[s]$. The time-evolution super-operator $\mathcal{L}[\vec{k}', s', \vec{k}, s]$ defines the component of the HOPS equation (Eq. \eqref{eq:NormNonLinearHops}) that connects amplitude in $\psi_t^{(\vec{k})}[s]\ket{\vec{k},s}=\Psi_t[\vec{k},s]\ket{\vec{k},s}$ to the time-evolution of the amplitude of $\psi_t^{(\vec{k}')}[s']\ket{\vec{k}',s'}=\Psi_t[\vec{k}',s']\ket{\vec{k}',s'}$.

The adaptive Hierarchy of Pure States (adHOPS) constructs a time-evolving reduced basis that takes advantage of dynamic localization. While the size of the HOPS basis scales rapidly with respect to system size, the interaction of the system and bath in an individual trajectory imposes localization in the basis of system states $\mathbb{S}$, with a corresponding localization in the basis of auxiliary vectors $\mathbb{A}$.\cite{varveloFormallyExactSimulations2021} At each time step of an adHOPS calculation, the adaptive algorithm constructs a subset of $\mathbb{S}$ and $\mathbb{A}$ (shown in Fig. \ref{fig:hops_basis}b), ensuring time-evolution with error no greater than a user-defined bound. By time-evolving the reduced forms of the HOPS wave function ($|\tilde{\Psi}_t\rangle$), adHOPS reduces the computational scaling of calculations while remaining formally exact. In a sufficiently extended aggregate, most states and auxiliary wave functions can be neglected at any given time point, and localization leads to $\mathcal{O}(1)$ scaling with respect to system size.\cite{varveloFormallyExactSimulations2021}

\subsection{The interpretation of HOPS Ensembles}
While stochastic unravellings of reduced density matrix equations-of-motion are often described as numerical conveniences,\cite{Molmer_optics_1993} quantum measurement theory provides a physical interpretation of the ensemble of physical wave functions ($\ket{\psi_t^{(\vec{0})}}$) at each time in a HOPS calculation.\cite{Gambetta_Non-Markovian_2002} We define a collection of bath measurement operators $\hat{M}_{\mathbf{a}} = \vert  \mathbf{0} \rangle\langle \mathbf{a}|$ where $\ket{\mathbf{0}}$ is the global ground state for all harmonic oscillators in all thermal environments, $\ket{\mathbf{a}} = \ket{a_0, ..., a_b,...,a_B}$, and $\ket{a_b}$ is a (non-normalized) coherent state which can be written in terms of harmonic oscillator number states ($\vert \eta_b\rangle$) as
\begin{equation}
    \vert a_b\rangle = \frac{1}{\sqrt{\pi}}e^{-\vert a_b\vert^2/2}\sum_{\eta_b}\frac{a_{b}^{\eta_b}}{\sqrt{\eta_b!}}\vert \eta_b\rangle.
\end{equation}
After the measurement, the total wave function is given by 
\begin{equation}
    \vert \Phi_{\mathbf{a}}(t)\rangle = \frac{1}{\sqrt{P(\mathbf{a},t)}} \hat{M}_{\mathbf{a}}\vert \Phi(t)\rangle = \vert \mathbf{0}\rangle \otimes \vert \psi_{\mathbf{a}}(t)\rangle 
\end{equation}
where 
\begin{equation}
    P(\mathbf{a},t) = \langle \Phi(t) \vert \hat{M}^\dagger_{\mathbf{a}} \hat{M}_{\mathbf{a}} \vert \Phi(t) \rangle
\end{equation}
is the probability of measuring the bath in state $\ket{\mathbf{a}}$ at time $t$ and 
\begin{equation}
    \vert \psi_{\mathbf{a}}(t) \rangle = \frac{1}{\sqrt{P(\mathbf{a},t)}} \langle \mathbf{a}|\Phi(t)\rangle
\end{equation}
is the conditional system wave function given the measurement on the bath state returns the value $\mathbf{a}$. Consistent with the physical picture originally proposed by Díosi and Strunz\cite{NMQSD} and the re-derivation in terms of quantum measurement theory by Gambetta and Wiseman,\cite{Gambetta_Non-Markovian_2002} NMQSD (and thus HOPS) calculates the time-evolution of the conditional system wave function ($\vert \psi_{\mathbf{a}}(t) \rangle$) when an environmental measurement at time $t$ is performed in the coherent state basis. Thus, the ensemble of physical wave functions constructed through the HOPS equations at each time are quantum-mechanical observables. 

There are two levels of ambiguity associated with the ensemble of physical wave functions described here. First, a different selection of measurement operators on the environment would correspond to a different equation-of-motion, yielding a different ensemble of physical wave functions at each time point. This is an inherent ambiguity in projecting out the system component of an entangled system-bath wave function and is best resolved by the selection of a specific measurement protocol. Second, in mapping the generic molecular aggregate Hamiltonian onto the open quantum system Hamiltonian expressed in Eq. \eqref{eq:Ham_Full} we have constructed a fictitious bath of harmonic oscillators that reproduces the thermal correlation function. As a result, we are not propagating the true environment that is entangled with the system, but rather an analogous environment which shares essential features with the original Hamiltonian. Given these two levels of ambiguity, we approach the analysis of wave function ensembles calculated with HOPS (e.g., as presented in Sec. \ref{sec:ChargeSeparation}) as a quantification of the behavior of the analogous Hamiltonian (Eq. \eqref{eq:Ham_Full}) when the environment is measured in minimum-uncertainty states with classical-like time-evolution (i.e., the coherent state basis). 

Finally, we note that while the wave function ensembles are observables, the corresponding trajectories are not.\cite{wiseman2008pure} Since the physical wave function $\vert \psi^{(\vec{0})}_t\rangle$ at time $t$ is the system state conditioned on an all-in-one measurement of the environment, any subsequent time-evolution would be disturbed by this measurement. As a result, the individual trajectories cannot be interpreted as a continuous observation of a single process except in the Markovian limit where fast bath relaxation ensures the effect of the projective measurement on the environment does not influence subsequent time-evolution.

\subsection{Structure of the Hierarchy}

\begin{figure*}
    \centering
    \includegraphics{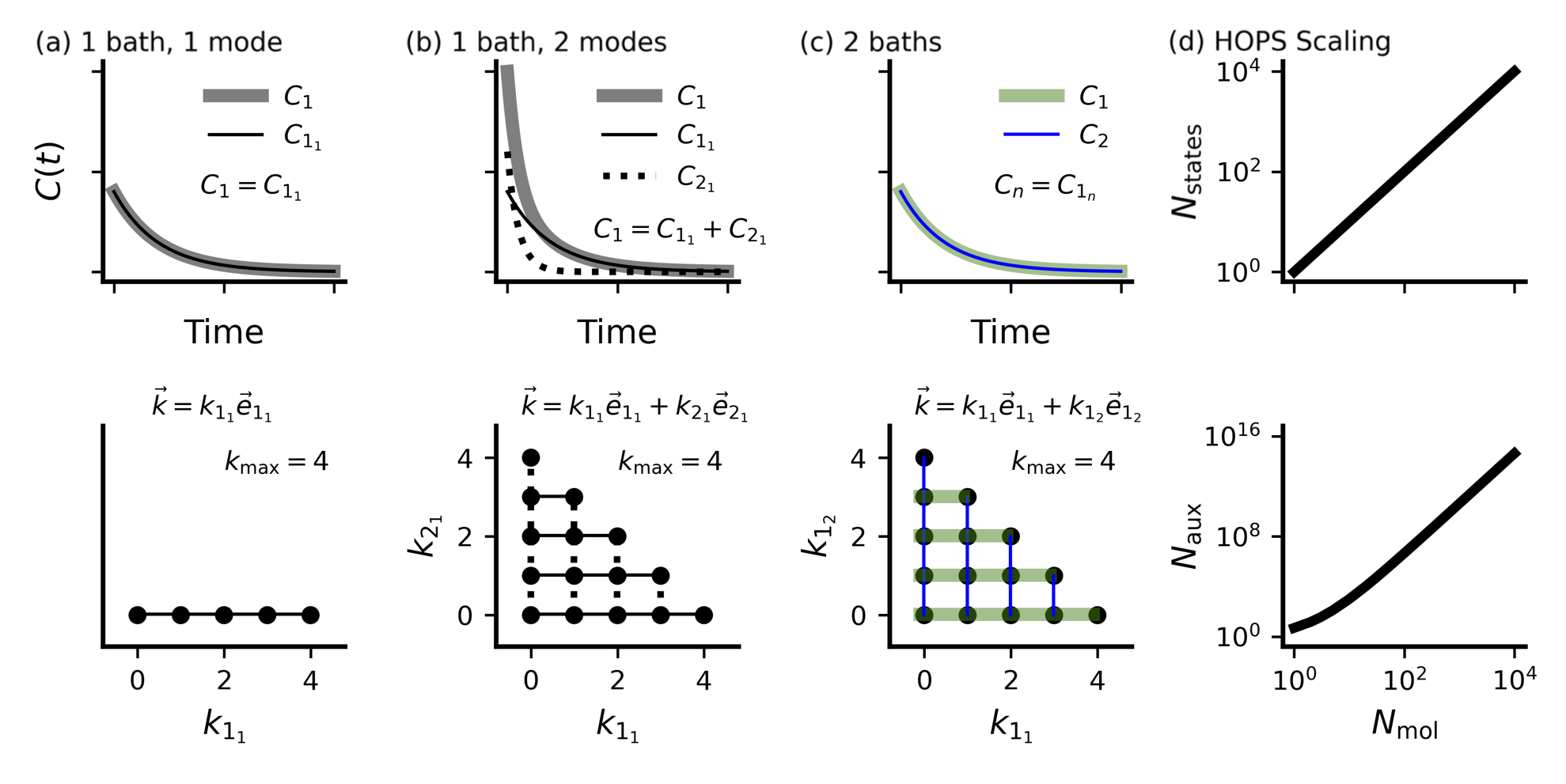}
    \caption{Structure of the hierarchy. For a system with a single thermal environment described by a correlation function with (a) $N_{\textrm{modes}}=1$  or (b) $N_{\textrm{modes}}=2$ exponential modes (top), the ensuing hierarchy is an $N_{\textrm{modes}}$-dimensional grid (bottom). (c) For a system with two independent thermal environments, each described by a correlation function with a single exponential mode (top), the ensuing hierarchy is an $(N_{\textrm{modes}}=2)$-dimensional grid (bottom). (d) Scaling of the state (top) and auxiliary (bottom) basis in a HOPS calculation of a singly-excited system where each molecule is coupled to a thermal environment described by a single-exponential correlation function. In all cases, $k_{\textrm{max}}=4$.}
    \label{fig:hier_structure}
\end{figure*}

The set of auxiliary vectors in the hierarchy form a grid of dimension $N_{\textrm{modes}}$, the total number of correlation function modes. Each auxiliary vector $\ket{\vec{k}}$ is described by its indices $\vec{k} = [k_{0_{0}},k_{1_0}\dots,k_{j_{0}},\dots k_{0_n},\dots k_{j_n},\dots k_{J_N}]$ where hierarchy depth $k_{j_n}$ is a non-negative integer associated with the $j^{\textrm{th}}$ exponential mode in the correlation function of bath $n$. To clarify the structure of the hierarchy and the indexing: Fig. \ref{fig:hier_structure}a shows the linear hierarchy arising from a single thermal environment with a single exponential correlation function mode, while Fig. \ref{fig:hier_structure}b  shows the two-dimensional hierarchy arising from a single thermal environment with two modes ($C_{1_1}$ and $C_{2_1}$). We note that a system with two thermal environments each with a single mode ($C_{1_1}$ and $C_{1_2}$) also exhibits a two-dimensional hierarchy (Fig. \ref{fig:hier_structure}c), though the underlying equation-of-motion is different because of the difference in the system-bath coupling operators ($\hat{L}$). 

To impose a finite size on the hierarchy, we select a maximum total depth ($k_{\textrm{max}}$) and truncate the hierarchy to only include  auxiliary vectors satisfying the triangular truncation condition $\sum_{n,j_n} k_{j_n} = \|\vec{k}\|_1 \leq k_{\textrm{max}}$ (visualized for $k_{\textrm{max}}=4$ in Fig. \ref{fig:hier_structure}). The number of auxiliary wave functions in a HOPS calculation with a triangular truncation scheme scales as $\binom{k_{\textrm{max}}+N_{\textrm{modes}}}{k_{\textrm{max}}}$, where $N_{\textrm{modes}}$ is the number of modes among all thermal environments. As shown in Fig. \ref{fig:hier_structure}d, the basis of a HOPS calculation scales rapidly with system size ($\sim N_{\textrm{modes}}^{k_{\textrm{max}}}$), rendering HOPS simulations of large aggregates intractable. 

\begin{figure}
    \centering
    \includegraphics{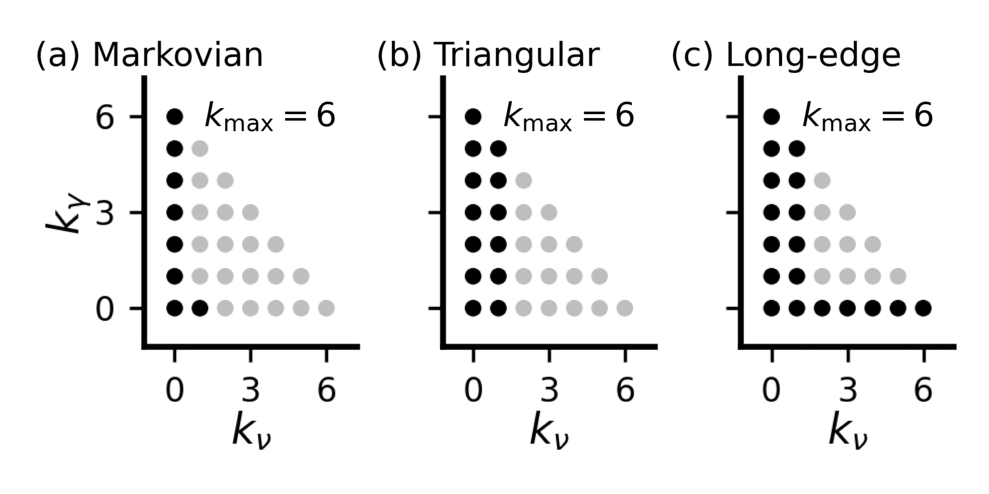}
    \caption{Static hierarchy filters. The change stemming from imposing the (a) Markovian filter, (b) triangular filter with $k_{\textrm{max}_2}=1$, and (c) long-edge filter with $k_{\textrm{depth}}=1$ on a hierarchy with filtered mode $\nu$ and unfiltered mode $\gamma$ for $k_{\textrm{max}}=6$. Auxiliary vectors that have been filtered out are represented with transparent circles. }
    \label{fig:static_filters}
\end{figure}

We can reduce the basis size of a HOPS calculation by modifying the triangular truncation condition $\|\vec{k}\|_1 \leq k_{\textrm{max}}$ with additional restrictions, which we refer to as static hierarchy filters. Generally speaking, these filters are appropriate when a subset of modes $\{\nu_n\}$ are characterized by faster exponential decay timescales than other modes. While a variety of static hierarchy filters have been considered previously,\cite{ zhangFlexibleSchemeTruncate2018} our code implements three: 
\begin{enumerate}
    \item \textbf{Markovian Filter (Fig. \ref{fig:static_filters}a)}: This filter restricts the associated modes ($\{\nu_n\}$) to auxiliary vectors of depth 1. We introduce additional truncation condition for $\{\nu_n\}$: $|\vec{k}\rangle \in \mathbb{A}$  if $\sum_{n,\nu_n} k_{\nu_n} = 0$ or $\vec{k} = \vec{e}_{\nu_n}$. 

    \item \textbf{Triangular Filter (Fig. \ref{fig:static_filters}b)}: This filter adds a secondary triangular truncation condition. We define $ k_{\textrm{max}_2} < k_{\textrm{max}}$ and introduce additional truncation condition for $\{\nu_n\}$: $|\vec{k}\rangle\in \mathbb{A}$  if $\sum_{n,\nu_n}k_{\nu_n} \leq k_{\textrm{max}_2}$.

    \item \textbf{Long-Edge Filter (Fig. \ref{fig:static_filters}c)}: This filter adds a secondary triangular truncation condition with the exception that the auxiliary vectors along the edges of the hierarchy ($\vec{k} = \|\vec{k}\|_1\vec{e}_{\nu_n}$) are always kept. We define $k_{\textrm{depth}} < k_{\textrm{max}}$ and introduce additional truncation condition for $\{\nu_n\}$: $|\vec{k}\rangle \in \mathbb{A}$ if  $\sum_{n,\nu_n}k_{\nu_n} \leq k_{\textrm{depth}}$ or $\vec{k} = \|\vec{k}\|_1\vec{e}_{j_n}$.

\end{enumerate}
All of these filters may be used in conjunction with adHOPS, but only the Markovian filter is incorporated within the adaptive algorithm. The triangular and long-edge filter are applied only after the adaptive basis is constructed, deteriorating the quality of the error estimates and increasing the size of the adaptive basis required to satisfy the user-specified error bound.

\subsection{Constructing Correlated Noise}
\label{sec:noise_construction}
To generate a sample of the noise, we employ a variation of the circulant embedding method for complex-valued correlation functions described in detail in Ref. \onlinecite{StocProcSampling}. The circulant embedding method is based on the fact that a mean-zero stationary Gaussian process $\mathbf{Z}$ with covariance matrix $\mathbf{R}$ can be represented as $\mathbf{Z} = \mathbf{BY}$, where $\mathbf{Y}$ is a vector of complex Gaussian random numbers and $\mathbf{R} = \mathbf{BB}^{T}$.  We extend $\mathbf{R}$ to a circulant matrix $\mathbf{R}^{\textrm{ext}}$ (see Ref. \onlinecite{StocProcSampling} ) with the spectral decomposition $\mathbf{R}^{\textrm{ext}} = \mathbf{Q}\Sigma \mathbf{Q}$, where $\mathbf{Q}$ is the Fourier matrix and $\Sigma$ is a diagonal matrix composed of the eigenvalues of $\mathbf{R}^{\textrm{ext}}$ calculated as the Fourier transform of the first column of $\mathbf{R}^{\textrm{ext}}$.    Consequently, we construct a matrix $\mathbf{B}^{\textrm{ext}} = \mathbf{Q}\Sigma^{1/2}$ and truncate to find $\mathbf{B}$.  This is faster than using Cholesky decomposition to compute $\mathbf{B}$ directly.\cite{StocProcSampling} 

We begin by sampling the correlation functions of each bath ($n$) at $N_{t}$ evenly-spaced times up to time $T_{\textrm{noise}}$,  $t_{m} = m\frac{T_{\textrm{noise}}}{N_{t}-1}$, $m = 0,1,\dots,N_{t} - 1$, to obtain the vectors $\mathbf{C}_n$ with $C_n[m] = C_{n}(t_{m})$.  We rotate each $\mathbf{C}_n$ in the complex plane such that it is real-valued at the final time-point and unchanged in the first time-point, obtaining a rotated vector $\mathbf{\tilde{C}}_n$ with $\tilde{C}_n[m] = C_{n}[m]e^{-i2\pi\phi_{n} m}$, where $\phi_{n} = \textrm{phase}(C_{n}[N_{t} - 1])/(2(N_{t}-1)\pi)$. Next, for each $n$, we embed $\mathbf{\tilde{C}}_{n}$ into a vector $\mathbf{\tilde{C}}_{n}^{\textrm{ext}}$ of size $2N_t - 2$, where $\tilde{C}_n^{\textrm{ext}}[2N_{t} - 2 - m] = \tilde{C}_{n}[m]^{*}$, $1 \leq m \leq N_{t}-2$. The vector $\tilde{\mathbf{C}}_{n}^{\textrm{ext}}$ corresponds to the first column of $\mathbf{R}^{\textrm{ext}}$,
allowing us to construct the eigenvalues of the correlation function, denoted $\Sigma_{n}[k]$, $k = 0,1,\dots,2N_{t}-3$, by taking the Fast Fourier Transform of $\mathbf{\tilde{C}}_{n}^{\textrm{ext}}$.

The set of $\Sigma_n[k]$ may contain negative eigenvalues as an artifact of the embedding, which is inconsistent with the fact that covariance matrices are non-negative definite. This can be avoided by increasing the number of samples $N_{t}$, but the value of $N_t$ required to ensure that the non-negative definite property is maintained after performing the embedding for general correlation functions is difficult to predict.\cite{StocProcSampling, Circulant_Embedding_Analysis} Additionally, it is sometimes convenient to use non-physical correlation functions which are not positive definite, such as the high-temperature limit of the Drude-Lorentz spectral density.
However, even in these cases, we can obtain a sampling corresponding to a physical correlation function by setting all negative eigenvalues to $0$. 

With the eigenvalues of the correlation function matrix determined, we define the sequence $W_n[k] = Z_n[k]\sqrt{\Sigma_n[k]/2}$ for $k=0,1,\dots,2N_{t}-3$ where $\mathbf{Z}_n$ are Fourier transformed vectors of uncorrelated noise sampled from a complex normal distribution with mean $0$ and variance $2$.  The sampled process $\tilde{z}_{n,t}$ is the inverse Fourier transform of vector $\mathbf{W}_n$ sampled at $t_{m}$ for $m = 0,1,\dots,2N_{t}-3$ and then truncated to include only the first $N_{t}$ terms.  Finally, we rotate the sample back to obtain the noise sample, $z_{n,t} = \tilde{z}_{n,t}e^{i2\pi \phi_{n} m}$.

\subsection{Low-Temperature Correction}
\label{main:LTC}
Here we introduce a low-temperature correction for HOPS where correlation function modes with a large decay frequency, such as the series of Matsubara modes that influence dynamics at low temperatures,\cite{Matsubara_Takeo_not_Miki_1955} are approximated as delta functions. Our approach is analogous to previous low-temperature corrections for methods such as HEOM, Fokker-Planck and Smoluchowski Equations, and Stochastic Schrodinger Equations.\cite{ishizaki_LTC_deriv, ikeda_tanimura_fokker-planck, Li_SSE_LTC}

A correlation function mode $\nu_n$ that decays on a timescale significantly faster than any other dynamics of the physical wave function can be approximated by a delta function
\begin{equation}
    C_{\nu_n}(t) = g_{\nu_n}e^{-\gamma_{\nu_n}t/\hslash} \approx \frac{\hslash g_{\nu_n}}{\gamma_{\nu_n}}\delta(t).
\end{equation}
The noise memory drift term associated with an ultrafast mode ($\xi_{\nu_n,t}$) is then given by
\begin{equation}
\label{eq:noise_mem_delta}
    \xi_{\nu_n,t} \approx \frac{1}{\hslash}\int_{0}^{t} ds \left(\frac{\hslash g_{\nu_n}}{\gamma_{\nu_n}}\right)^*\delta(t-s)\braket{\hat{L}^{\dagger}_{n}}_s = \left(\frac{g_{\nu_n}}{\gamma_{\nu_n}}\right)^*\braket{\hat{L}^{\dagger}_{n}}_t.
\end{equation}
We find the first order auxiliary wave function $|\psi^{(\vec{e}_{\nu_n})}_t\rangle$ via the terminator approximation, \cite{HOPS} 
\begin{equation}
\label{eq:terminator}
    |\psi_t^{(\vec{k})}\rangle \approx \sum_{n,{j_n}}\frac{k_{j_n}\gamma_{j_n}}{\vec{k}\cdot \vec{\gamma}}\hat{L}_n|\psi^{(\vec{k} - \vec{e}_{j_n})}_t\rangle \rightarrow |\psi^{(\vec{e}_{\nu_n})}_t\rangle \approx \hat{L}_n |\psi^{(\vec{0})}_t\rangle
\end{equation}
which becomes exact in the limit that $\gamma_{\nu_n}$ goes to infinity.
%$\vec{k}\cdot\vec{\gamma}\rightarrow \infty$. 
The low-temperature-corrected normalized nonlinear HOPS equation is
\begin{flalign}
\begin{aligned}
\label{eq:HOPS_LTC}
    \hslash \frac{ d\vert \psi^{(\Vec{k})}_t \rangle}{dt} 
=  \big(-i\hat{H}_S - \Vec{k} \cdot \Vec{\gamma} -\Gamma_t + \sum_{n} \hat{L}_{n} (z^*_{n,t}+ \sum_j\xi_{{j_n},t})\big)\vert \psi^{(\Vec{k})}_t \rangle &\\ 
+ \sum_n (\Xi_{n,t}\hat{L}_n - \delta_{\vec{k},\vec{0}}\hat{T}_{n,t} - \tilde{\Gamma}_{n,t}) \vert \psi^{(\Vec{k})}_t \rangle &\\
+ \sum_{n,{j_n}} k_{j_n} \gamma_{j_n} \hat{L}_{n}  \vert \psi^{(\Vec{k} -\Vec{e}_{j_n})}_t \rangle &\\
- \sum_{n,{j_n}} \Big(\frac{g_{j_n}}{\gamma_{j_n}}\Big)(\hat{L}^{\dagger}_{n} - \langle\hat{L}^{\dagger}_{n}\rangle_{t}) \vert \psi^{(\Vec{k}+\Vec{e}_{j_n})}_t\rangle &
\end{aligned}
\end{flalign}
where
\begin{equation}
    \Xi_{n,t} = \sum_{\nu_n} \xi_{\nu_n,t} \approx G_n^*\braket{\hat{L}^{\dagger}_{n}}_t  
\end{equation}
collects the low-temperature-corrected noise memory drift terms, 
\begin{equation}
\label{eq:sum_terminator1}
    \hat{T}_{n,t}|\psi^{(\vec{0})}\rangle = G_n(\hat{L}^{\dagger}_{n} - \langle\hat{L}^{\dagger}_{n}\rangle_{t})\hat{L}_n |\psi^{(\vec{0})}_t\rangle
\end{equation}
describes the influence of the terminator-approximated first-order auxiliary wave functions on the physical wave function, 
\begin{equation}
    \tilde{\Gamma}_{n,t} = \textrm{Re}[G_n]\big(2\langle\hat{L}^\dag_n\rangle_t\langle\hat{L}_n\rangle_t - \langle\hat{L}^{\dagger}_{n}\hat{L}_n\rangle_t\big)
\end{equation}
is the low-temperature correction to the normalization correction factor, 
and 
\begin{equation}
\label{eq:LTC_coeff}
    G_n = \sum_{\nu_n} \frac{g_{\nu_n}}{\gamma_{\nu_n}}.
\end{equation}
Thus, the low-temperature correction allows us to account for the effect of ultrafast modes without the explicit calculation of additional auxiliary wave functions or noise memory drift. See Appendix \ref{app:LowTempCorrection} for a detailed derivation and the low-temperature corrections of other HOPS equations-of-motion and Appendix \ref{app:LTC_test} for sample calculations demonstrating the effectiveness of the low-temperature correction.

\subsection{Effective Integration of the Noise}

To account for fast fluctuations in the noise $\mathbf{z}_{t}$ without inconveniently small time steps, we introduce an effective integration of the noise. When there exists a separation of timescales between the stochastic noise and all other terms in the equation-of-motion (e.g.,  due to the low-temperature correction), then by taking a moving average over the noise, we can effectively integrate its dynamics on a finer timescale and converge calculations with a larger time step.

The time-evolution super-operator $\mathcal{L}$ (Eq. \eqref{eq:HOPS_superoperator}) is the sum of $\mathcal{L}_0$, which is not explicitly dependent on noise, and $\mathcal{L}_{z_{t}}$, the explicitly noise-dependent component of the time-evolution.
When the fluctuations of the noise ($\mathbf{z}_t$) are faster than all timescales of both the system dynamics ($\hat{H}_S$) and flux between auxiliary wave functions, the short-time dynamics of a HOPS trajectory are dominated by $\mathcal{L}_{z_{t}}$.
To capture the influence of rapid fluctuations in the noise, 
we sub-divide the time step of integration $(\Delta t)$ into $N_\tau$ finer time steps $(\tau)$ and replace $\mathcal{L}_{z_{t}}$ with a moving average, $\bar{\mathcal{L}}_{z_{t}}$, defined by its effect on each auxiliary wave function
\begin{equation}
    \label{eq:eff_int_main_text}
    \bar{\mathcal{L}}_{z_{t}} |\psi_t^{(\vec{k})}\rangle = \frac{1}{\hslash}\big(-\Gamma'_t + \sum_{n}(z'^{*}_{n,t} + \sum_{j_n}\xi_{j_n,t})\hat{L}_n\big)|\psi_t^{(\vec{k})}\rangle
\end{equation}
where
\begin{equation}
 z'^*_{n, t} = \frac{1}{N_\tau}\sum_{m=0}^{N_\tau-1}z^*_{n, t+m\tau}
\end{equation}
and
\begin{flalign}
\begin{aligned}
    \Gamma'_t = &\sum_{n} \braket{\hat{L}_{n}}_{t} \textrm{Re}[z'^*_{n,t}+ \sum_{j_n}\xi_{{j_n},t}] \\
    - &\sum_{n, j_n} \textrm{Re}[\braket{\psi^{(\Vec{0})}_{t} |\hat{L}^{\dagger}_{N}| \psi^{(\Vec{e}_{j_n})}_{t}}] \\
    + &\sum_{n, j_n} \braket{\hat{L}^{\dagger}_n}_{t} \textrm{Re}[\braket{\psi^{(\Vec{0})}_{t} | \psi^{(\Vec{e}_{j_n})}_{t}}].
\end{aligned}
\end{flalign}

In Appendix \ref{app:EffectNoiseIntegration}, we prove that the moving average of the noise is equivalent to assuming a separation of timescales and performing an integration that uses a finer time step to integrate the noise-dependent dynamics (associated with $\mathcal{L}_{z_{t}}$) than the remainder of the equation-of-motion ($\mathcal{L}_0$). In Appendix \ref{app:eff_int_test}, we provide sample calculations that demonstrate significant reduction in error associated with large integration time steps under the effective integration of the noise. To avoid generating noise that will not be used in our calculations, we set the effective integration time step $\tau$ to the noise sampling time step (see Sec. \ref{sec:noise_construction}).

\section{Adaptive Basis Construction}
\label{sec:adaptive_basis}
The core of adaptive HOPS (adHOPS) is the algorithm for constructing the adaptive basis that maintains a formally exact equation-of-motion while enabling size-invariant scaling. Our approach to constructing the adaptive basis prioritizes ensuring that the corresponding time-evolution satisfies a user-specified error bound at each time point. The challenge, then, is to ensure the error bound at each time point without constructing the full basis (or anything scaling with the full basis). Ref. \onlinecite{varveloFormallyExactSimulations2021} provided the first derivation of a size-invariant scaling algorithm for the adaptive basis construction, restricted to the case where system-bath coupling operators ($\hat{L}$) are site-projection operators. Our derivation here provides a more condensed notation and tighter error bounds; we also generalize the formulation to allow for arbitrary diagonal system-bath coupling operators, extending the adHOPS algorithm to describe additional physical processes including multi-particle dynamics. 

\subsection{Overview}
Given the HOPS wave function $|\Psi_{t}\rangle$, consisting of the physical and all auxiliary wave functions, we seek an approximate HOPS wave function $|\tilde{\Psi}_{t}\rangle$ in a reduced basis that satisfies an upper bound on the derivative error,
\begin{equation}
    E = \left \Vert \frac{d|\Psi_{t}\rangle}{dt} - \frac{d\tilde{|\Psi}_{t}\rangle}{dt} \right \Vert_{2} = \sqrt{\sum_{\vec{k}, s} \left \vert \frac{d\Psi_{t}[\vec{k},s]}{dt}- \frac{d\tilde{\Psi}_{t}[\vec{k},s]}{dt} \right\vert^2}.
\end{equation}
The derivative error ($E$) is local to each time step, and, for sufficiently small error bound ($\delta$), we find the approximate trajectory reproduces the dynamics calculated with the full basis. The formal error bound on an adHOPS trajectory is given by $t\delta$ at time $t$; in practice, however, the error lies below the bound, exhibiting a rapid growth during early (coherent) dynamics often followed by much slower growth. We treat the derivative error bound as a phenomenological convergence parameter that depends on both the Hamiltonian parameters and observable of interest. \cite{varveloFormallyExactSimulations2021,varvelo2023formally,gera_simulating_2023}

We determine the basis for time-evolving $|\tilde{\Psi}_{t}\rangle$ by minimizing both the number of auxiliary vectors in the hierarchy and the size of the state space at time $t$. $|\tilde{\Psi}_{t}\rangle$ exists in the direct product basis $\mathbb{A}_{t} \bigotimes \mathbb{S}_{t}$, where $\mathbb{A}_{t}\subseteq \mathbb{A}$ and $\mathbb{S}_{t} \subseteq \mathbb{S}$ are, respectively, the subset of auxiliary vectors and states included in the basis at time $t$ as determined during the previous step of the time-evolution (time-evolving $t - \Delta t \rightarrow t$). When constructing the basis in which the derivative of $|\tilde{\Psi}_{t}\rangle$ will be determined to time-evolve to $|\tilde{\Psi}_{t+\Delta t}\rangle$, we first construct the auxiliary basis $\mathbb{A}_{t + \Delta t}$, then the state basis $\mathbb{S}_{t + \Delta t}$, minimizing $\mathbb{A}_{t + \Delta t}$ and $\mathbb{S}_{t + \Delta t}$ separately, rather than guaranteeing a minimal combined basis $\mathbb{A}_{t + \Delta t}\bigotimes\mathbb{S}_{t + \Delta t}$. We determine bases $\mathbb{A}_{t + \Delta t}$ and $\mathbb{S}_{t + \Delta t}$ as reduced sets that satisfy derivative error bounds $\delta_A$ and $\delta_S$, such that 
\begin{equation}\label{eq:stepbystepbound1}
    \left \Vert \frac{d|\tilde{\Psi}_{t}\rangle}{dt} - \frac{d|\hat{\Psi}_{t}\rangle}{dt} \right \Vert_{2} \leq \delta = \sqrt{\delta_A^2 + \delta_S^2}
\end{equation}
where $|\hat{\Psi}_{t}\rangle$ is $|\tilde{\Psi}\rangle_t$ projected into basis $\mathbb{A}_{t + \Delta t} \bigotimes \mathbb{S}_{t + \Delta t}$.

Both the auxiliary and state basis constructions are broken into two steps. The first step of the auxiliary basis construction is to find the stable auxiliary basis $(\mathbb{A}_{t + \Delta t}^{s})$, the portion of $\mathbb{A}_{t + \Delta t}$ corresponding to the reduced subset of $\mathbb{A}_{t}$ that introduces error $E_{\mathbb{A}_{t}^s}$ no greater than $\delta_A/2$. The next step in the construction of $\mathbb{A}_{t + \Delta t}$ is finding the reduced set of boundary auxiliary vectors ($|\vec{k}_b\rangle \in \mathbb{A}\setminus\mathbb{A}_{t}$)\bibnote{Here we use standard set notation for set subtraction where $\mathbb{A}\setminus \mathbb{A}_t$ represents the set of elements in $\mathbb{A}$ and not in $\mathbb{A}_t$.} satisfying error bound $\delta_A' = \sqrt{\delta_A^2-E^2_{\mathbb{A}_{t}^s}}$, guaranteeing that the total error associated with the construction of $\mathbb{A}_{t+\Delta t}$ is bounded by $\delta_A$. The state basis is constructed in exactly the same manner as the auxiliary basis, with the stipulation that we consider only the flux into auxiliary wave functions in the basis $\mathbb{A}_{t + \Delta t}$ found above, since the error associated with all other flux terms have already been accounted for.  The state basis is given its own error tolerance $\delta_{S}$, such that total derivative error associated with constructing the new basis satisfies error bound $\delta = \sqrt{\delta_A^2 + \delta_S^2}$. While the first adHOPS calculations reported in Ref. \onlinecite{varveloFormallyExactSimulations2021} used the same error bound for the auxiliary and state basis, subsequent work in Ref. \onlinecite{varvelo2023formally} demonstrated the utility of treating the state and auxiliary error bounds as independent convergence parameters. 

\subsection{Notation}
In the following section, we find it convenient to express flux between elements $\Psi_t[\vec{k}, s]|\vec{k},s\rangle$ in terms of auxiliary wave functions $|\psi^{(\vec{k})}_t\rangle = \sum_{s\in \mathbb{S}_t} \Psi_t[\vec{k},s]|s\rangle$ to consolidate all amplitudes associated with each auxiliary vector $|\vec{k}\rangle$. When $|\vec{k}\rangle$ is not in the basis $\mathbb{A}_t$, the associated auxiliary wave function $|\psi^{(\vec{k})}_t\rangle$ is $0$ and both $|\vec{k}\rangle$  and $|\psi^{(\vec{k})}_t\rangle$ are said to be unpopulated. For the same reason, we define state wave function $|\phi^{(s)}_t\rangle=\sum_{\vec{k}\in \mathbb{A}_t} \Psi_t[\vec{k},s]|\vec{k}\rangle$, which is $0$ when $|s\rangle$ is not in the basis  $\mathbb{S}_t$. 

In the super-operator notation of Eq. \eqref{eq:HOPS_superoperator}, the equation-of-motion may be written as
\begin{flalign}
\begin{aligned}\label{hopssuperoperator}
    \frac{d\Psi_t[\vec{k},s]}{dt} = \sum_{s' \in \mathbb{S}_{t}}\Big(&\mathcal{L}[\vec{k},s, \vec{k},s']\Psi_t[\vec{k},s']\\ 
    &+ \sum_{m} \mathcal{L}[\vec{k},s, \vec{k}+\vec{e}_m,s']\Psi_t[\vec{k}+\vec{e}_m,s']\\
    &+ \sum_m  \mathcal{L}[\vec{k},s, \vec{k}-\vec{e}_m,s']\Psi_t[\vec{k}-\vec{e}_m,s'] \Big)   
\end{aligned}
\end{flalign}
where index $\vec{k}$ refers to an auxiliary vector, indices $s$ and $s'$ refer to states in the state basis, and index $m$ refers to a correlation function mode associated with system-bath projection operator $\hat{L}_m$ (which is more convenient here than indexing by independent environment $n$).

\subsection{The Adaptive Algorithm}
Assuming that the system-bath projection operators are diagonal, we can write Eq. (\ref{hopssuperoperator}) as
\begin{flalign}
\begin{aligned}\label{hopsdiagonalL}
    \frac{d\Psi_t[\vec{k},s]}{dt} =&\sum_{s'}\mathcal{L}[\vec{k},s, \vec{k},s']\Psi_t[\vec{k},s']\\ 
    + &\sum_{m} \mathcal{L}[\vec{k},s, \vec{k}+\vec{e}_m,s]\Psi_t[\vec{k}+\vec{e}_m,s]\\
    + &\sum_{m}  \mathcal{L}[\vec{k},s, \vec{k}-\vec{e}_m,s]\Psi_t[\vec{k}-\vec{e}_m,s]
\end{aligned}
\end{flalign}
where the first term on the right-hand side represents transitions between states, and the second and third terms on the right-hand side represent fluxes from auxiliary wave functions one step higher or lower in the hierarchy, respectively.

The calculation of an auxiliary basis that satisfies Eq. (\ref{eq:stepbystepbound1}) is confounded by the presence of flux terms between previously-populated auxiliary vectors removed during the construction of $\mathbb{A}_{t+\Delta t}$.  Determining a minimum auxiliary basis that satisfies the derivative error bound $\delta_A$ would involve calculating the derivative error associated with each combination of deleted auxiliary vectors, which is computationally impractical.  Thus, we calculate the derivative error associated with neglecting each auxiliary vector individually. We calculate the derivative error associated with neglecting individual states in the same manner. We define $E_{\mathbb{A}_{t}^s}[\vec{k}]$ and $E_{\mathbb{S}^s_t}[s]$ as the derivative errors arising from removing auxiliary vector $|\vec{k}\rangle\in\mathbb{A}_t$ and state $s\in\mathbb{S}_t$ when integrating from $t \rightarrow t + \Delta t$. The derivative errors from continuing to neglect auxiliary vector $|\vec{k}_b\rangle \in \mathbb{A} \setminus \mathbb{A}_{t}$ and state $|s_b\rangle \in \mathbb{S} \setminus \mathbb{S}_{t}$ are $E_{\mathbb{A}_t^b}[\vec{k}_{b}]$ and $E_{\mathbb{S}_t^b}[s_{b}]$, respectively. We find an upper bound on the derivative error of Eq. (\ref{eq:stepbystepbound1}):
\begin{equation}\label{errorboundtotal}
\begin{split}
    \left \Vert \frac{d|\tilde{\Psi}\rangle}{dt} - \frac{d|\hat{\Psi}\rangle}{dt} \right \Vert_{2}^2 \leq &\sum_{\vec{k} \in \mathbb{A}_{t} \setminus \mathbb{A}_{t+\Delta t}}E^{2}_{\mathbb{A}_{t}^s}[\vec{k}]  + \sum_{\vec{k}_{b} \in \mathbb{A} \setminus (\mathbb{A}_{t} \cup \mathbb{A}_{t+\Delta t})}E^{2}_{\mathbb{A}_t^b}[\vec{k}_{b}] \\
    + &\sum_{s \in \mathbb{S}_{t} \setminus \mathbb{S}_{t+\Delta t}} 
    E^{2}_{\mathbb{S}_{t}^s}[s] + 
    \sum_{s_{b} \in \mathbb{S}\setminus (\mathbb{S}_t \cup \mathbb{S}_{t+\Delta t})}E^{2}_{\mathbb{S}_t^b}[s_{b}]
\end{split}
\end{equation}
which in turn satisfies derivative error bound $\delta^2 = \delta_{S}^2 + \delta_{A}^2$.

For each of these four groups of errors, we find a reduced basis by removing the maximum number of elements guaranteed not to violate the associated error bound. To do this, we sort the errors of a given group from smallest to largest and discard as many elements as possible, in ascending order of introduced error, while satisfying the error bound.

\subsubsection{Adaptive Auxiliary Basis}
The derivative error introduced by excluding $|\vec{k}\rangle\in\mathbb{A}$ from the auxiliary basis $\mathbb{A}_{t+\Delta t}$ is given by
\begin{flalign}
\begin{aligned}\label{GeneralAuxiliaryRemoveError}
E^{2}_{\mathbb{A}_t}[\vec{k}] &= \sum_{s \in \mathbb{S}_{t}}\Bigg(\left \vert \left(\mathcal{L}_{\mathbb{A}_t\otimes\mathbb{S}_t} \Psi_t\right)[\vec{k},s] + \frac{\Psi_{t}[\vec{k},s]}{\Delta t}\right \vert^{2}\\
&+ \sum_{m} \left \vert \mathcal{L}[\vec{k}+\vec{e}_{m},s, \vec{k},s]\Psi_{t}[\vec{k},s] \right\vert^{2}\\
&+ \sum_m\left \vert  \mathcal{L}[\vec{k} - \vec{e}_{m},s, \vec{k},s]\Psi_{t}[\vec{k},s]\right \vert^{2}\Bigg )\\
&+ \sum_{s_b \in \mathbb{S} \setminus \mathbb{S}_{t}} \left \vert \sum_{s \in \mathbb{S}_{t}}\mathcal{L}[\vec{k},s_b,\vec{k},s]\Psi_{t}[\vec{k},s]\right \vert^{2} .
\end{aligned}
\end{flalign}
The first term is the squared error arising from the flux into $|\psi^{(\vec{k})}_t\rangle$ constructed as the sum of the flux when the equation-of-motion is restricted to the basis $\mathbb{A}_t\otimes\mathbb{S}_t$ ($ \mathcal{L}_{\mathbb{A}_t\otimes\mathbb{S}_t} |\Psi_t\rangle$) and the deletion flux ($\Psi_{t}[\vec{k},s]/\Delta t$) that is implicitly added to the time-evolution to account for $|\psi^{(\vec{k})}_t\rangle \rightarrow 0$ when $|\vec{k}\rangle$ is removed from the basis (where $\Delta t$ is the integration time step). The second and third terms are the squared errors arising from fluxes up and down, respectively, into neighboring auxiliary wave functions $|\psi^{(\vec{k}')}_t\rangle$ for $|\vec{k}'\rangle=|\vec{k}\pm\vec{e}_m\rangle \in \mathbb{A}$. Finally, the fourth term only includes contribution from populated states $\ket{s}$ to unpopulated states $\ket{s_b}$ in the auxiliary wave function $|\psi_t^{(\vec{k})}\rangle$ arising from couplings within the system Hamiltonian (i.e., $\ket{\vec{k},s} \rightarrow \ket{\vec{k},s_b}$ when $\ket{s} \in \mathbb{S}_t$ and $\ket{s_b} \in \mathbb{S}\setminus\mathbb{S}_t$), because all other fluxes of this form are either 0 or were included in the first term. We present an explicit form of $E^2_{\mathbb{A}_t}[\vec{k}]$ in Appendix \ref{app:explicit_adap_aux}.

\subsubsection{Adaptive State Basis}
The derivative error for excluding $|s\rangle \in \mathbb{S}$ from $\mathbb{S}_{t+\Delta t}$ is the sum of errors associated with fluxes into and out of the associated state wave function $|\phi^{(s)}_t\rangle=\sum_{\vec{k}} \Psi_t[\vec{k},s]|\vec{k}\rangle$.  Since we use the auxiliary basis $\mathbb{A}_{t + \Delta t}$ during the construction of the state basis, it is convenient to define two kinds of basis elements: first, the stable auxiliary vectors ($|\vec{k}\rangle \in \mathbb{A}_{t + \Delta t}^{s} = \mathbb{A}_t \cap \mathbb{A}_{t+\Delta t}$) consisting of all auxiliary vectors that carried over from $\mathbb{A}_{t}$, and, second, the boundary auxiliary vectors newly added to the basis ($|\vec{k}_b\rangle \in\mathbb{A}_{t + \Delta t}^{b} = \mathbb{A}_{t+\Delta t} \setminus \mathbb{A}_t$). Taking into account this partitioning, the squared derivative error associated with excluding $|s\rangle$ from $\mathbb{S}_{t+\Delta t}$ is given as
\begin{equation}
\begin{aligned}\label{GeneralStateRemoveError}
E^{2}_{\mathbb{S}_t}[s] = \sum_{\vec{k} \in \mathbb{A}_{t + \Delta  t}^{s}}\Bigg{(} &\left\vert (\mathcal{L}_{\mathbb{A}_{t+\Delta t}^s\otimes\mathbb{S}_t} \Psi_t)[\vec{k},s] + \frac{\Psi_{t}[\vec{k},s]}{\Delta t}\right \vert^{2}\\
+ \sum_{m: \vec{k} + \vec{e}_{m} \in \mathbb{A}_{t + \Delta t}^{b}}&\left \vert \mathcal{L}[\vec{k} + \vec{e}_{m},s,\vec{k},s]\Psi_{t}[\vec{k},s] \right \vert^{2}\\
+ \sum_{m: \vec{k} - \vec{e}_{m} \in \mathbb{A}_{t + \Delta t}^{b}}&\left \vert \mathcal{L}[\vec{k} - \vec{e}_{m},s,\vec{k},s]\Psi_{t}[\vec{k},s] \right \vert^{2}
\\
+\sum_{s' \in \mathbb{S}\setminus \{s\}}&\left \vert \mathcal{L}[\vec{k},s',\vec{k},s]\Psi_{t}[\vec{k},s] \right\vert^{2}\Bigg{)}.
\end{aligned}
\end{equation}
The outer sum is restricted to $\vec{k} \in \mathbb{A}_{t + \Delta  t}^{s}$ because only stable auxiliary vectors can contribute non-zero flux. The first term contains the flux into $\ket{\phi_t^{(s)}}$ constructed as a sum over the flux when the equation-of-motion is restricted to the basis $\mathbb{A}_{t+\Delta t}^s\otimes\mathbb{S}_t$ ($ \mathcal{L}_{\mathbb{A}_{t+\Delta t}^s\otimes\mathbb{S}_t} |\Psi_t\rangle$) and the deletion flux ($\Psi_{t}[\vec{k},s]/\Delta t$) that is implicitly added to the time-evolution to account for $\ket{\phi_t^{(s)}} \rightarrow 0$ when $\ket{s}$ is removed from the basis (where $\Delta t$ is the integration time step). The second and third terms are the squared errors arising from fluxes up and down, respectively, restricted to flux into boundary auxiliary wave function $\ket{\psi_t^{(\vec{k}_b)}}$ for $\ket{\vec{k}_b} = \ket{\vec{k}\pm\vec{e}_m} \in \mathbb{A}_{t+\Delta t}^b$ because flux into the stable auxiliary wave functions is contained in the first term of Eq. \eqref{GeneralStateRemoveError} and the derivative error arising from neglecting flux into $|\psi^{(\vec{k'})}_t\rangle$ for $|\vec{k'}\rangle\in\mathbb{A}\setminus\mathbb{A}_{t+\Delta t}$ has already been accounted for during the calculation of $\mathbb{A}_{t+\Delta t}$. Finally, the fourth term accounts for all flux out of the state $|s\rangle$ due to couplings in the system Hamiltonian. We present an explicit form of $E^{2}_{\mathbb{S}_{t}}[s]$ in Appendix \ref{app:explicit_adap_state}.

\subsection{N-Particle Extension}
The error calculations above account for any system-bath couplings of the form
\begin{equation}
    \hat{L}_m = \sum_s \hat{L}_m[s,s] |s\rangle\langle s|
\end{equation}
which are diagonal in the state basis, as opposed to assuming each system-bath coupling operator is a site-projection operator ($\hat{L}_m = |s_m\rangle\langle s_m|$) as presented in Ref. \onlinecite{varveloFormallyExactSimulations2021}. 
Although the adaptive algorithm for arbitrary diagonal system-bath coupling operators is more computationally demanding, its increased flexibility allows adHOPS simulations to cover a broader range of physics required to describe multi-excitation processes such as charge separation and triplet fusion. 

\subsection{Adaptive Linear Absorption}
The adaptive construction of the linear absorption equation provides for a relative rather than absolute user-specified error bound. \cite{gera_simulating_2023} The HOPS linear absorption equation (Sec. \ref{app:LinearAbsorption}) is propagated using a modified nonlinear HOPS equation where the physical wave function is not normalized. As a result, we modify the adaptive algorithm in this case to account for the changing magnitude of the physical wave function by introducing new error bounds that are scaled relative to the magnitude of the physical wave function $\left( \Delta_{A/S}(t) = \delta_{A/S} \cdot \sqrt{\langle \psi^{(\vec{0})}(t) \vert \psi^{(\vec{0})}(t) \rangle} \right)$. This approach ensures that a consistent relative accuracy is maintained even as the physical wave function magnitude changes. We have also extended this approach to adaptive HOPS calculations that use the non-normalized nonlinear HOPS equation (Sec. \ref{app:NonLinearHOPS}).

\subsection{Early Time Basis Construction}
The adaptive error algorithm described above works efficiently once a reasonable basis has been determined, but defining the basis at early time is challenging. To ensure a sufficiently large basis at early time, we iteratively call the adaptive basis algorithm at early time points: during each iteration we construct a new basis and then perform the set union between this basis and the previous basis. The number of iterations done at each time point and the number of time points where the early time algorithm is used are user-defined variables. In our experience 2-3 iterations for the first 5-10 time points are sufficient to capture the early time basis reliably.  

\subsection{Discard Fraction ($f_{dis}$)}
\label{sec:discard_fraction}
During the calculation of the boundary auxiliary basis (Appendix \ref{app:explicit_adap_aux_bound}), the vast majority of flux terms that must be summed over are extremely small. Matching these minuscule flux terms to the proper auxiliary wave functions is computationally inefficient for large systems. Instead, we employ an additional boundary auxiliary error bound defined by the discard fraction ($f_{dis}\in[0,1]$). After the flux terms are constructed, but before they are matched to unpopulated auxiliary wave functions $|\psi^{(\vec{k}_b)}_t\rangle$ for $|\vec{k}_b\rangle\in\mathbb{A}\setminus\mathbb{A}_t$ according to Eq. \eqref{eq:boundauxfluxerror}, 
we discard the maximum number of fluxes such that their summed magnitude is less than $f_{dis}\delta_A'$. The remaining flux terms are fed back into Eq. \eqref{eq:boundauxfluxerror}, and the derivative error stemming from the discarded flux terms is taken into account to ensure the original error bound $\delta_A'= \sqrt{\delta_A^2-E^2_{\mathbb{A}_{t}^s}}$ is satisfied. 

\subsection{Active Mode Basis}
To ensure computational expense in large aggregates does not scale with the number of modes in the calculation, we construct an additional basis of active modes ($\{m\}$). A mode ($m$) is active if $\sum_{\vec{k}\in\mathbb{A}_t}|\langle \psi^{(\vec{k})}_t|\hat{L}_m|\psi^{(\vec{k})}_t\rangle| > 0$ or there exists any $|\vec{k}\rangle\in\mathbb{A}_t$ for which $k_{m}>0$. In the adaptive basis construction, any explicit sums over $m$ are restricted to the set of active modes. 

\subsection{Update Time}
Because updating the state and auxiliary bases at each time point is computationally expensive, we reduce calculation time by only updating the basis at each update time ($u_t$) of the simulation. Like $\delta_A$ and $\delta_S$, we treat $u_t$ as a phenomenological convergence parameter. During early time basis construction, update time is always equal to the integration time step.

\subsection{Low-Temperature Correction}
The adaptive basis construction algorithm described here is compatible with the low-temperature correction described in Sec. \ref{main:LTC}. Assuming that system-bath projection operators are diagonal, new terms introduced by the low-temperature correction (Eq. \eqref{eq:HOPS_LTC}, second line) are accounted for by the first terms of Eqs. \eqref{GeneralAuxiliaryRemoveError} and \eqref{StateRemovalError1} (the flux-in error calculations). As a result, the adaptive algorithm accounts for the low-temperature correction without further modification. 

\subsection{Performance}
The adHOPS basis size is sensitive to the specific choice of Hamiltonian and integration parameters. The dependence of computational cost on integration parameters is relatively straight-forward: it increases with smaller time steps, smaller adaptive error bounds ($\delta_{A}, \delta_{S}$), and shorter update times. In other words, the computational cost increases with precision of the numerical integration for the equation of motion. The computational cost of adHOPS as a function of Hamiltonian parameters, however, involves complicated trade-offs between electronic delocalization extents and the non-Markovian influence of the bath. For example, as the coupling between electronic states becomes large, the size of the state basis at long time will go towards the full state basis, but the corresponding depth of the hierarchy ($\textrm{k}_{\textrm{max}}$) required to describe the reorganization process will decrease because the effective coupling to the bath decreases with increasing delocalization extent. On the other hand, as the system-bath coupling becomes large, the depth of the hierarchy required to describe the reorganization process increases, but the corresponding exciton delocalization extent becomes smaller, creating a densely coupled auxiliary basis spanning the modes immediately connected to the excited molecule(s). In the intermediate regime where all Hamiltonian parameters are of a similar magnitude, the balance of these two effects is difficult to predict.

\section{Implementation}

\begin{figure}
    \centering
    \includegraphics{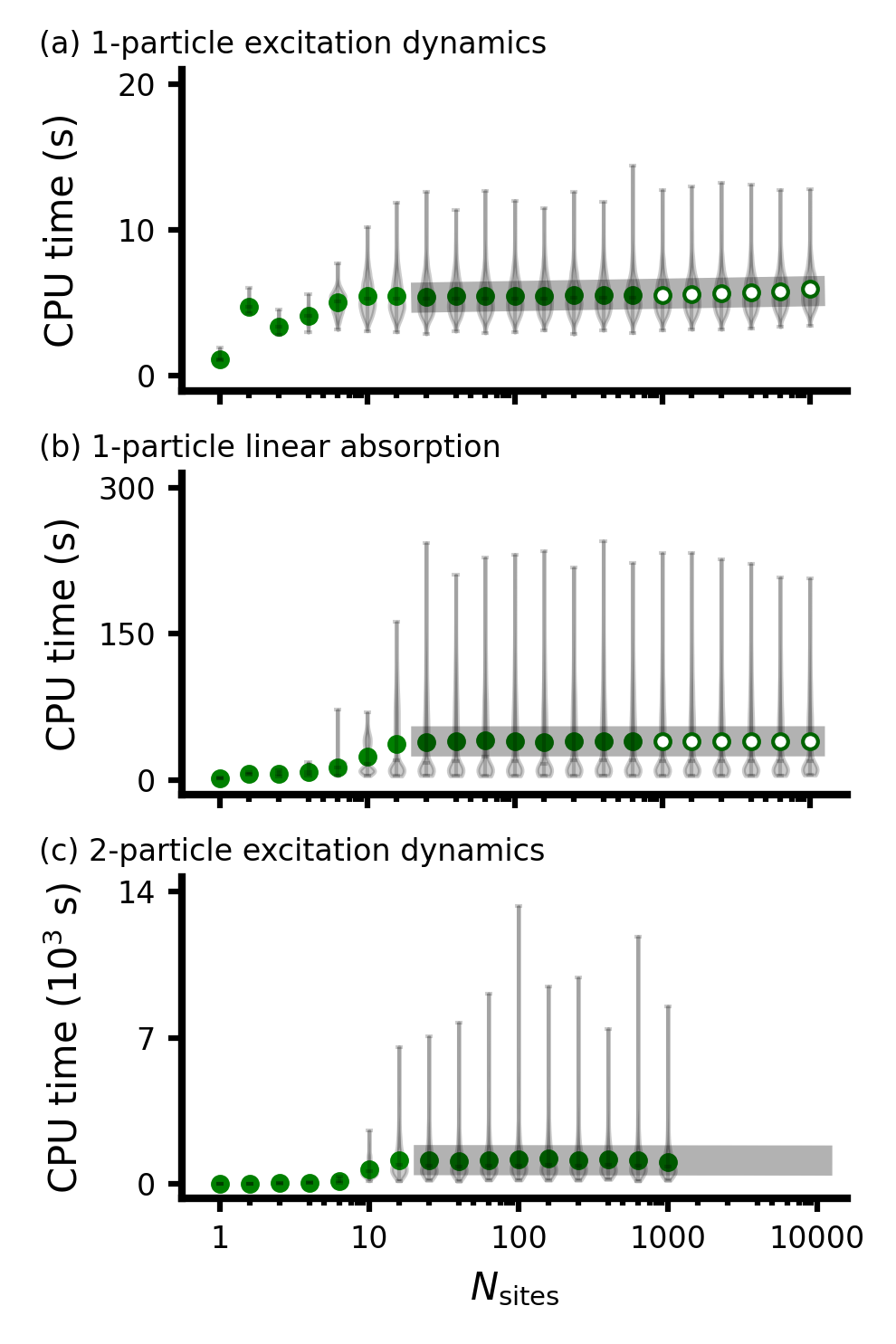}
    \caption{Size invariance of MesoHOPS. CPU time to propagate single-excitation dynamics with the (a) normalized nonlinear and (b) linear absorption equations-of-motion, and (c) two-excitation dynamics with the normalized nonlinear equation-of-motion, for 2 ps in a linear chain consisting of $N_{\textrm{sites}}$ molecules. The green circles show mean CPU time (open circles indicate that noise was only  calculated for the first 1000 sites). Grey violin plots represent the distribution, with horizontal grey lines on each violin plot corresponding to the minimum, median, and maximum. CPU times scale as $N^0$ in large chains, with log-log fits (thick grey lines) exhibiting a slope deviating from 0 by 0.01 or less. Hamiltonian details are listed in Appendix \ref{sec:Size_invariance_sys}. Convergence parameters are given in Table \ref{tab:convergence_params}. All calculations run on AMD EPYC 7763 64-Core Processor ("Milan").}
    \label{fig:SizeInvariance}
\end{figure}

MesoHOPS is an open-source Python package.\cite{mesohops_1_4} The  version of MesoHOPS used here (1.4) depends on the NumPy,\cite{numpy} SciPy,\cite{scipy} and Numba\cite{numba} libraries and supports the linear (Eq. \eqref{HOPS_linear}), nonlinear (Eq. \eqref{eq:NonLinearHops}), normalized nonlinear (Eq. \eqref{eq:NormNonLinearHops}), and linear absorption (Eq. \eqref{eq:C(t)_final_normalized_a}) equations-of-motion detailed in Appendix \ref{app:EoM}. In addition, MesoHOPS supports adHOPS calculations using the nonlinear, normalized nonlinear, and linear absorption equations that exhibit size-invariant scaling ($\mathcal{O}(1)$) as a function of system size in large aggregates. Fig. \ref{fig:SizeInvariance} plots the CPU time for adaptive calculations as a function of the number of molecules in a linear chain model (details given in Appendix \ref{app:SizeInvariance}) when the system is singly-excited (panels a and b, normalized nonlinear and linear absorption equations-of-motion) and doubly-excited (panel c, normalized nonlinear equation-of-motion). In all cases, MesoHOPS reaches a size-invariant regime where the average CPU time does not meaningfully increase with longer chains around the 10s of sites. In the current implementation, there are two limiting factors associated with memory: first, pre-calculating the noise trajectory for each environment, and, second, storing the full system Hamiltonian for the double-excitation manifold. To avoid excessive memory allocations we only calculate the noise ($z_{n,t}$) for the first 1000 sites of any linear chain, sufficient to encompass all dynamics during the 2 ps trajectory, and we limit the size of the 2-particle calculations to 1000-site chains. Looking forward, we expect these memory limitations can be removed by constructing the noise and the 2-particle system Hamiltonian on-the-fly for calculations involving more than $10^3$ molecules.

\section{Results and Discussion}
\label{sec:ChargeSeparation}

\begin{figure}
    \centering
    \includegraphics{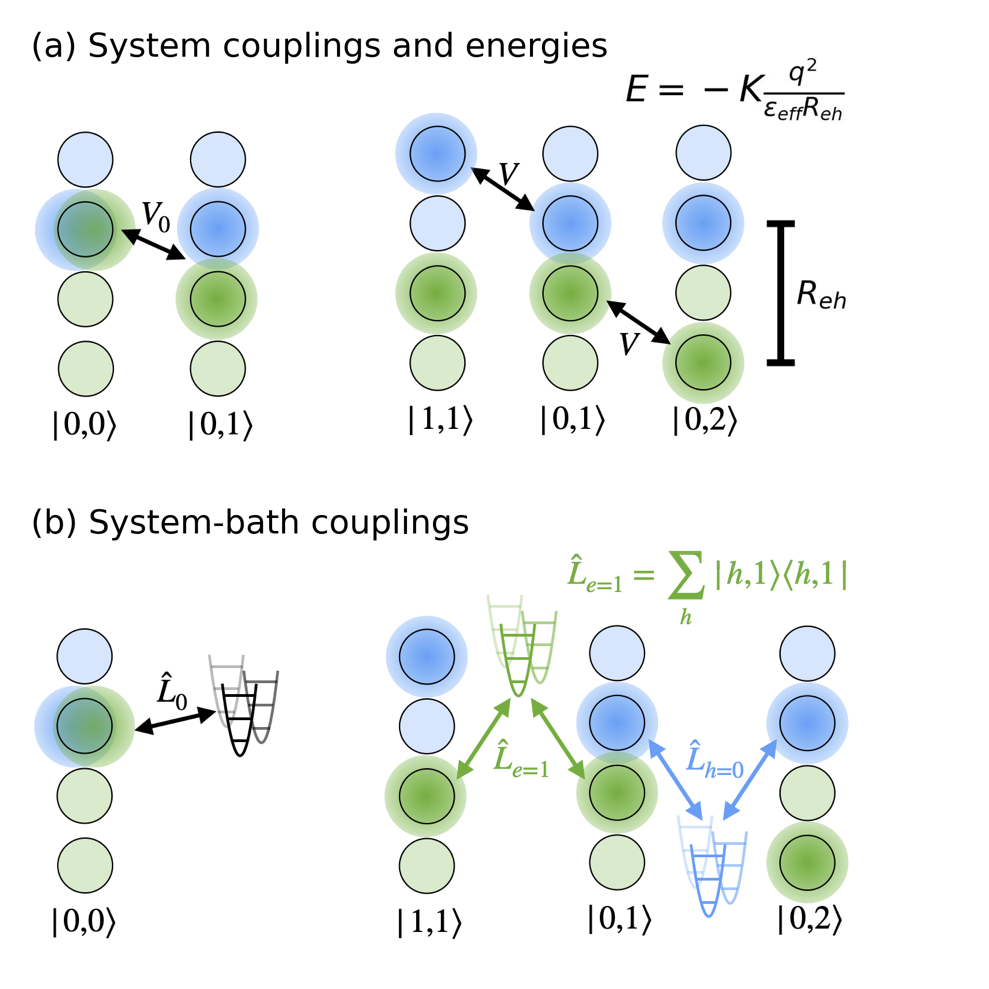}
    \caption{Schematic of the mobile-hole OPV model. (a) The high-energy exciton state may dissociate to the tightly-bound interfacial charge transfer state. Charge transfer states exhibit nearest-neighbor couplings to other charge transfer states where  either the electron or hole is separated by 1 molecular step. The Coulombic attraction that determines the potential energy of a given charge transfer state is purely dependent on electron-hole separation $R_{eh}$. (b) The exciton state is coupled to a unique thermal environment. All charge transfer states are coupled to the two independent environments corresponding to its occupied acceptor and donor molecule.}
    \label{fig:OPV_states}
\end{figure}

Charge separation in organic photovoltaics (OPVs) is an ultrafast and efficient process despite the strong Coulombic interaction between charge carriers, but the mechanism remains controversial.\cite{Gelinas_2014_OPV,Provencher_OPV_2014, jailaubekov2013hot, Grancini_OPV_2013, Muntwiler_OPV_2008, Park_OPV_2009, Kato_OPV, balzer_OPV_2022, Ostroverkhova_OPV_2016} Previous work has suggested that material dimensionality and disorder can provide an entropic drive towards the energetically unfavorable process of charge separation.\cite{monahan2015entropy, Hood_OPV_2016, Shi_entropy_disorder_2017} Here, we demonstrate that simultaneous electron and hole mobility can also introduce an entropic drive towards charge separation in a one-dimensional bulk heterojunction.

We explore the charge separation dynamics of a one-dimensional bulk heterojunction model with both a fixed and mobile hole. Our model consists of a linear chain of $N_{don}$ donor and $N_{acc}$ acceptor molecules. As shown in Fig. \ref{fig:OPV_states}a, the system begins in the exciton state, $|0,0\rangle$, in which the electron and hole are localized on the electron donor molecule at the interface. The exciton couples ($V_0$) to the interfacial charge transfer state $|0,1\rangle$,  where the hole remains on the exciton-hosting donor molecule, and the electron is 1 molecular unit away. Each charge transfer state $|h,e\rangle$ (with the hole and electron $h$ and $e$ steps away from their initial position, respectively) is equally coupled ($V$) to nearest-neighbor states $|h\pm 1, e\rangle$ and $|h, e\pm 1\rangle$. The Coulombic attraction between the electron and hole is dependent only on charge separation distance
\begin{equation}
\hat{R}_{eh} = \sum_{h,e} (e+h)|h,e\rangle \langle h,e|
\end{equation}
in units of the molecular spacing.
The corresponding system Hamiltonian is
\begin{flalign}
\begin{aligned}
\label{eq:2_particle_OPV_H_sys}
    \hat{H}_S = &V_0(|0,0\rangle\langle 0,1| + h.c.) \\
    + &E_1\sum_{h=0}^{N_{don}-1} \sum_{e=1}^{N_{acc}}\bigg(\frac{1}{h+e}\bigg)|h,e\rangle\langle h,e|\\
    + &\sum_{\substack{h,h' \in (0,N_{don}-1) \\ e,e' \in (1,N_{acc})}}\mathcal{V}_{h,h',e,e'}(|h,e\rangle\langle h',e'| + h.c.) 
\end{aligned}
\end{flalign}
where 
\begin{equation}
    \mathcal{V}_{h,h',e,e'} = \begin{cases}
        V & \textrm{if } |h-h'| + |e-e'| = 1\\
        0 & \textrm{otherwise}
    \end{cases}
\end{equation}  
and the values of $e\in (1,N_{acc})$ and $h\in (0,N_{don}-1)$ are restricted to account for the separated electron and hole being confined to the acceptor and donor materials, respectively.
We include $N_{don}+N_{acc}+1$ independent thermal environments (Fig. \ref{fig:OPV_states}b): each electron acceptor ($e$) and electron donor ($h$) molecule is coupled to an independent environment via
\begin{equation}
    \hat{L}_e = \sum_{h}|h,e\rangle\langle h,e|
\end{equation}
and 
\begin{equation}
    \hat{L}_h = \sum_{e}|h,e\rangle\langle h,e|
\end{equation}
respectively. In the fixed-hole case, we exclude the environment coupled to the hole (on the sole donor site $h=0$). Furthermore, we assume in both cases that the exciton couples to an independent vibrational environment via
\begin{equation}
    \hat{L}_{0} = |0,0\rangle\langle 0,0|
\end{equation}
as opposed to coupling to the same environment as the hole ($h=0$) at the interfacial CT state (see Appendix \ref{app:exciton_correlation} for details).
Following parameterization from Ref. \onlinecite{Kato_OPV}, $V_0 = 1200$ cm$^{-1}$, $E_1 = -2400$ cm$^{-1}$, $V = 800$ cm$^{-1}$, and the thermal environments are defined by Drude-Lorentz spectral densities with $\lambda = 160\textrm{ cm}^{-1}$ and $\gamma_0 = 270\textrm{ cm}^{-1}$. We run all simulations at room temperature ($T=300$ K), using the low-temperature correction to account for the first 5 Matsubara modes. 

\begin{figure*}
    \centering
    \includegraphics{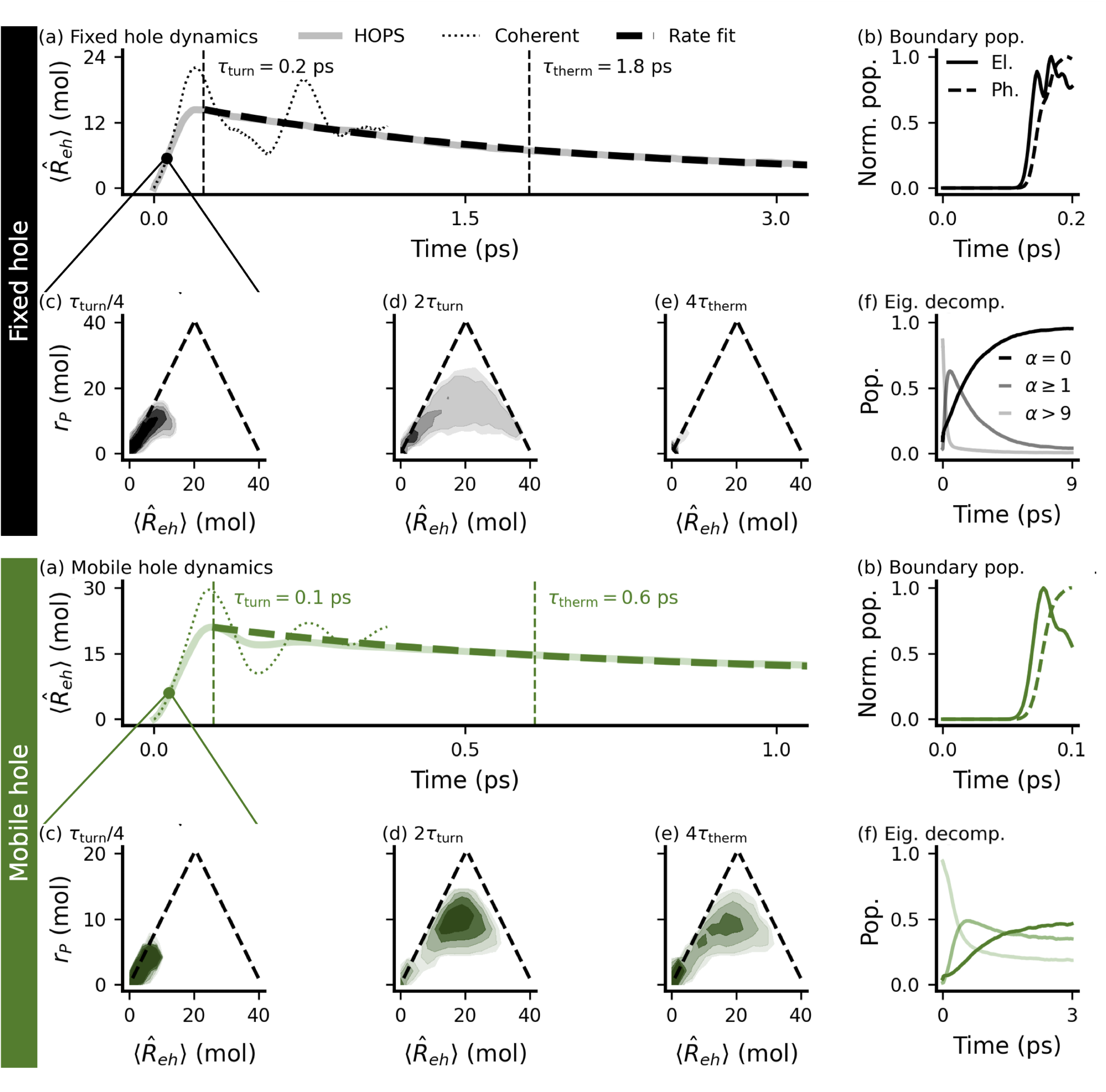}
    \caption{Mechanism of charge separation in an OPV (fixed-hole above, mobile-hole below). (a) The average distance between electron and hole: HOPS (solid line) and fully-coherent dynamics (dotted line). An exponential fit to the HOPS dynamics (dashed line) defines the thermalization timescale ($\tau_{\textrm{therm}}$) after $\langle \hat{R}_{eh}\rangle_t$ reaches its peak at $t=\tau_{\textrm{turn}}$. (b) The population of the last electron acceptor site in terms of the electron  and the expected occupation of the phonon given by $\langle\hat{X}_{N_{acc}}\rangle_t = -\frac{2\lambda\gamma_{0}}{\hslash}\int_0^te^{e^{-\gamma_0 s/\hslash}}\langle\hat{L}_{N_{acc}}\rangle_{t-s}ds$.\cite{Kato_OPV} (c-e) Contour plot showing the ensemble distribution of $\langle \hat{R}_{eh}\rangle_t$ and N-particle inverse participation ratio $r_P$ at $\tau_{\textrm{turn}}/4$, $2\tau_{\textrm{turn}}$, and  $4\tau_{\textrm{therm}}$, respectively. The dashed lines represent the limiting combinations of delocalization extent and charge separation distance (see Appendix \ref{app:DelocLimits}). (f) HOPS eigenstate population dynamics. The fixed-hole simulations were run as set out in Sec. \ref{sec:ChargeSeparation} with $N_{don}=1$, $N_{acc}=40$. The mobile-hole simulations were run with $N_{don}=N_{acc}=20$. The boundaries of each region in the contour plots, from least to most transparent, represent 0.2\%, 0.4\%, 1\%, 2\%, and 4\% of trajectories, respectively. Convergence parameters are given in Table \ref{tab:convergence_params}.}
    \label{fig:mech_overview}
\end{figure*}

In our model, charge separation proceeds via a rapid coherent dissociation of the exciton to a spectrum of delocalized, charge separated states, followed by incoherent relaxation into an equilibrium distribution. In Fig. \ref{fig:mech_overview}a, the charge separation dynamics at early time, as characterized by the average electron-hole distance, are replicated by purely-coherent (i.e., Schr\"{o}dinger equation) dynamics of electronic states (dotted lines). Once the average distance peaks at $\tau_{\textrm{turn}} = 180$ fs ($90$ fs in the mobile-hole case), the dynamics become incoherent, with $\langle \hat{R}_{eh}\rangle$ relaxing to a thermalized value with timescale $\tau_{\textrm{therm}}$ = 1.8 ps (610 fs in the mobile-hole case). In this parameter regime, the timescale on which the dynamics become incoherent is much longer than the bath reorganization timescale of 20 fs. In Fig. \ref{fig:mech_overview}b, we show that the turnover from coherent to incoherent dynamics occurs after the charge carriers reflect off the boundary and the expectation value of the collective phonon coordinate of the boundary acceptor molecule
 \begin{equation}
     \hat{X}_n = \sum_{q_n} \Lambda_{q_n}(\hat{a}^\dagger_{q_n} + \hat{a}_{q_n})
 \end{equation}
reaches its maximum value. While the transition to incoherent dynamics might suggest that the charge carriers become localized by interactions with the bath on timescales longer than $\tau_{\textrm{turn}}$, Fig. \ref{fig:mech_overview}d shows the majority of the charge carriers at $t=2\tau_{\textrm{turn}}$ are substantially delocalized in the HOPS ensemble, as characterized by an N-particle inverse participation ratio\bibnote{This definition of charge carrier delocalization reduces to the inverse participation ratio (IPR) in the 1-particle case. For two particles, it corresponds to the number of sites occupied by a single particle associated with a separable two-particle wave function where both particles have the same delocalization extents.} 
\begin{equation}
    r_P(t) = \left(\left\vert\sum_{e,h}\langle e,h|\psi^{(\vec{0})}_t\rangle\right|^4\right)^{-1/N_{part}}
\end{equation}
for both the fixed ($N_{part}=1$, top) and mobile ($N_{part}=2$, bottom) hole case.
On the other hand, localization does occur on longer timescales, but is driven by relaxation into the interfacial charge transfer state ($\ket{ 0, 1 }$), rather than by dynamic localization through interaction with the thermal environments. After thermalization, Fig. \ref{fig:mech_overview}e shows population remains in the delocalized and separated states for the mobile-hole case (bottom), whereas in the fixed-hole case almost all population has relaxed to the low-energy charge transfer state localized at the interface (top), consistent with the eigenstate population dynamics (Fig. \ref{fig:mech_overview}f). 
Thus, we suggest the slow relaxation to the thermal equilibrium proceeds through incoherent transport between delocalized eigenstates.

\begin{figure}
    \centering
    \includegraphics{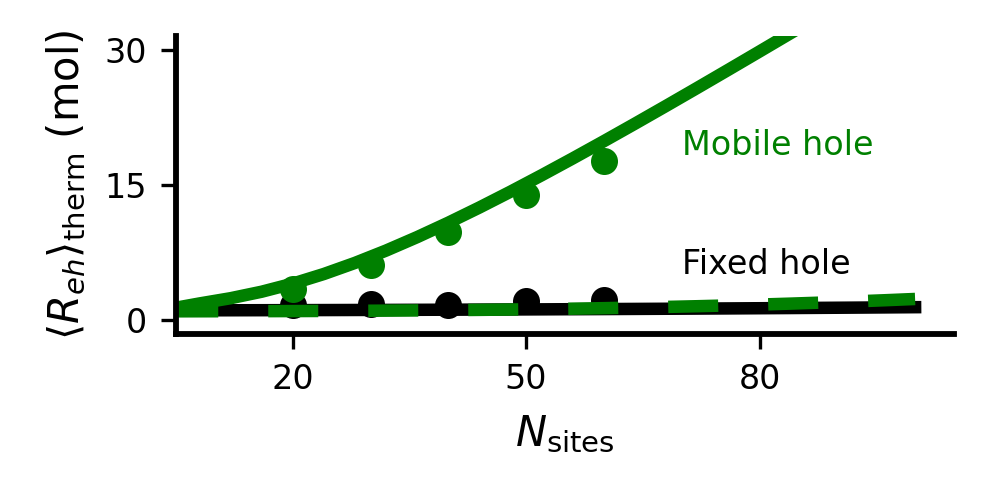}
    \caption{Thermal charge separation dependence on OPV chain length. Circles represent equilibrium populations from adHOPS simulations. The solid lines represent the Boltzmann distribution. The dashed line provides an additional comparison to the Boltzmann distribution of localized states ($V=V_0=0$). Convergence parameters are given in Table \ref{tab:convergence_params}.}
    \label{fig:model_extension}
\end{figure}

The model parameters presented here do not support substantial dynamic localization of charge carriers and, in the mobile-hole case, result in a charge separation efficiency that depends on chain length. The fast early time charge separation and slow transport of the electron back to the interface could suggest a quantum-classical ratchet mechanism wherein the system-bath interactions localize the electron after long-range charge separation, slowing relaxation into the bound charge transfer state ($\ket{0,1}$). \cite{Kato_OPV}
However, our analysis shows that this mechanism is not sufficient to describe the dynamics: the charge carriers are nearly maximally delocalized given the chain geometry (limits shown as dashed lines in Fig. \ref{fig:mech_overview}c-e, Appendix \ref{app:DelocLimits}) until they relax into the low-energy state at the interface. Furthermore, Fig. \ref{fig:model_extension} shows the average electron-hole separation in the thermal distribution calculated by HOPS (dots) is well-described for both the fixed-hole (black) and mobile-hole (green) cases by a Boltzmann distribution of the electronic eigenstates (solid lines) across different chain lengths ($N_{sites}$), consistent with the relatively weak system-bath coupling ($\lambda = 160 \textrm{ cm}^{-1}$) compared to the electronic coupling between states ($V = 800 \textrm{ cm}^{-1}$). Finally, we note that if the charge carriers fully localized, and thus could be thought of as hopping between neighboring molecules, the average electron-hole separation in the thermal distribution of the two-particle calculations would be much smaller and have a different functional form with increasing chain length due to the enhanced localization of the low-energy states, as shown by the dashed green line in Fig. \ref{fig:model_extension}. Thus, we do not observe small polaron formation - rather, occupation of delocalized system eigenstates at long time is essential to support efficient charge separation for the mobile hole case in this parameter regime.

Here, we have demonstrated that efficient charge separation is possible even for a one-dimensional heterojunction when both the electron and hole are mobile. The entropic drive towards charge separation caused by the simultaneous electron and hole mobility - there are $N$ states of equal electronic energy wherein the electron and hole are $N$ molecular units apart - is similar to that arising from increased material dimensionality which has been discussed previously.\cite{Hood_OPV_2016, Shi_entropy_disorder_2017} We expect the charge separation mechanism described above to extend to higher-dimensional materials in similar parameter regimes where the electronic coupling will be effectively enhanced by the increased number of nearest neighbors. We note, however, that this mechanism is presumably not general:
\textit{ab initio} simulations in various molecular materials suggest variable extents of electron delocalization (from a few to tens of molecules),\cite{Giannini_OSC_localization_2019} and other calculations indicate that the inclusion of static disorder,\cite{balzer_OPV_2022} parameterization of the energy gap between the exciton and interfacial CT state,\cite{tamura2013ultrafast} presence of shared thermal environments for exciton/CT states,\cite{Hestand2018review} and choice of spectral density\cite{popp2021quantum} all influence the dynamics of charge separation.
In particular, it is possible that stronger system-bath coupling could realize the dynamic localization (i.e. small polaron formation) required for a quantum-classical ratchet mechanism.
Looking forward, the ability of adHOPS to efficiently simulate a wide variety of materials in different parameter regimes will provide an important benchmarking tool. More excitingly, though, the favorable scaling of adHOPS makes it possible to imagine running full device simulations from light absorption to charge separation using a formally exact method.

\section{Conclusion}

In this paper, we introduce an updated version of the open-source MesoHOPS library with an improved adaptive algorithm allowing for reduced-scaling simulations of any systems described by arbitrary diagonal system-bath couplings. We also developed a low-temperature correction and effective integration of the noise that simplify calculations in the presence of ultrafast vibrational relaxation. By exploiting intuitive physics, adHOPS has size-invariant scaling, a perfectly parallel algorithm, and a well-defined physical interpretation that make it a powerful tool for formally exact simulations of mesoscale open quantum systems. To demonstrate the new capabilities of the MesoHOPS code, we simulated a mobile-hole model of charge separation at a one-dimensional heterojunction and characterized the mechanism of transport. As expected, early time electron-hole separation proceeds coherently in this parameter regime, but the subsequent incoherent transport is not associated with dynamic localization of the charge carriers: rather, localization is the result of relaxation into the lowest-energy charge transfer state with the electron and hole bound at the interface. Looking forward, we expect that the adHOPS algorithm, and our MesoHOPS library, will provide useful tools both for benchmarking approximate methods in difficult parameter regimes and understanding the corresponding mechanisms.

\section{Acknowledgements}
The authors thank Alexia Hartzell for review and the table-of-contents figure, as well as Jacob Krich and Oliver Kühn for their feedback on the manuscript. The authors acknowledge support from the Robert A. Welch Foundation (Grant N-2026-20200401) and start-up funds from the University of Texas at Austin. DIGBR and TG acknowledge support from a US National Science Foundation CAREER Award (Grant CHE-2145358). Additionally, JKL acknowledges support from a Moody Fellowship. The authors acknowledge the Texas Advanced Computing Center (\href{http://www.tacc.utexas.edu}{TACC}) at The University of Texas at Austin and Southern Methodist University's \href{https://southernmethodistuniversity.github.io/hpc_docs/index.html}{Center for Research Computing} for providing HPC resources that have contributed to the research results reported within this paper.

\section{Author Declarations}
\subsection{Conflict of Interest}
The authors have no conflicts to disclose.

\subsection{Author Contributions}
\textbf{Brian Citty}: Methodology (lead); Software (lead); Writing - original draft (equal); Writing - review (equal). \textbf{Jacob K. Lynd}: Investigation (lead); Visualization (lead); Writing - original draft (equal); Writing - review (equal); Methodology (supporting); Software (supporting); Project administration (supporting). \textbf{Tarun Gera}: Software (supporting); Investigation (supporting); Visualization (supporting); Writing - original draft (supporting); Writing - review (supporting). \textbf{Leo Varvelo}: Software (supporting); Writing - review (supporting). \textbf{Doran I. G. B. Raccah}: Project administration (lead); Supervision (lead); Writing - original draft (equal); Writing - review (equal); Methodology (supporting); Software (supporting); Visualization (supporting).

\section{Data Availability}
All calculations were performed using MesoHOPS v1.4.\cite{mesohops_1_4} All input and figure scripts, along with the corresponding summary data, are available in a supplementary Zenodo archive.\cite{mesohops_paper_zenodo_archive}

\appendix 
\section{HOPS Equations of Motion}
\label{app:EoM}
\subsection{Linear Eq. of Motion}
The linear HOPS equation-of-motion is a general solution to the Non-Markovian Quantum State Diffusion equation (NMQSD), an exact equation-of-motion for open quantum systems, when $C_n(t) = \sum_{j_n} g_{j_n}e^{-\gamma_{j_n}t/\hslash}$.\cite{NMQSD,HOPS} It is given as 
\begin{flalign}
\begin{aligned}
\label{HOPS_linear_nonscaled}
    \hslash\frac{d\vert \psi^{(\Vec{k})}_t \rangle}{dt}  = (-i\hat{H}_S -\vec{k}\cdot\vec{\gamma} + \sum_n\hat{L}_nz^*_{n,t})\vert \psi^{(\Vec{k})}_t \rangle&\\ + \sum_{n,j_n}k_{j_n} g_{j_n}\hat{L}_n\vert \psi^{(\Vec{k}-\vec{e}_{j_n})}_t \rangle - \sum_{n,j_n}\hat{L}^\dag_n\vert \psi^{(\Vec{k}+\vec{e}_{j_n})}_t \rangle&
\end{aligned}
\end{flalign}
where $\vec{\gamma}$ is the vectorized list of exponential decay constants for the correlation functions.

We make the auxiliary wave function transformation
\begin{equation}
    \ket{\psi^{(\Vec{k})}_{t}} \rightarrow  \prod_{n,j_n}\left(\frac{g_{j_n}}{\gamma_{j_n}}\right)^{k_{j_n}}\ket{\psi^{(\Vec{k})}_t}.
\end{equation}
to prevent the auxiliary wave
functions that define the edges of the hierarchy ($|\psi^{(\vec{k})}_t\rangle$ for $\vec{k}=\|\vec{k}\|_1\vec{e}_{j_n}$) from diverging with increasing hierarchy depth.  This transforms the linear equation-of-motion to
\begin{flalign}
\begin{aligned}
\label{HOPS_linear}
    \hslash\frac{d \vert \psi^{(\Vec{k})}_t \rangle}{dt} = (-i\hat{H}_S -\vec{k}\cdot\vec{\gamma} + \sum_n\hat{L}_nz^*_{n,t})\vert \psi^{(\Vec{k})}_t \rangle&\\ + \sum_{n,j_n}k_{j_n} \gamma_{j_n}\hat{L}_n\vert \psi^{(\Vec{k}-\Vec{e}_{j_n})}_t \rangle - \sum_{n,j_n}\left(\frac{g_{j_n}}{\gamma_{j_n}}\right)\hat{L}^\dag_n\vert \psi^{(\Vec{k}+\Vec{e}_{j_n})}_t \rangle.&
\end{aligned}
\end{flalign}

The reduced density matrix $\hat{\rho}_{t}$ is
\begin{equation}
    \hat{\rho}_{t} = \mathbb{E}_z\big[ \ket{\psi_{t}^{(\vec{0})}}\bra{\psi_{t}^{(\vec{0})}}\big]
\end{equation}
where the expectation is taken with respect to the ensemble of trajectories $|\psi_{t}^{(\vec{0})}\rangle$ defined by realizations of the Gaussian stochastic processes $z_{n,t}$ with properties $\mathbb{E}_z[z_{n,t}] = 0$, $\mathbb{E}_z[z_{n,t}z_{n,s}] = \mathbb{E}_z[z_{m,t}z_{n,s}] = 0$, $\mathbb{E}_z[z_{n,t}^{*}z_{m,s}] = \delta_{m,n}C_{n}(t-s)$.  Similarly, operator expectation $\braket{\hat{O}}_t$ for an operator $\hat{O}$ is calculated
\begin{equation}
    \braket{\hat{O}}_t = \mathbb{E}_z[\bra{\psi_{t}^{(\vec{0})}}\hat{O}\ket{\psi_{t}^{(\vec{0})}}].
\end{equation}

In the linear HOPS equation-of-motion, the majority of trajectories go to 0 at long times, while a infinitesimal fraction of trajectories have diverging norms as time goes to infinity. Thus, the statistical convergence of a linear HOPS ensemble is poor outside of the weak system-bath coupling regime.\cite{HOPS}

\subsection{Nonlinear Eq. of Motion}
\label{app:NonLinearHOPS}
We improve the poor statistical convergence of linear HOPS by using a Girsanov transform to weight the probability of a trajectory appearing in a given ensemble by the norm of its physical wave function at a given point in time.\cite{NMQSD,HOPS} The resulting nonlinear equation-of-motion is given by
\begin{flalign}
\begin{aligned}
\label{eq:NonLinearHops}
\hslash \frac{d\vert \psi^{(\Vec{k})}_t \rangle}{dt} 
=  \big(-i\hat{H}_S - \Vec{k} \cdot \Vec{\gamma} + \sum_{n} \hat{L}_{n} (z^*_{n,t}+ \sum_{j_n}\xi_{{j_n},t})\big)\vert \psi^{(\Vec{k})}_t \rangle &\\ 
+ \sum_{n,{j_n}} k_{j_n} \gamma_{j_n} \hat{L}_{n}  \vert \psi^{(\Vec{k} -\Vec{e}_{j_n})}_t \rangle &\\
- \sum_{n,{j_n}} \left(\frac{g_{j_n}}{\gamma_{j_n}}\right)(\hat{L}^{\dagger}_{n} - \langle\hat{L}^{\dagger}_{n}\rangle_{t}) \vert \psi^{(\Vec{k}+\Vec{e}_{j_n})}_t\rangle &
\end{aligned}
\end{flalign}
where
\begin{equation}
    \langle\hat{L}^{\dagger}_{n}\rangle_{t} = \frac{\langle \psi^{(\vec{0})}_t \vert \hat{L}^{\dagger}_{n}\vert \psi^{(\vec{0})}_t \rangle}{\braket{\psi^{(\vec{0})}_t \vert \psi^{(\vec{0})}_t}}
\end{equation}
and
\begin{equation}
    \xi_{j_n,t} = \frac{1}{\hbar}\int_{0}^{t} ds C^{*}_{j_n}(t-s) \braket{\hat{L}^{\dagger}_{n}}_s
\end{equation}
is the noise memory drift. 

The reduced density matrix $\hat{\rho}_{t}$ is now calculated by taking the expectation over normalized wave functions,
\begin{equation}
    \hat{\rho}_{t} = \mathbb{E}_z\bigg[ \frac{\ket{\psi_{t}^{(\vec{0})}}\bra{\psi_{t}^{(\vec{0})}}}{\braket{\psi_{t}^{(\vec{0})} | \psi_{t}^{(\vec{0})}}}\bigg],
\end{equation}
and the expectation value for an operator $\hat{O}$ is
\begin{equation}
    \langle\hat{O}\rangle_t = \mathbb{E}_z\bigg[\frac{\langle \psi^{(\vec{0})}_{t} \vert \hat{O}\vert \psi^{(\vec{0})}_{t} \rangle}{\braket{\psi^{(\vec{0})}_{t} \vert \psi^{(\vec{0})}_{t}}}\bigg].
\end{equation}

\subsection{Normalized Nonlinear Eq. of Motion}

To propagate HOPS trajectories in which the physical wave function and expectation values of observables are normalized during propagation, we define a new set of auxiliary wave functions that are divided by the norm of the physical wave function, such that
\begin{equation}
   \vert  \phi^{(\vec{k})}_t \rangle = \frac{\vert \psi_{t}^{(\vec{k})}\rangle}{\sqrt{\braket{\psi^{(\vec{0})}_{t} | \psi^{(\vec{0})}_{t}}}}.
\end{equation}
The time-evolution of $\vert \phi^{(\Vec{k})}_t\rangle$ differs from the nonlinear HOPS equation because of the time-derivative of the norm of the physical wave function 
\begin{equation}
    \frac{d\ket{\phi^{(\Vec{k})}_{t}}}{dt} = \frac{\frac{d}{dt} \ket{\psi^{(\Vec{k})}_{t}}} {\sqrt{\braket{\psi^{(\vec{0})}_{t} | \psi^{(\vec{0})}_{t}}}} - \ket{\phi^{(\Vec{k})}_{t}} \frac{\frac{d}{dt} [\braket{\psi^{(\vec{0})}_{t} | \psi^{(\vec{0})}_{t}}]}{2 \braket{\psi^{(\vec{0})}_{t} | \psi^{(\vec{0})}_{t}}}.
\end{equation}

Taking the time derivative of the inner product $\braket{\psi^{(\vec{0})}_{t} | \psi^{(\vec{0})}_{t}}$ gives a purely real term
\begin{equation}
    \frac{\frac{d}{dt} \braket{\psi^{(\vec{0})}_{t} | \psi^{(\vec{0})}_{t}}} {2 \braket{\psi^{(\vec{0})}_{t} | \psi^{(\vec{0})}_{t}}} = \frac{\braket{\psi^{(\vec{0})}_{t} | \frac{d}{dt}\psi^{(\vec{0})}_{t}}} {2 \braket{\psi^{(\vec{0})}_{t} | \psi^{(\vec{0})}_{t}}} + c.c. = \textrm{Re}\Bigg[\frac{\braket{\psi^{(\vec{0})}_{t} | \frac{d}{dt}\psi^{(\vec{0})}_{t}}} {\braket{\psi^{(\vec{0})}_{t}| \psi^{(\vec{0})}_{t}}}\Bigg]
\end{equation}
which is solved with the nonlinear equation-of-motion (Eq. \eqref{eq:NonLinearHops}) to produce the normalization correction factor
\begin{flalign}
\begin{aligned}
    \Gamma_t = &\sum_{n} \braket{\hat{L}_{n}}_{t} \textrm{Re}[z^*_{n,t}+\sum_{j_n}\xi_{j_n,t}] \\
    &- \sum_{n, j_n} \textrm{Re}\left[\left(\frac{g_{j_n}}{\gamma_{j_n}}\right) \braket{\phi^{(\vec{0})}_{t} |\hat{L}^{\dagger}_{n}| \phi^{(\vec{e}_{j_n})}_{t}}\right] \\
    &+ \sum_{n, j_n} \braket{\hat{L}^{\dagger}_n}_{t} \textrm{Re}\left[\left(\frac{g_{j_n}}{\gamma_{j_n}} \right)\braket{\phi^{(\vec{0})}_{t} | \phi^{(\vec{e}_{j_n})}_{t}}\right].
\end{aligned}
\end{flalign}

Incorporating the normalization correction factor yields the normalized nonlinear equation-of-motion
\begin{flalign}
\begin{aligned}
\label{eq:adhops_final}
\hbar \frac{d\vert \phi^{(\Vec{k})}_{t}\rangle}{dt} 
=& \big(-iH_S - \Vec{k} \cdot \Vec{\gamma} -\Gamma_t + \sum_{n} \hat{L}_{n} (z^*_{n,t} + \sum_{j_n}\xi_{j_n,t})\big) \vert \phi^{(\Vec{k})}_{t} \rangle
\\ 
&+ \sum_{n, j_n} k_{j_n} \gamma_{j_n} \hat{L}_{n} \vert  \phi^{(\Vec{k} -\Vec{e}_{j_n})}_{t}\rangle \\ 
&- \sum_{n, j_n} \left(\frac{g_{j_n}}{\gamma_{j_n}}\right) (\hat{L}^{\dagger}_{n} - \braket{\hat{L}^{\dagger}_{n}}_{t}) \vert \phi^{(\Vec{k}+\Vec{e}_{j_n})}_{t}\rangle.
\end{aligned}
\end{flalign}
For simplicity, in sections other than this appendix, we express the normalized wave functions, when present, as $|\psi_{t}^{(\vec{k})}\rangle$ rather than $|\phi_{t}^{(\vec{k})}\rangle$.

In the normalized nonlinear formulation of HOPS the reduced density matrix $\hat{\rho}_{t}$ is 
\begin{equation}
    \hat{\rho}_{t} = \mathbb{E}_z\bigg[ \ket{\phi_{t}^{(\vec{0})}}\bra{\phi_{t}^{(\vec{0})}}\bigg]
\end{equation}
and the expectation value for an operator $\hat{O}$ is
\begin{equation}
    \braket{\hat{O}}_t = \mathbb{E}_z[\bra{\phi_{t}^{(\vec{0})}}\hat{O}\ket{\phi_{t}^{(\vec{0})}}].
\end{equation}

\subsection{Linear Absorption Eq. of Motion}
\label{app:LinearAbsorption}
We calculate the linear absorption spectrum using the dipole autocorrelation function, 
\begin{eqnarray}
C(t)
=\mathrm{Tr}\left\lbrace\hat{\mu}_{\textrm{eff}}\,{e}^{-i\hat{H}t/\hslash}\big(\hat{\mu}_{\textrm{eff}}{|}g\rangle\langle{g}|\otimes\hat{\rho}_{\mathrm{B}}\big) e^{i\hat{H}t/\hslash}\right\rbrace
\end{eqnarray}
where $\hat{\mu}_{\textrm{eff}}=\sum_{n=1}^N (\bfmu_n\cdot \bfepsilon) \,|n\rangle\langle{g}|+ h.c.$ is the collective dipole moment operator, $\ket{n}$ is the first excited state of the $n^{\textrm{th}}$ pigment, and $\ket{g}$ is the global ground state. The initial total density matrix, $\hat{\rho}_0=|g\rangle\langle{g}|\otimes\hat{\rho}_{\mathrm{B}}$ is factorized into the system density matrix ${|}g\rangle\langle{g}|$  and the density matrix of the thermal bath $\hat{\rho}_{\mathrm{B}}=e^{-\beta{\hat{H}}_{\mathrm{B}}}/\mathrm{Tr}_{\mathrm{B}}\left\lbrace{e}^{-\beta{\hat{H}}_{\mathrm{B}}}\right\rbrace$. 
HOPS cannot directly calculate the dipole-dipole autocorrelation function because it requires the system wave function to be in a pure state. Ref. \onlinecite{HOPS_Open_Quantum_System_Reponse} introduced the pure state decomposition method to describe the mixed state ($\ket{n}\bra{g}$) required for linear absorption which was subsequently generalized for linear absorption with a local initial excitation condition and combined with the adHOPS framework.\cite{Sim_Absorb_Spectra_HOPS,gera_simulating_2023} 

Here we provide a brief summary of the HOPS linear absorption equation-of-motion in its local formulation. We begin by decomposing the collective transition dipole moment operator into a sum over different local excitation operators ($\hat{\sigma}_a$) with real-valued weights $A_a$,
\begin{equation}
    \hat{\mu}_\mathrm{eff}=\sum_a A_a \hat{\sigma}_a.
\end{equation}
There is a corresponding decomposition of the dipole-dipole autocorrelation function
\begin{equation}
C(t) = \sum_a A_a C_a(t) \label{eq:Ct_ISD}
\end{equation}
where
\begin{equation}
    C_a(t) = \mathrm{Tr}\left\lbrace\hat{\mu}_{\mathrm{eff}}\,{e}^{-i\hat{H}t/\hslash}\big( \hat{\sigma}_\mathrm{a}{|}g\rangle\langle{g}|\otimes\hat{\rho}_{\mathrm{B}}\big) e^{i\hat{H}t/\hslash}\right\rbrace.\label{eq:ISD_ct}
\end{equation}
Next we define a set of pure states 
\begin{equation}
\label{eq:pure_state_m}
    \ket{v_{\eta,a}} = \frac{1}{\sqrt{2}}\Big(\eta\ket{g} + 
\ket{\psi_a}\Big)
\end{equation}
where $\eta\in \{\pm 1, \pm i\}$ and $\ket{\psi_a} = \hat{\sigma}_a \ket{g}$, allowing us to rewrite the mixed state density matrix 
\begin{equation}
\label{eq:pure_decomp}
\hat{\rho}_a(t=0) = \hat{\sigma}_\mathrm{a} \ket{g}\bra{g}=
 \sum_{\eta\in \{\pm 1, \pm i\}} \frac{\eta}{2} \ \ket{v_{\eta, a}}\bra{v_{\eta,a }}
\end{equation}
as a sum over pure states. A detailed derivation of Eq. \eqref{eq:pure_decomp} can be found in the appendix of Ref. \onlinecite{gera_simulating_2023}.

By substituting Eq. \eqref{eq:pure_decomp} into Eq. \eqref{eq:ISD_ct}, we find 
\begin{equation}
    C_a(t) = \sum_{\eta\in \{\pm 1, \pm i\}}   \mathrm{Tr}\left\lbrace\hat{\mu}_{\mathrm{eff}}\,{e}^{-i\hat{H}t/\hslash}\Big( \frac{\eta}{2} \ket{v_{\eta, a}}\bra{v_{\eta,a }}\otimes\hat{\rho}_{\mathrm{B}}\Big) e^{i\hat{H}t/\hslash}\right\rbrace
\end{equation}
which can also be written as 
\begin{equation}
\label{eq:Ct_eta_a_hops}
    C_a(t) = \sum_{\eta\in \{\pm 1, \pm i\}}   \mathbb{E}_z\Bigg[\frac{\frac{\eta}{2}\braket{v_{\eta, a}(t)|\hat{\mu}_{\mathrm{eff}}|v_{\eta, a}(t)}}{\braket{v_{\eta, a}(t)|v_{\eta, a}(t)}}\Bigg]
\end{equation}
where the time-evolution of the pure state $\ket{v_{\eta, a}}$ may be calculated using the nonlinear HOPS equation-of-motion as long as both the ground and excited electronic states are included in the system Hilbert space. This result can be further simplified by noting that neither the system Hamiltonian nor the system-bath coupling operators connect the ground and excited electronic states, so the components of $\ket{v_{\eta}}$ time-evolve separately as
\begin{equation}
\label{eq:pure_state_a}
    \ket{v_{\eta,a}(t)} = \frac{1}{\sqrt{2}}\Big(\eta\ket{g}e^{-iE_g t/\hslash} + 
\ket{\psi_a(t)}\Big).
\end{equation}
Combining Eq. \eqref{eq:pure_state_a} with Eq. \eqref{eq:Ct_eta_a_hops} and accounting for cancellations due to the sum over $\eta$, we find 
\begin{equation}
\label{eq:C(t)_final_normalized_a}
C_a(t)=   \mu_{\mathrm{tot}}\mathbb{E}_z\Bigg[\frac{\braket{\psi_\mathrm{ex} \vert \psi_a(t)}}{\frac{1}{2}\left(||\psi_a(t)||^2 + 1\right)}\Bigg]e^{i E_{g} t/\hslash}
\end{equation}
where
\begin{equation}
\label{eq:psi_ex}
    \ket{\psi_\mathrm{ex}} = \frac{1}{{\mu}_\mathrm{tot}} \hat{\mu}_\mathrm{eff} \ket{g}
\end{equation}
and $\mu_\mathrm{tot}= \sqrt{\sum_{n=1}^N (\bfmu_n \cdot \bfepsilon)^2 }$. \\
The linear absorption HOPS equation propagates $\ket{\psi_{a}(t)}$ using the nonlinear HOPS equation in which $\hat{H}_S$ includes only the first excitation manifold and the expectation value of the system-bath coupling operator in Eq. \eqref{eq:NonLinearHops} is redefined as
\begin{equation}
    \langle\hat{L}^{\dagger}_{n}\rangle_{t} = \frac{\langle \psi^{(\vec{0})}(t) \vert \hat{L}^{\dagger}_{n}\vert \psi^{(\vec{0})}(t) \rangle}{\braket{\psi^{(\vec{0})}(t) \vert \psi^{(\vec{0})}(t)}+1}
\end{equation}
to account for the influence of the ground state. Using Eq. \eqref{eq:C(t)_final_normalized_a} in conjunction with the adaptive framework (Dyadic adaptive HOPS - DadHOPS) allows us to efficiently simulate the total dipole correlation function for large molecular aggregates.\cite{gera_simulating_2023} 

\section{Deriving the Low-Temperature Correction}
\label{app:LowTempCorrection}
\subsection{Linear Equation of Motion}
In the linear HOPS equation-of-motion, there is no noise memory drift term, so the low-temperature correction is given fully by the terminator approximation of the Markovian auxiliary wave functions $|\psi^{(\vec{e}_{\nu_n})}_t\rangle$ associated with the set of ultrafast modes $\{\nu_n\}$. Adding in a set of low-temperature-corrected ultrafast modes augments the linear HOPS equation-of-motion:
\begin{flalign}
\begin{aligned}
    \hslash\frac{d\vert \psi^{(\Vec{k})}_t \rangle}{dt}  = (-i\hat{H}_S -\vec{k}\cdot\vec{\gamma} + \sum_n\hat{L}_nz^*_{n,t})\vert \psi^{(\Vec{k})}_t \rangle&
    \\ + \sum_{n,j_n}k_{j_n} \gamma_{j_n}\hat{L}_n\vert \psi^{(\Vec{k}-\vec{e}_{j_n})}_t \rangle - \sum_{n,j_n}\Big(\frac{g_{j_n}}{\gamma_{j_n}}\Big)\hat{L}^\dag_n\vert \psi^{(\Vec{k}+\vec{e}_{j_n})}_t \rangle&
    \\ - \delta_{\vec{k}, \vec{0}}\sum_{n,\nu_n}\Big(\frac{g_{\nu_n}}{\gamma_{\nu_n}}\Big)\hat{L}^\dag_n\vert \psi^{(\vec{e}_{\nu_n})}_t\rangle.
\end{aligned}
\end{flalign}
By substituting in Eq. \eqref{eq:terminator}, we assert that
\begin{equation}
    \sum_{n,\nu_n}\Big(\frac{g_{\nu_n}}{\gamma_{\nu_n}}\Big)\hat{L}^\dag_n\vert \psi^{(\vec{e}_{\nu_n})}_t\rangle = \sum_{n,\nu_n}\Big(\frac{g_{\nu_n}}{\gamma_{\nu_n}}\Big)\hat{L}^\dag_n\hat{L}_n\vert \psi^{(\Vec{0})}_t \rangle = \sum_n \hat{L}^\dag_n\hat{L}_nG_n|\psi^{(\Vec{0})}_t\rangle
\end{equation}
where $G_n$ is defined in Eq. \eqref{eq:LTC_coeff}. Thus, the low-temperature-corrected linear HOPS equation-of-motion is
\begin{flalign}
\begin{aligned}
    \hslash\frac{d\vert \psi^{(\Vec{k})}_t \rangle}{dt}  = (-i\hat{H}_S -\vec{k}\cdot\vec{\gamma} + \sum_n\hat{L}_nz^*_{n,t})\vert \psi^{(\Vec{k})}_t \rangle&
    \\ + \sum_{n,j_n}k_{j_n} \gamma_{j_n}\hat{L}_n\vert \psi^{(\Vec{k}-\vec{e}_{j_n})}_t \rangle - \sum_{n,j_n}\Big(\frac{g_{j_n}}{\gamma_{j_n}}\Big)\hat{L}^\dag_n\vert \psi^{(\Vec{k}+\vec{e}_{j_n})}_t \rangle&
    \\ - \delta_{\vec{k}, \vec{0}}\sum_n \hat{L}^\dag_n\hat{L}_nG_n|\psi^{(\Vec{0})}_t\rangle.
\end{aligned}
\end{flalign}
Note that the inclusion of the $\delta_{\vec{k}, \vec{0}}$ ensures that the low-temperature correction only affects the physical wave function.

\subsection{Nonlinear Equation of Motion}
In the nonlinear equation-of-motion, the low-temperature correction has two components: the delta function approximation to the noise memory drift terms associated with ultrafast modes, and the terminator approximation to the Markovian auxiliary wave functions associated with the same modes. Adding in a set of low-temperature-corrected modes, the nonlinear HOPS equation-of-motion becomes
\begin{flalign}
\begin{aligned}
\hslash \frac{d\vert \psi^{(\Vec{k})}_t \rangle}{dt} 
=  \big(-i\hat{H}_S - \Vec{k} \cdot \Vec{\gamma} + \sum_{n} \hat{L}_{n} (z^*_{n,t}+ \sum_{j_n}\xi_{{j_n},t} + \sum_{\nu_n}\xi_{{\nu_n},t})\big)\vert \psi^{(\Vec{k})}_t \rangle &\\ 
+ \sum_{n,{j_n}} k_{j_n} \gamma_{j_n} \hat{L}_{n}  \vert \psi^{(\Vec{k} -\Vec{e}_{j_n})}_t \rangle &\\
- \sum_{n,{j_n}} \Big(\frac{g_{j_n}}{\gamma_{j_n}}\Big)(\hat{L}^{\dagger}_{n} - \langle\hat{L}^{\dagger}_{n}\rangle_{t}) \vert \psi^{(\Vec{k}+\Vec{e}_{j_n})}_t\rangle &
\\
- \delta_{\vec{k},\vec{0}}\sum_{n,{\nu_n}} \Big(\frac{g_{\nu_n}}{\gamma_{\nu_n}}\Big)(\hat{L}^{\dagger}_{n} - \langle\hat{L}^{\dagger}_{n}\rangle_{t}) \vert \psi^{(\Vec{k}+\Vec{e}_{\nu_n})}_t\rangle &.
\end{aligned}
\end{flalign}
By substituting in Eqs. \eqref{eq:noise_mem_delta}, \eqref{eq:terminator}, and  \eqref{eq:LTC_coeff}, we simplify the equation-of-motion:
\begin{flalign}
\begin{aligned}
\hslash \frac{d\vert \psi^{(\Vec{k})}_t \rangle}{dt} 
=  \big(-i\hat{H}_S - \Vec{k} \cdot \Vec{\gamma} + \sum_{n} \hat{L}_{n} (z^*_{n,t}+ \sum_{j_n}\xi_{{j_n},t})\big)\vert \psi^{(\Vec{k})}_t \rangle &\\ 
+ \sum_{n,{j_n}} k_{j_n} \gamma_{j_n} \hat{L}_{n}  \vert \psi^{(\Vec{k} -\Vec{e}_{j_n})}_t \rangle &\\
- \sum_{n,{j_n}} \Big(\frac{g_{j_n}}{\gamma_{j_n}}\Big)(\hat{L}^{\dagger}_{n} - \langle\hat{L}^{\dagger}_{n}\rangle_{t}) \vert \psi^{(\Vec{k}+\Vec{e}_{j_n})}_t\rangle &
\\
- \delta_{\vec{k},\vec{0}}\sum_{n} G_n(\hat{L}^{\dagger}_{n} - \langle\hat{L}^{\dagger}_{n}\rangle_{t}) \hat{L}_n\vert \psi^{(\Vec{0})}_t\rangle &\\
+\sum_{n}G_n^*\langle\hat{L}^\dag_n\rangle_t\hat{L}_n\vert \psi^{(\Vec{k})}_t \rangle &.
\end{aligned}
\end{flalign}

For simplicity, we compress all terms associated with the low-temperature-corrected ultrafast modes into a set of constants
\begin{equation}
    \Xi_{n,t} = G_n^*\braket{\hat{L}^{\dagger}_{n}}_t
\end{equation}
and operators
\begin{equation}
\label{eq:sum_terminator2}
    \hat{T}_{n,t}|\psi^{(\vec{0})}_t\rangle =  G_n(\hat{L}^{\dagger}_{n} - \langle\hat{L}^{\dagger}_{n}\rangle_{t})\hat{L}_n |\psi^{(\vec{0})}_t\rangle
\end{equation}
to give the final low-temperature-corrected nonlinear HOPS equation-of-motion
\begin{flalign}
\begin{aligned}
\hslash \frac{d\vert \psi^{(\Vec{k})}_t \rangle}{dt} 
=  \big(-i\hat{H}_S - \Vec{k} \cdot \Vec{\gamma} + \sum_{n} \hat{L}_{n} (z^*_{n,t}+ \sum_{j_n}\xi_{{j_n},t})\big)\vert \psi^{(\Vec{k})}_t \rangle &\\ 
+ \sum_{n,{j_n}} k_{j_n} \gamma_{j_n} \hat{L}_{n}  \vert \psi^{(\Vec{k} -\Vec{e}_{j_n})}_t \rangle &\\
- \sum_{n,{j_n}} \Big(\frac{g_{j_n}}{\gamma_{j_n}}\Big)(\hat{L}^{\dagger}_{n} - \langle\hat{L}^{\dagger}_{n}\rangle_{t}) \vert \psi^{(\Vec{k}+\Vec{e}_{j_n})}_t\rangle &
\\
+ \sum_{n} \big(\Xi_{n,t}\hat{L}_n -\delta_{\vec{k},\vec{0}}\hat{T}_{n,t}\big)\vert\psi^{(\Vec{k})}_t\rangle &.
\end{aligned}
\end{flalign}

\subsection{Normalized Nonlinear Equation of Motion}
The derivation of the low-temperature correction to the nonlinear equation-of-motion above is valid for the normalized nonlinear HOPS equation-of-motion as well. However, the low-temperature correction also alters the normalization correction factor presented in Eq. \eqref{eq:normcorr}. The normalization correction factor in the presence of ultrafast modes 
\begin{flalign}
\begin{aligned}
    \Gamma_t' = &\sum_{n} \braket{\hat{L}_{n}}_{t} \textrm{Re}[z^*_{n,t}+ \sum_{j_n}\xi_{{j_n},t}+\sum_{\nu_n}\xi_{{\nu_n},t}] \\
    - &\sum_{n, j_n} \textrm{Re}\left[\Big(\frac{g_{j_n}}{\gamma_{j_n}}\Big)\braket{\psi^{(\Vec{0})}_{t} |\hat{L}^{\dagger}_{n}| \psi^{(\Vec{e}_{j_n})}_{t}}\right] \\
    - &\sum_{n, \nu_n} \textrm{Re}\left[\Big(\frac{g_{\nu_n}}{\gamma_{\nu_n}}\Big)\braket{\psi^{(\Vec{0})}_{t} |\hat{L}^{\dagger}_{n}| \psi^{(\Vec{e}_{\nu_n})}_{t}}\right] \\
    + &\sum_{n, j_n} \braket{\hat{L}^{\dagger}_n}_{t} \textrm{Re}\left[\Big(\frac{g_{j_n}}{\gamma_{j_n}}\Big)\braket{\psi^{(\Vec{0})}_{t} | \psi^{(\Vec{e}_{j_n})}_{t}}\right] \\
    + &\sum_{n, \nu_n} \braket{\hat{L}^{\dagger}_n}_{t} \textrm{Re}\left[\Big(\frac{g_{\nu_n}}{\gamma_{\nu_n}}\Big)\braket{\psi^{(\Vec{0})}_{t} | \psi^{(\Vec{e}_{\nu_n})}_{t}}\right]
\end{aligned}
\end{flalign}
can be written as a sum of $\Gamma_t$, the expression in Eq. \eqref{eq:normcorr}, and the associated low-temperature correction 
\begin{flalign}
\begin{aligned}
    \tilde{\Gamma}_t = &\sum_{n,\nu_n} \braket{\hat{L}_{n}}_{t} \textrm{Re}\left[\Big(\frac{g_{\nu_n}}{\gamma_{\nu_n}}\Big)^*\braket{\hat{L}^{\dagger}_n}_{t}\right] \\
    - &\sum_{n, \nu_n} \textrm{Re}\left[\Big(\frac{g_{\nu_n}}{\gamma_{\nu_n}}\Big)\braket{\psi^{(\Vec{0})}_{t} |\hat{L}^{\dagger}_{n}\hat{L}_n| \psi^{(\Vec{0})}_{t}}\right] \\
    + &\sum_{n, \nu_n} \braket{\hat{L}^{\dagger}_n}_{t} \textrm{Re}\left[\Big(\frac{g_{\nu_n}}{\gamma_{\nu_n}}\Big)\braket{\psi^{(\Vec{0})}_{t} | \hat{L}_n|\psi^{(\Vec{0})}_{t}}\right]
\end{aligned}
\end{flalign}
Which simplifies by Eq. \eqref{eq:LTC_coeff} to
\begin{equation}
    \tilde{\Gamma}_t = \sum_n\textrm{Re}[G_n]\big(2\langle\hat{L}^\dag_n\rangle_t\langle\hat{L}_n\rangle_t - \langle\hat{L}^{\dagger}_{n}\hat{L}_n\rangle_t\big).
\end{equation}
We thus express the low-temperature-corrected normalized nonlinear equation-of-motion
\begin{flalign}
\begin{aligned}
\label{eq:HOPS_LTCapp}
    \hslash \frac{d\vert \psi^{(\Vec{k})}_t \rangle}{dt}  
=  \big(-i\hat{H}_S - \Vec{k} \cdot \Vec{\gamma} -\Gamma_t + \sum_{n} \hat{L}_{n} (z^*_{n,t}+ \sum_{j_{n}}\xi_{{j_n},t})\big)\vert \psi^{(\Vec{k})}_t \rangle &\\ 
+ \sum_{n,{j_n}} k_{j_n} \gamma_{j_n} \hat{L}_{n}  \vert \psi^{(\Vec{k} -\Vec{e}_{j_n})}_t \rangle &\\
- \sum_{n,{j_n}} \Big(\frac{g_{j_n}}{\gamma_{j_n}}\Big)(\hat{L}^{\dagger}_{n} - \langle\hat{L}^{\dagger}_{n}\rangle_{t}) \vert \psi^{(\Vec{k}+\Vec{e}_{j_n})}_t\rangle &\\
+ \sum_n (\Xi_{n,t}\hat{L}_n - \delta_{\vec{k},\vec{0}}\hat{T}_{n,t} - \tilde{\Gamma}_{n,t}) \vert \psi^{(\Vec{k})}_t \rangle &.
\end{aligned}
\end{flalign}

\subsection{Linear Absorption Equation of Motion}

The linear absorption equation-of-motion is identical to the nonlinear equation-of-motion, except that expectation values of operators are defined
\begin{equation}
    \langle \hat{O} \rangle_t = \frac{\braket{\psi^{(\Vec{0})}_{t} | \hat{O}|\psi^{(\Vec{0})}_{t}}}{1 + \braket{\psi^{(\Vec{0})}_{t} | \psi^{(\Vec{0})}_{t}}}.
\end{equation}
Because the derivation of the low-temperature correction does not depend on the form of the expectation value of observable operators, substituting the expectation values calculated above into the nonlinear low-temperature correction is sufficient to generate a low-temperature correction for the linear absorption equation-of-motion.

\subsection{Testing the Low-Temperature Correction}
\label{app:LTC_test}
\begin{figure}
    \centering
    \includegraphics{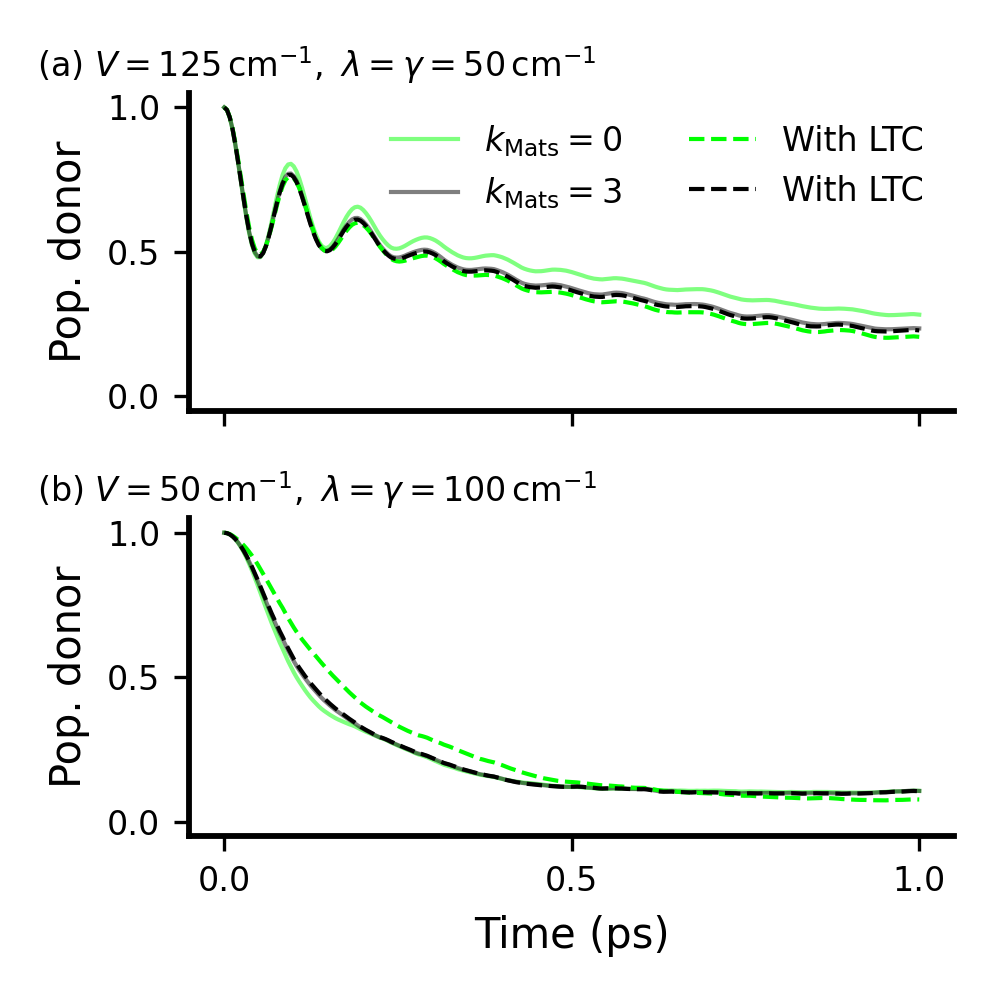}
    \caption{Low temperature correction (LTC) for a dimer at $T=45$ K. Population dynamics with respect to the number of Matsubara modes $k_{\textrm{Mats}}$ with and without the low-temperature correction (dashed and solid lines, respectively). Convergence parameters are given in Table \ref{tab:convergence_params}.}
    \label{fig:LTC_dimer}
\end{figure}

\begin{figure}
    \centering
    \includegraphics{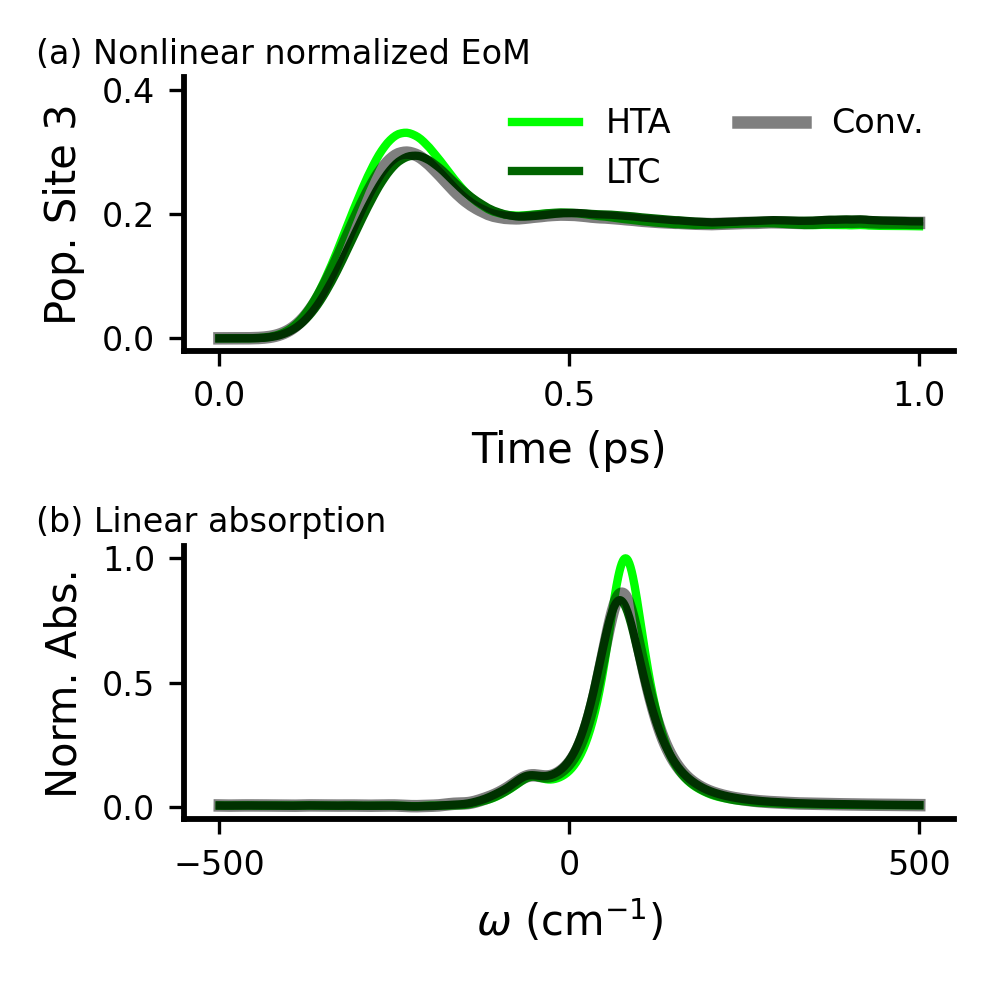}
    \caption{Low temperature correction (LTC) in a four-site linear chain. (a) Population of the final site and (b) absorption spectrum for a linear chain with $V=\gamma_0=\lambda=50 \textrm{ cm}^{-1}$ at $T=45$ K. Convergence parameters are given in Table \ref{tab:convergence_params}.}
    \label{fig:LTC_Linchain}
\end{figure}

The low-temperature correction provides a more efficient description of Matsubara modes when their relaxation is faster than all other timescales. We first explore a dimer model at $T=45$ K 
\begin{equation}
    \hat{H}_S = E(|0\rangle\langle 0|- |1\rangle\langle 1|) + V(|0\rangle\langle 1| + |1\rangle\langle 0| )
\end{equation}
where $E=V$ and each site is coupled to an independent bath ($\hat{L}_n = |n\rangle\langle n|$) defined by a Drude-Lorentz spectral density with reorganization energy $\lambda_n$ and inverse timescale $\gamma_{0_n}$. In Fig. \ref{fig:LTC_dimer}, we compare the population dynamics of the donor site, with and without the low-temperature correction. In Fig. \ref{fig:LTC_dimer}a we find the low-temperature correction provides a converged result even when no Matsubara modes are explicitly included. On the other hand, when the bath memory time in the high-temperature approximation (HTA) is comparable to the timescale of the first Matsubara mode, the low-temperature approximation breaks down (Fig. \ref{fig:LTC_dimer}b) due to the delta-function approximation. However, explicitly including Matsubara modes alongside the low-temperature correction quickly converges the calculation. Extending this treatment to a four-site linear chain of equivalent pigments 
\begin{equation}
    \hat{H}_S = \sum_{n=0}^2 V|n\rangle\langle n+1| + h.c.
\end{equation}
we find that the low-temperature correction provides a converged simulation for both population dynamics and linear absorption spectra when no explicit Matsubara modes are included in the calculation (Fig. \ref{fig:LTC_Linchain}). Thus, the application of the low-temperature correction can decrease computational cost, but necessitates a careful consideration of the parameter regime and convergence behavior.  

\section{Deriving the Effective Integration of the Noise}
\label{app:EffectNoiseIntegration}
To derive the expression for effective noise integration, we exploit the fact that, at a small enough timescale $\tau$, the time-evolution of the wave function,
\begin{equation}
    \label{eq:small_time_evolution_HOPS}
    |\Psi_{t + \tau}\rangle = |\Psi_{t}\rangle + \int_{0}^{\tau}(\mathcal{L}_{0} + \mathcal{L}_{z_{t + s}})|\Psi_{t + s}\rangle ds
\end{equation}
is dominated by the fluctuations of the noise $z_{t}^{*} + \xi_{t}$ present in $\mathcal{L}_{z_{t + s}}$.  Dividing the time step $\Delta t$ into $N_\tau$ smaller intervals $\tau = \Delta t/N_\tau$, and defining $\mathcal{L}_{0}|\Psi_{t}\rangle$ as the terms in the right-hand side of Equation \ref{eq:NormNonLinearHops} not containing the noise $z_{t}^{*} + \xi_{t}$ explicitly, we make the approximation
\begin{equation}\label{eq:approxL0}
    \int_{0}^{\tau}\mathcal{L}_{0}|\Psi_{t + s}\rangle ds \approx \tau \mathcal{L}_{0}|\Psi_{t}\rangle.
\end{equation}  
We define $\mathcal{L}_{z_{t}}|\Psi_{t}\rangle$ to contain the remaining explicitly noise-dependent terms such that for the non-linear normalized HOPS equation
\begin{flalign}
\begin{aligned}
\label{eq:noise_dependent_L}
    \mathcal{L}_{z_{t}}|\Psi_{t}\rangle = &\frac{1}{\hslash}\sum_{n} \hat{L}_{n} (z^*_{n,t} + \sum_{j_n}\xi_{j_n,t}) |\Psi_{t}\rangle\\
    - &\frac{1}{\hslash}\sum_{n} \braket{\hat{L}_{n}}_{t} \textrm{Re}[z^*_{n,t}+\sum_{j_n}\xi_{j_n,t}] |\Psi_{t}\rangle
\end{aligned}
\end{flalign}
where the second line arises from the components of the normalization correction factor $\Gamma_t$.
We approximate the noise-dependent time-evolution as 
\begin{flalign}
\begin{aligned}\label{eq:approxLzt}
    \int_{0}^{\tau}\mathcal{L}_{z_{t+s}}|\Psi_{t + s}\rangle ds &\approx \frac{1}{\hslash}\Bigg(\sum_{n}\hat{L}_{n}\int_{0}^{\tau}\Big(z_{n,t+s}^{*} + \sum_{j_n}\xi_{j_n,t+s}\Big)ds\\
    &\quad - \sum_{n}\langle \hat{L}_{n} \rangle_{t}\int_{0}^{\tau}\textrm{Re}\Big[z^{*}_{n,t + s} + \sum_{j_n}\xi_{j_n,t+s}\Big]ds\Bigg)|\Psi_{t}\rangle
\end{aligned}
\end{flalign}
where all terms not dependent on $z^{*}_t + \xi_t$ are approximated by the left-point and pulled out of the integral.

Using the Eqs. \eqref{eq:approxL0} and \eqref{eq:approxLzt}, we find an $\mathcal{O}(\tau)$ approximation to the time-evolution (Eq. \ref{eq:small_time_evolution_HOPS}) with forward-Euler integration from $t$ to $t + \tau$,
\begin{flalign}
\begin{aligned}
    |\Psi_{t + \tau}\rangle &\approx \Bigg(1 + \tau \mathcal{L}_{0} + \frac{1}{\hslash}\sum_{n}\hat{L}_{n}\int_{0}^{\tau}\Big(z^{*}_{n,t + s} + \sum_{j_n}\xi_{j_n,t + s}\Big)ds\\
    &\quad - \frac{1}{\hslash}\sum_{n}\langle \hat{L}_{n} \rangle_{t}\int_{0}^{\tau}\textrm{Re}\Big[z^{*}_{n,t+s} + \sum_{j_n}\xi_{j_n,t+s}\Big]ds\Bigg)|\Psi_{t}\rangle\\
    &+ \mathcal{O}(\tau^{2})
\end{aligned}
\end{flalign}
and then from $t+\tau$ to $t+2\tau$:
\begin{flalign}
\begin{aligned}
&|\Psi_{t + 2\tau}\rangle \approx \\
    &\Bigg(1 + 2\mathcal{L}_{0}\tau + \frac{1}{\hslash}\sum_{n}\hat{L}_{n}\bigg(\int_{0}^{\tau} + \int_{\tau}^{2\tau}\bigg)\Big(z^{*}_{n,t+s} + \sum_{j_n}\xi_{j_n,t+s}\Big)ds\\
    &\quad - \frac{1}{\hslash}\sum_{n}\langle \hat{L}_{n} \rangle_{t}\bigg(\int_{0}^{\tau} + \int_{\tau}^{2\tau}\bigg)\textrm{Re}\Big[z^{*}_{n,t+s} + \sum_{j_n}\xi_{j_n,t+s}\Big]ds \Bigg)|\Psi_{t}\rangle\\
    &+ \mathcal{O}(\tau^{2}).
\end{aligned}
\end{flalign}
By induction, $|\Psi_{t + \Delta t}\rangle$ is given as
\begin{flalign}
\begin{aligned}
\label{eq:Induction_Effective_noise_Psi}
&|\Psi_{t + \Delta t}\rangle \approx \\
     &\Bigg(1 + \Delta t\mathcal{L}_{0} + \frac{1}{\hslash}\sum_{n}\hat{L}_{n}\sum_{m=0}^{N_\tau-1}\int_{m\tau}^{(m+1)\tau}\Big(z^{*}_{n,t+s} + \sum_{j_n}\xi_{j_n,t+s}\Big)ds\\
    &\quad - \frac{1}{\hslash}\sum_{n}\langle \hat{L}_{n} \rangle_{t}\sum_{m=0}^{N_\tau-1}\int_{m\tau}^{(m+1)\tau}\textrm{Re}\Big[z^{*}_{n,t+s} + \sum_{j_n}\xi_{j_n,t+s}\Big]ds\Bigg)|\Psi_{t}\rangle\\
    &+ \mathcal{O}(\tau^{2}).
\end{aligned}
\end{flalign}
Using the left-side point to approximate the integral in Eq. \eqref{eq:Induction_Effective_noise_Psi}
\begin{flalign}
\begin{aligned}
\label{eq:eff_int_left_side}
&\sum_{m=0}^{N_\tau-1}\int_{m\tau}^{(m+1)\tau}\Big(z^{*}_{n,t+s} + \sum_{j_n}\xi_{j_n,t+s}\Big)ds \\ \approx &\frac{\Delta t}{N_\tau}\sum_{m=0}^{N_\tau-1}\Big(z^{*}_{n,t+m\tau} + \sum_{j_n}\xi_{j_n,t+m\tau}\Big) = \Delta t\Big(z'^{*}_{n,t} + \sum_{j_n}\xi_{j_n,t}'\Big)
\end{aligned}
\end{flalign}
gives the total time evolution as 
\begin{flalign}
\begin{aligned}
\label{eq:Induction_Effective_noise_Psi_with_avg_noise}
|\Psi_{t + \Delta t}\rangle \approx &\Bigg(1 + \Delta t\mathcal{L}_{0} + \frac{\Delta t}{\hslash}\sum_{n}\hat{L}_{n} \Big(z'^{*}_{n,t} + \sum_{j_n}\xi_{j_n,t}'\Big)\\
&\quad - \frac{\Delta t}{\hslash}\sum_{n}\langle \hat{L}_{n} \rangle_{t}\textrm{Re}\left[\Big(z'^{*}_{n,t} + \sum_{j_n}\xi_{j_n,t}'\Big)\right]\Bigg)|\Psi_{t}\rangle\\
&+ \mathcal{O}(\tau^{2}).
\end{aligned}
\end{flalign}
where 
\begin{equation}
    z_{n,t}'^* = \frac{1}{N_\tau}\sum_{m=0}^{N_\tau-1}z_{n,t+m\tau}
\end{equation}
is the left-hand moving average of the noise, and
\begin{equation}
    \xi_{j_n,t}'^* = \frac{1}{N_\tau}\sum_{m=0}^{N_\tau-1}\xi_{j_n,t+m\tau}
\end{equation}
is the left-hand moving average of the noise memory drift. The moving average of the noise ($z_{n,t}'^*$) can be calculated directly, but the same is not true for the noise memory drift.

To evaluate the moving average of the noise memory drift ($\xi_{j_n,t}'$), we use the time-derivative
\begin{equation}
    \frac{d\xi_{j_n,t}}{dt} = \frac{1}{\hslash}\bigg(\langle \hat{L}^\dag\rangle_tC_{j_n}^*(0)- \gamma_{j_n}^*\xi_{j_n,t}\bigg)
\end{equation}
and the left-side Euler integral to time-evolve the noise memory drift on the timescale $\tau$ 
\begin{equation}
    \xi_{j_n,t+\tau} = \xi_{j_n,t} + \frac{\tau}{\hslash}\bigg(\langle\hat{L}^\dag\rangle_tC_{j_n}^*(0) - \gamma_{j_n}^*\xi_{j_n,t}\bigg).
\end{equation}
Under the separation of timescales, $\langle\hat{L}^\dag\rangle_{t+\tau}\approx\langle\hat{L}^\dag\rangle_{t}$, such that
\begin{flalign}
    \begin{aligned}
        \xi_{j_n,t+2\tau} \approx \xi_{j_n+\tau,t} + \frac{\tau}{\hslash}\bigg(\langle\hat{L}^\dag\rangle_tC_{j_n}^*(0) - \gamma_{j_n}^*\xi_{j_n,t+\tau}\bigg)\\
    =    \xi_{j_n,t} + \left(2\frac{\tau}{\hslash}-\gamma^*_{j_n}\left(\frac{\tau}{\hslash}\right)^2\right)\bigg(\langle\hat{L}^\dag\rangle_tC_{j_n}^*(0) - \gamma_{j_n}^*\xi_{j_n,t}\bigg).
    \end{aligned}
\end{flalign}
Taking an $\mathcal{O}(\tau)$ approximation, we find by induction that
\begin{equation}
    \xi_{j_n,t+m\tau} \approx \xi_{j_n,t} + m\frac{\tau}{\hslash}\bigg(\langle\hat{L}^\dag\rangle_tC_{j_n}^*(0) - \gamma_{j_n}^*\xi_{j_n,t}\bigg),
\end{equation}
and the moving average of the noise memory drift is
\begin{flalign}
    \begin{aligned}
    \label{eq:approx_avg_noise_memory_drift}
    \xi_{j_n,t}' &= \frac{1}{N_\tau}\sum_{m=0}^{N_\tau-1}\xi_{j_n,t+m\tau} \\
    &\approx \xi_{j_n,t} + \frac{\tau}{N_\tau}\sum_{m=0}^{N_\tau-1}m\frac{1}{\hslash}\bigg(\langle\hat{L}^\dag\rangle_tC_{j_n}^*(0) - \gamma_{j_n}^*\xi_{j_n,t}\bigg) \\
    &= \xi_{j_n,t} + \frac{(N_\tau-1)\tau}{2}\frac{d}{dt}\xi_{j_n,t} \approx \xi_{j_n,t} + \frac{\Delta t}{2}\frac{d}{dt}\xi_{j_n,t}.
    \end{aligned}
\end{flalign}
Noting the presence of $\Delta t$ in Eq. \eqref{eq:Induction_Effective_noise_Psi_with_avg_noise}, we drop the $\Delta t$ dependent component of Eq. \eqref{eq:approx_avg_noise_memory_drift}, which leads to a term of order $\Delta t^2$ in the time-evolution. The noise-memory drift term is then given by
\begin{equation}
    \xi_{j_n,t}' = \xi_{j_n,t}.
\end{equation}

The resulting time-evolution under the effective integration of the noise is given by
\begin{equation}
\label{eq:eff_int_full_deriv}
    |\Psi_{t + \Delta t}\rangle = (1 + \Delta t(\mathcal{L}_{0} + \bar{\mathcal{L}}_{z_{t}}))|\Psi_{t}\rangle
\end{equation}
where $\bar{\mathcal{L}}_{z_{t}}$ is the same noise-dependent component of the non-linear normalized HOPS equation given in Eq. \eqref{eq:noise_dependent_L}, except that the noise is replaced with the left-hand moving average over the finer time-step ($z_{n,t}'^*$). Thus, the effectively-integrated normalized nonlinear equation-of-motion is
\begin{flalign}
\begin{aligned}
\label{eq:adhops_effective_integration}
\hslash\frac{d|\psi^{(\Vec{k})}_{t}\rangle}{dt} 
=& \big(-iH_{S} - \Vec{k} \cdot \Vec{\gamma} -\Gamma_t' + \sum_{n} \hat{L}_{n} (z'^*_{n,t} + \sum_{j_n}\xi_{j_n,t})\big) |\psi^{(\Vec{k})}_{t}\rangle
\\ 
&+ \sum_{n, j_n} k_{j_n} \gamma_{j_n} \hat{L}_{n}  |\psi^{(\Vec{k} -\Vec{e}_{j_n})}_{t}\rangle\\ 
&- \sum_{n, j_n} \Big(\frac{g_{j_n}}{\gamma_{j_n}}\Big) (\hat{L}^{\dagger}_{n} - \braket{\hat{L}^{\dagger}_{n}}_{t}) |\psi^{(\Vec{k}+\Vec{e}_{j_n})}_{t}\rangle
\end{aligned}
\end{flalign}
where $\Gamma_t'$ is the normalization correction factor calculated with the moving average of the noise. This effective integration also holds for the non-normalized nonlinear, linear absorption, and linear equations-of-motion, none of which introduce any terms not analyzed in this derivation.

\subsection{Testing the Effective Integration of the Noise}
\label{app:eff_int_test}
When the stochastic noise fluctuates more rapidly than the system wave function can evolve, an effective integration of the noise can increase the converged time step of the simulation. Here we demonstrate that, for a small chain of molecules where the low-temperature correction is relevant, effective integration of the noise increases the allowable integration time step by a factor of 10. 

\begin{figure}
    \centering
    \includegraphics{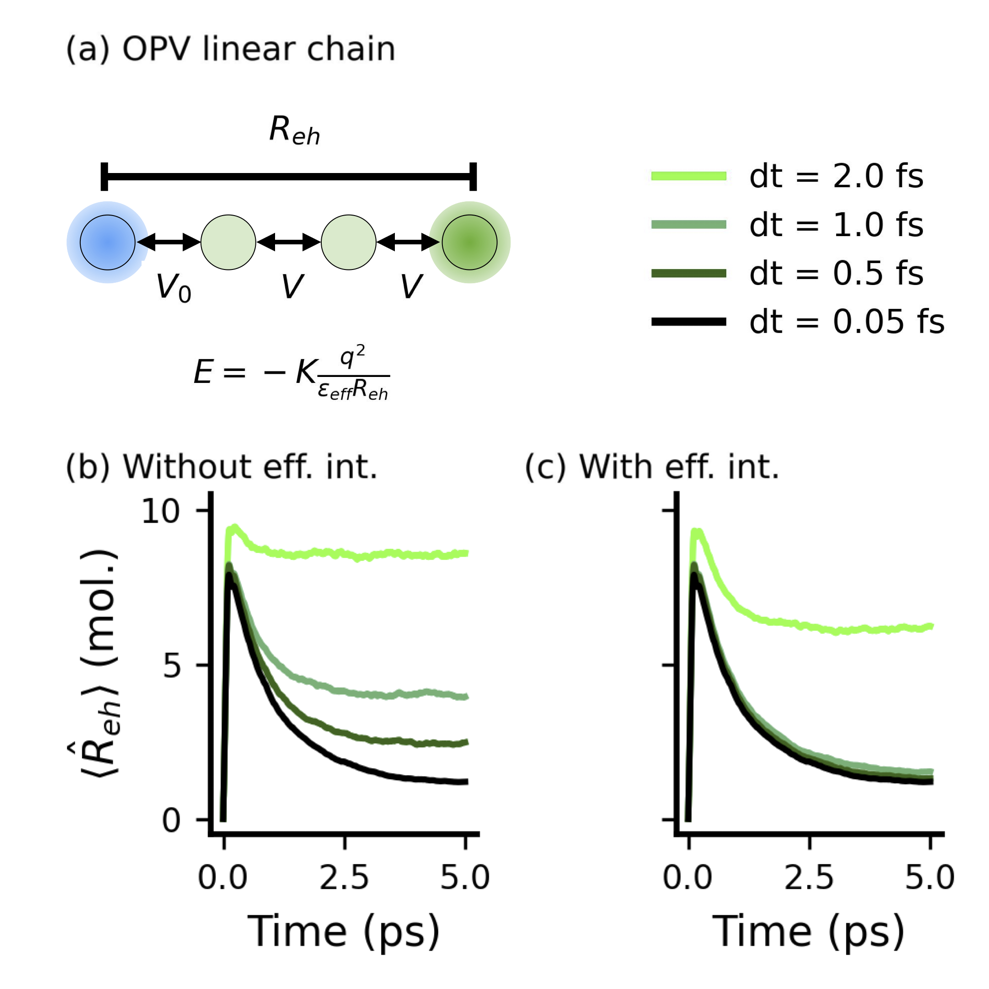}
    \caption{Effective integration of the noise in a fixed-hole OPV model. (a) Schematic of the model system: the exciton state is coupled to the interfacial charge transfer state by $V_0$, coupling between neighboring charge transfer states is $V$, and the Coulombic potential of a charge transfer state with a given acceptor molecule occupied by the electron (green) is inverse to its distance from the hole (blue) $R_{eh}$. (b) Electron-hole separation dynamics in a linear chain with $N_{acc}=20$ at various integration time steps $dt$. (c) The same dynamics with an effective integration of the noise. Convergence parameters are given in Table \ref{tab:convergence_params}.}
    \label{fig:OPV_time step}
\end{figure}

The effective integration of the noise decreases sensitivity of charge separation dynamics to the time step of integration. We model charge separation in a linear chain model of a bulk heterojunction, as described in Sec. \ref{sec:ChargeSeparation}, with $N_{don}=1$ and $N_{acc}=20$ such that the hole is fixed (a schematic of the model is shown in Fig. \ref{fig:OPV_time step}a). As laid out previously, the Coulombic interaction at unit distance $(E_1)$ is $-2400$ cm$^{-1}$, nearest-neighbor coupling between charge transfer states $(V)$ is 800 cm$^{-1}$, the energy of the exciton state is 0, and the coupling between the exciton state and the first charge transfer state, $(V_0)$, is 1200 cm$^{-1}$. Each environment is characterized by a Drude-Lorentz spectral density (Eq. \eqref{eq:drude_lorentz}) with reorganization energy $\lambda$ = $160$ cm$^{-1}$ and inverse bath reorganization timescale $\gamma_0$ = $270$ cm$^{-1}$. The exponential form of the bath correlation function is calculated at room temperature ($T = 300$ K) with an additional five Matsubara modes included within the low-temperature correction ($k_{\textrm{Mats}}=5$). The system-bath interaction associated with the hole in the charge transfer states is neglected, as it is responsible for an identical fluctuation in the vertical excitation energies of all charge transfer states. Fig. \ref{fig:OPV_time step}b shows the change in dynamics of the expectation value of the electron-hole distance with respect to the integration time steps ($dt$), with converged results requiring $dt=0.05$ fs. Fig. \ref{fig:OPV_time step}c shows that when using effective noise integration the expectation value of the electron-hole distance in the charge separation model is less sensitive to the time step, with even $dt=1.0$ fs providing a reliable calculation. 

\section{Explicit Derivative Error Calculations}
\label{app:explicit_adap}
\subsection{Auxiliary Basis}
\label{app:explicit_adap_aux}
Expanding the terms in Eq. (\ref{GeneralAuxiliaryRemoveError}) using the  normalized nonlinear HOPS equation (Eq. \eqref{eq:NormNonLinearHops}), the squared derivative error from excluding $|\vec{k}\rangle$ becomes
\begin{flalign}
\begin{aligned}\label{AuxiliaryRemovalError1}
E^{2}_{\mathbb{A}_t}[\vec{k}] &= \sum_{s \in \mathbb{S}_{t}}\left \vert \left(\mathcal{L}_{\mathbb{A}_t\otimes\mathbb{S}_t} \Psi_t\right)[\vec{k},s] + \frac{\Psi_{t}[\vec{k},s]}{\Delta t}\right \vert^{2}\\
&+ \frac{1}{\hslash^2}\sum_{s \in \mathbb{S}_{t}}\sum_{m: \vec{k} + \vec{e}_{m} \in \mathbb{A}} \left \vert (k_{m} + 1)\gamma_{m}\hat{L}_{m}\Psi_{t}[\vec{k},s] \right\vert^{2}\\
&+ \frac{1}{\hslash^2}\sum_{s \in \mathbb{S}_{t}}\sum_{m: \vec{k} - \vec{e}_{m} \in \mathbb{A}}\left \vert  \frac{g_{m}}{\gamma_{m}}(\hat{L}_{m} - \braket{ \hat{L}_{m}}_t )\Psi_{t}[\vec{k},s]\right \vert^{2}\\
&+  \frac{1}{\hslash^2}\sum_{s_b\in \mathbb{S} \setminus \mathbb{S}_{t}}\left \vert \sum_{s \in\mathbb{S}_{t}}  -i\hat{H}_S[s_b,s]\Psi_{t}[\vec{k},s] \right \vert^{2}
\end{aligned}
\end{flalign}
where we note that the numerical construction of $\left(\mathcal{L}_{\mathbb{A}_t\otimes\mathbb{S}_t} |\Psi_t\rangle\right)$is performed as an additional derivative evaluation using the HOPS equation expressed in the reduced basis $\mathbb{A}_t\otimes\mathbb{S}_t$.

\subsubsection{Stable Auxiliary Basis}
Starting from Eq. \eqref{AuxiliaryRemovalError1}, we calculate the error contributions associated with excluding $\ket{\vec{k}} \in \mathbb{A}_t$ from the basis $\mathbb{A}_{t+\Delta t}$ as described below. The two components of the flux-in error calculation,
\begin{equation}
    E^{2}_{\mathbb{A}_t^s,\textrm{in}}[\vec{k},s] = \left \vert \left(\mathcal{L}_{\mathbb{A}_t\otimes\mathbb{S}_t} \Psi_t\right)[\vec{k},s]  + \frac{\Psi_{t}[\vec{k},s]}{\Delta t}\right \vert^{2}
\end{equation}
are calculated together because they contribute to the same amplitudes and, as a result, can cancel. The flux-up error calculation,
\begin{flalign}
\begin{aligned}\label{fluxup}
    \hslash^2E^{2}_{\mathbb{A}^{s}_{t},\textrm{up}}[\vec{k},m] &= \sum_{s \in \mathbb{S}_t}\delta_{\vec{k} + \vec{e}_{m} \in \mathbb{A}} \left \vert (k_{m} + 1)\gamma_{m}\hat{L}_{m}\Psi_{t}[\vec{k},s] \right\vert^{2}\\
    &= F_{\mathbb{A},\textrm{up}}[\vec{k},m] \vert (k_{m} + 1)\gamma_{m} \vert^{2} \sum_{s \in \mathbb{S}_{t}} \left \vert \hat{L}_{m}[s,s]\Psi_{t}[\vec{k},s] \right\vert^{2}
\end{aligned}
\end{flalign}
 is simplified by introducing $F_{\mathbb{A},\textrm{up}}[\vec{k},m] = \delta_{\vec{k} + \vec{e}_{m} \in \mathbb{A}}$, a filter restricting the sum over modes to only include fluxes to valid auxiliary vectors.  The flux-down error calculation,
\begin{flalign}
\begin{aligned}\label{fluxdown} 
    \hslash^2E^{2}_{\mathbb{A}^{s}_{t},\textrm{down}}[\vec{k},m] &= \sum_{s \in \mathbb{S}_t}\delta_{\vec{k} - \vec{e}_{m} \in \mathbb{A}}\left \vert  \frac{g_{m}}{\gamma_{m}}(\hat{L}_{m} - \braket{ \hat{L}_{m}}_t )\Psi_{t}[\vec{k},s]\right \vert^{2}\\
    &= F_{\mathbb{A},\textrm{down}}[\vec{k},m]\left \vert  \frac{g_{m}}{\gamma_{m}} \right \vert^{2} \sum_{s \in \mathbb{S}_{t}}\left \vert \hat{D}_{m,t}[s,s]\Psi_{t}[\vec{k},s]\right \vert^{2}
\end{aligned}
\end{flalign}
is similarly simplified by introducing filter $F_{\mathbb{A},\textrm{down}}[\vec{k},m] = \delta_{\vec{k} - \vec{e}_{m} \in \mathbb{A}}$. Additionally, we have introduced
\begin{equation}
    \hat{D}_{m,t} = (\hat{L}_{m} - \braket{ \hat{L}_{m}}_t )
\end{equation}
which is a diagonal operator as long as $\hat{L}_m$ is diagonal.
Finally, the state-flux error calculation,
\begin{equation}
\begin{aligned}
    \hslash^2E^{2}_{\mathbb{A}_t^s,\textrm{state}}[\vec{k}] &= \sum_{s_{b} \in \mathbb{S} \setminus \mathbb{S}_{t}}\left \vert \sum_{s \in \mathbb{S}_t}-i\hat{H}_{S}[s_b,s]\Psi_{t}[\vec{k},s]\right \vert^{2}\\
    &= \sum_{s_{b} \in \mathbb{S} \setminus \mathbb{S}_{t}}\left \vert \left(\hat{H}_S\psi_t^{(\vec{k})}\right)[s_b]\right\vert^2
\end{aligned}
\end{equation}
is evaluated by the action of system Hamiltonian on the auxiliary wave function $\ket{\psi_t^{(\vec{k})}}$ creating flux to states outside of the current basis $\mathbb{S}_t$. The total squared derivative error (\ref{AuxiliaryRemovalError1}) is
\begin{flalign}
\begin{aligned}
    E^{2}_{\mathbb{A}_t^s}[\vec{k}] = &\sum_{s \in \mathbb{S}_{t}}E^{2}_{\mathbb{A}_t^s,\textrm{in}}[\vec{k},s] \\+& \sum_{m}E^{2}_{\mathbb{A}^{s}_{t},\textrm{up}}[\vec{k},m]\\
    +& \sum_{m}E^{2}_{\mathbb{A}^{s}_{t},\textrm{down}}[\vec{k},m]\\
    +& E^{2}_{\mathbb{A}_t^s,\textrm{state}}[\vec{k}]
\end{aligned}
\end{flalign}

\subsubsection{Boundary Auxiliary Basis}
\label{app:explicit_adap_aux_bound}
Starting from Eq. \eqref{AuxiliaryRemovalError1}, the derivative error associated with excluding auxiliary vector ($|\vec{k}_b\rangle \in \mathbb{A} \setminus \mathbb{A}_t$) from $\mathbb{A}_{t+\Delta t}$ is only the sum of flux-up and flux-down terms arising from neighboring populated auxiliary wave functions (i.e., $\ket{\psi_t^{(\vec{k})}} \rightarrow \ket{\psi_t^{(\vec{k}_b)}}$ for $\vec{k}\in\mathbb{A}_t$), because $|\vec{k}_b\rangle$ is unpopulated.
As a result, we re-use the flux-up and flux-down squared error components constructed in Eqs. \eqref{fluxup} and \eqref{fluxdown} with new filter functions:
\begin{equation}
    \hslash^2E^{2}_{\mathbb{A}_t^b,\textrm{up}}[\vec{k},m] =  F_{\mathbb{A}_t^b,\textrm{up}}[\vec{k},m]\Big\vert (k_{m} + 1)\gamma_{m} \Big\vert^{2} \sum_{s \in \mathbb{S}_t} \left \vert \hat{L}_{m}[s,s]\Psi_{t}[\vec{k},s] \right\vert^{2}
\end{equation}
and
\begin{equation}
     \hslash^2E^{2}_{\mathbb{A}_t^b,\textrm{down}}[\vec{k},m] = F_{\mathbb{A}_t^b,\textrm{down}}[\vec{k},m]\left \vert  \frac{g_{m}}{\gamma_{m}} \right \vert^{2} \sum_{s \in \mathbb{S}_t}\left \vert \hat{D}_{m,t}[s,s]\Psi_{t}[\vec{k},s]\right \vert^{2},
\end{equation}
where $F_{\mathbb{A}_t^b,\textrm{up}} = \delta_{\vec{k} + \vec{e}_{m} \in \mathbb{A} \setminus \mathbb{A}_{t}}$ and $F_{\mathbb{A}_t^b,\textrm{down}} = \delta_{\vec{k} - \vec{e}_{m} \in \mathbb{A} \setminus \mathbb{A}_{t}}$

Thus, the squared derivative error associated with excluding $|\vec{k}_b\rangle \in \mathbb{A}\setminus\mathbb{A}_t$ from $\mathbb{A}_{t+\Delta t}$ is given by
\begin{equation}
    \label{eq:boundauxfluxerror}E^{2}_{\mathbb{A}_t^b}[\vec{k}_b] = \sum_{m,\vec{k}: \vec{k} + \vec{e}_{m} = \vec{k}_{b}}E^{2}_{\mathbb{A}_t^b,\textrm{up}}[\vec{k},m] + \sum_{m,\vec{k}: \vec{k} - \vec{e}_{m} = \vec{k}_{b}}E^{2}_{\mathbb{A}_t^b,\textrm{down}}[\vec{k},m].
\end{equation}
Even in this simplified form, the memory cost of tracking every auxiliary vector $|\vec{k}_{b}\rangle$ grows with the size of the adaptive auxiliary basis $\mathbb{A}_{t}$, but the vast majority of error elements are many orders of magnitude smaller than the most important elements. In some cases, we compute an approximation to the entries of $E^{2}_{\mathbb{A}_t^b}$, detailed in Sec. \ref{sec:discard_fraction}, that still ensures the user-defined error bound while substantially decreasing the number of individual flux elements that must be considered.

\subsection{State Basis}
\label{app:explicit_adap_state}
Expanding the terms in Eq. (\ref{GeneralStateRemoveError}) using the HOPS Eq. (\ref{eq:NormNonLinearHops}), the squared derivative error associated with removing  state $|s\rangle \in \mathbb{S}$ from $\mathbb{S}_{t+\Delta t}$ is
\begin{flalign}
\begin{aligned}\label{StateRemovalError1}
E^{2}_{\mathbb{S}^s_{t}}[s] &=  \sum_{\vec{k} \in \mathbb{A}_{t+\Delta t}^{s}}\left \vert (\mathcal{L}_{\mathbb{A}_{t+\Delta t}^s\otimes\mathbb{S}_t}\Psi_t) [\vec{k},s] + \frac{\Psi_{t}[\vec{k},s]}{\Delta t}\right \vert^{2}\\
&+ \frac{1}{\hslash^2}\sum_{\vec{k} \in \mathbb{A}_{t+\Delta t}^{s}}\sum_{m: \vec{k} + \vec{e}_{m}\in \mathbb{A}_{t+\Delta t}^{b}} \left \vert  \gamma_{m}(k_{m}+1)\hat{L}_{m}\Psi_{t}[\vec{k},s]\right \vert^{2}\\
&+ \frac{1}{\hslash^2}\sum_{\vec{k} \in \mathbb{A}_{t+\Delta t}^{s}}\sum_{m: \vec{k} - \vec{e}_{m}\in \mathbb{A}_{t+\Delta t}^{b}} \left \vert \frac{g_{m}}{\gamma_{m}}\hat{D}_{m,t} \Psi_{t}[\vec{k},s] \right \vert^{2}\\
&+ \frac{1}{\hslash^2}\sum_{\vec{k} \in \mathbb{A}_{t+\Delta t}^{s}}\sum_{s' \in \mathbb{S}\setminus \{s\}}\left \vert \hat{H}_{S}[s',s]\Psi_{t}[\vec{k},s] \right \vert^{2} 
\end{aligned}
\end{flalign}
where we note that the numerical construction of $(\mathcal{L}_{\mathbb{A}_{t+\Delta t}^s\otimes\mathbb{S}_t} |\Psi_t\rangle)$ is performed as an additional derivative evaluation using the HOPS equation expressed in the reduced basis $\mathbb{A}_t\otimes\mathbb{S}_t$ with a HOPS wave function $|\Psi_t\rangle$ that has been updated to account for the reduced basis such that
\begin{equation}
    (\mathcal{L}_{\mathbb{A}_{t+\Delta t}^s\otimes\mathbb{S}_t}\Psi_t) [\vec{k},s]  = \begin{cases}
        \frac{d}{dt}\Psi_t[\vec{k},s] \textrm{ if } \vec{k} \in \mathbb{A}_{t+\Delta t}^s\\
        0 \textrm{ otherwise}.
    \end{cases}
\end{equation}

\subsubsection{Stable State Basis}
Starting from Eq. \eqref{StateRemovalError1}, we calculate the error contributions associated with excluding $\ket{s} \in \mathbb{S}_t$ from the basis $\mathbb{S}_{t+\Delta t}$ as described below. The two components of the flux-in error calculation
\begin{equation}
    E^{2}_{\mathbb{S}^s_{t},\textrm{in}}[\vec{k},s] = \left \vert \frac{d\Psi_{t}[\vec{k},s]}{dt} + \frac{\Psi_{t}[\vec{k},s]}{\Delta t}\right \vert^{2}
\end{equation}
are calculated together since they are able to cancel. The flux-up
\begin{flalign}
\begin{aligned}\label{fluxupstate}
    \hslash^2E^{2}_{\mathbb{S}^s_{t},\textrm{up}}[\vec{k},s] &= \sum_{m}\delta_{\vec{k} + \vec{e}_{m} \in \mathbb{A}_{t+\Delta t}^{b}} \left \vert (k_{m}+1)\gamma_{m}\hat{L}_{m}\Psi_{t}[\vec{k},s] \right\vert^{2}\\
    &= \left \vert \Psi_{t}[\vec{k},s] \right \vert^{2} \sum_{m}F_{\mathbb{S}^s_{t},\textrm{up}}[\vec{k},m] \Big \vert (k_{m}+1)\gamma_{m}\hat{L}_{m}[s,s] \Big \vert^{2}
\end{aligned}
\end{flalign}
and flux-down
\begin{flalign}
\begin{aligned}\label{fluxdownstate} 
    \hslash^2E^{2}_{\mathbb{S}^s_{t},\textrm{down}}[\vec{k},s] &= \sum_{m}\delta_{\vec{k} - \vec{e}_{m} \in \mathbb{A}_{t+\Delta t}^{b}}\left \vert  \frac{g_{m}}{\gamma_{m}}\hat{D}_{m,t}\Psi_{t}[\vec{k},s]\right \vert^{2}\\
    &= \left \vert \Psi_{t}[\vec{k},s] \right \vert^{2}\sum_{m} F_{\mathbb{S}^s_{t},\textrm{down}}[\vec{k},m] \left \vert  \frac{g_{m}}{\gamma_{m}}\hat{D}_{m,t}[s,s]\right \vert^{2}
\end{aligned}
\end{flalign}
error calculations are simplified by introducing the filters  $F_{\mathbb{S}^s_{t},\textrm{up}}[\vec{k},m] = \delta_{\vec{k} + \vec{e}_{m} \in \mathbb{A}^{b}_{t + \Delta t}}$ and $F_{\mathbb{S}^s_{t},\textrm{down}}[\vec{k},m] = \delta_{\vec{k} - \vec{e}_{m} \in \mathbb{A}_{t + \Delta t}^{b}}$. The squared state-flux error is given by
\begin{equation}
    \hslash^2E^{2}_{\mathbb{S}^s_{t},\textrm{state}}[s] = \sum_{\vec{k} \in \mathbb{A}_{t + \Delta t}^{s}}\sum_{s'\in\mathbb{S} \setminus \{s\}} \left \vert\hat{H}_{S}[s',s]\Psi_{t}[\vec{k},s] \right \vert^{2}.
\end{equation}
 The total squared derivative error (Eq. \eqref{StateRemovalError1}) is given by
\begin{flalign}
\begin{aligned}
    E^{2}_{\mathbb{S}^s_{t}}[s] = &\sum_{\vec{k} \in \mathbb{A}_{t+\Delta t}^{s}}E^{2}_{\mathbb{S}^s_{t},\textrm{in}}[\vec{k},s]\\ 
    + &\sum_{\vec{k} \in \mathbb{A}_{t+\Delta t}^{s}}E^{2}_{\mathbb{S}^s_{t},\textrm{up}}[\vec{k},s]\\
    + &\sum_{\vec{k} \in \mathbb{A}_{t+\Delta t}^{s}}E^{2}_{\mathbb{S}^s_{t},\textrm{down}}[\vec{k},s]\\
    + &E^{2}_{\mathbb{S}^s_{t},\textrm{state}}[s].
\end{aligned}
\end{flalign}

\subsubsection{Boundary State Basis}
Starting from Eq. \eqref{StateRemovalError1}, the squared derivative error arising from excluding a state $\ket{s_b} \in \mathbb{S}\setminus\mathbb{S}_{t}$ from the basis $\mathbb{S}_{t+\Delta t}$ is given by
\begin{flalign}
 \begin{aligned}
    E^{2}_{\mathbb{S}_t^b}[s_{b}] &= \sum_{\vec{k} \in \mathbb{A}_{t+\Delta t}^{s}}\left \vert \sum_{s \in \mathbb{S}_{t}} \mathcal{L}[\vec{k},s_{b},\vec{k},s]\Psi_{t}[\vec{k},s] \right \vert^{2}\\
    &= \frac{1}{\hslash^2}\sum_{\vec{k} \in \mathbb{A}_{t + \Delta t}^{s}}\left \vert \sum_{s \in \mathbb{S}_{t}} -i\hat{H}_{S}[s_{b},s]\Psi_{t}[\vec{k},s] \right \vert^{2}
 \end{aligned}
\end{flalign}
where we have simplified the outer sum because boundary auxiliary vectors $|\vec{k}_b\rangle\in\mathbb{A}^b_{t+\Delta t}$ are unpopulated.

\section{Size Invariance Proof-of-Concept System}
\label{app:SizeInvariance}
\label{sec:Size_invariance_sys}
The calculation presented in Fig. \ref{fig:SizeInvariance}a-b is a nearest-neighbour coupled linear chain of identical pigments with a system Hamiltonian given by
\begin{equation}
    \hat{H}_S = \sum_{n=0}^{N_{\textrm{sites}}-1} V|n\rangle\langle n+1| + h.c.
\end{equation}
 with nearest-neighbor couplings $V=50\textrm{ cm}^{-1}$. Each pigment (occupied by the excitation in state $|n\rangle$) is coupled to an independent environment with system-environment coupling operator
 \begin{equation}
     \hat{L}_n = |n\rangle \langle n|
 \end{equation}
and a Drude-Lorentz spectral density (Eq. \eqref{eq:drude_lorentz}) with 
$\lambda=\gamma_0=50\textrm{ cm}^{-1}$. The temperature is 300 K. Each environment is given an extra Markovian-filtered correlation function mode described by $g_{\textrm{Mark}_n} = -i\textrm{Im}[g_{0_n}]$ and $\gamma_{\textrm{Mark}_n} = 500 \textrm{ cm}^{-1}$ to ensure that $C_n(0)$ is real. The initial system state is $|0\rangle$.

The 2-particle calculation in Fig. \ref{fig:SizeInvariance}c takes place in the same system, doubly-excited, such that the system Hamiltonian is
\begin{equation}
     \hat{H}_S = \sum_{n=0}^{N_{\textrm{sites}}-1}\sum_m^n V\Big(|n,m\rangle\langle n\pm1,m| + |n,m\rangle\langle n,m\pm1|\Big) + h.c.
\end{equation}
where state $|n,m\rangle$ represents sites $n$ and $m$ both being singly-excited, state $|n,n\rangle$ represents site $n$ being doubly-excited, and the nearest-neighbor coupling $V$ is the same as the singly-excited case. Each pigment $n$ is coupled to an independent environment with system-environment coupling operator
\begin{equation}
    \hat{L}_n = 2|n,n\rangle\langle n,n| + \sum_{m\neq n} |n,m\rangle\langle n,m|
\end{equation}
and the same correlation function as the singly-excited case. The initial system state is $|N_{\textrm{sites}}-1, 0\rangle$.

\section{Effect of Correlated Excitonic and Charge-Transfer System-Environment Couplings in OPV model}
\label{app:exciton_correlation}
\begin{figure}
    \centering
    \includegraphics{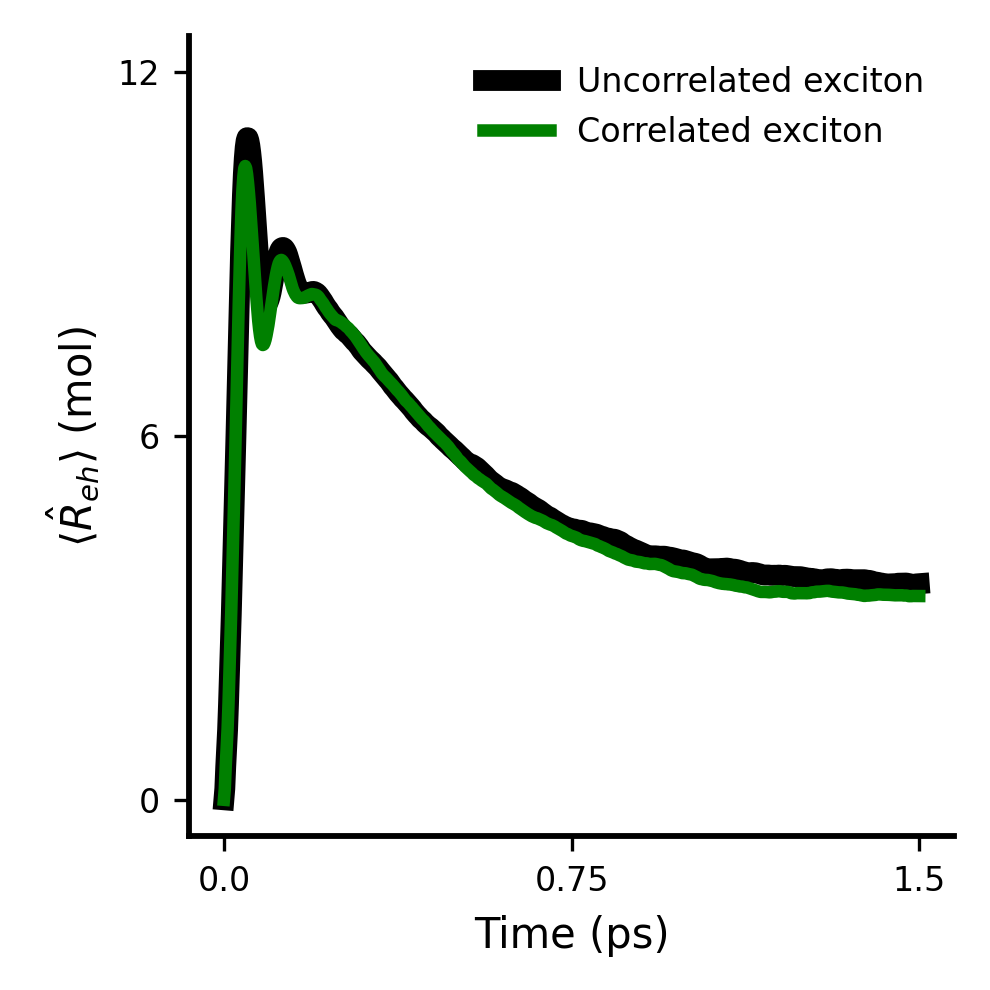}
    \caption{Charge separation in an $N_{acc}=N_{don}=10$ mobile-hole chain when the exciton state is coupled to a unique environment (black) and the same environment as the hole on the donor molecule at the boundary (green). Convergence parameters are given in Table \ref{tab:convergence_params}.}
    \label{fig:exciton_correlation}
\end{figure}

The relationship between system-environment couplings of the charge-transfer and exciton states associated with the same molecule depends on the chemical identity of the system.\cite{Andrzejak_2018} In the absence of a specific molecular parameterization, our generic model describes the thermal environment composed of low-frequency vibrations coupled to the exciton state as uncorrelated to the thermal environment associated with charge-transfer states, as laid out in Sec. \ref{sec:ChargeSeparation}. To test the influence of this approximation, we compare our original uncorrelated exciton model to a correlated exciton model wherein the exciton state is coupled to the same environment as the charge-transfer states with $h=0$, as previously set forth in other models.\cite{Hestand2018review} In this correlated model, we exclude the environment described by $\hat{L}_{0}$, and redefine
\begin{equation}
\hat{L}_{h=0} = 2|0,0\rangle\langle 0,0| + \sum_e |0,e\rangle\langle 0,e|.    
\end{equation}
Fig. \ref{fig:exciton_correlation} shows that the charge separation dynamics are the same for the correlated and uncorrelated models when $N_{acc}=N_{don}=10$.  

\section{Limit of Delocalization in the OPV Model}
\label{app:DelocLimits}

\begin{figure}
    \centering
    \includegraphics{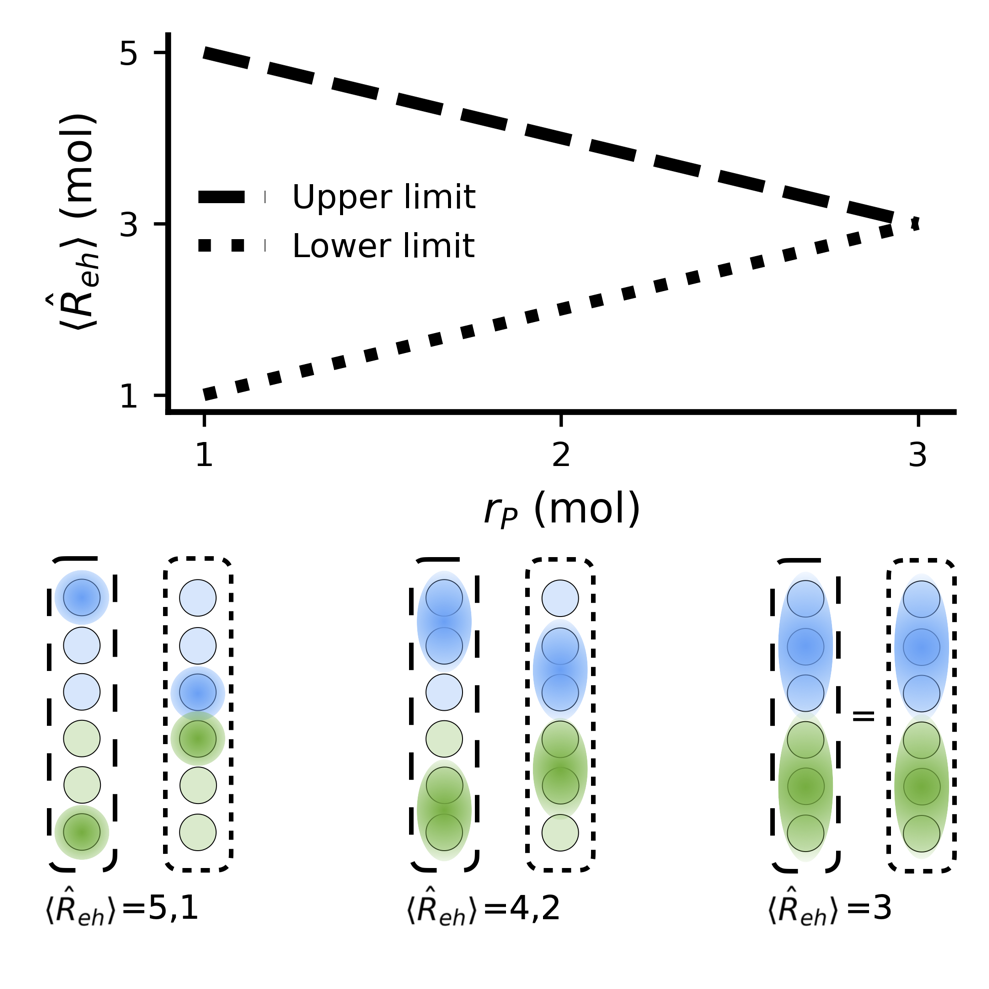}
    \caption{Geometric limitations on $\langle \hat{R}_{eh}\rangle$ for delocalized states. The maximum and minimum $\langle\hat{R}_{eh}\rangle$ for evenly-delocalized states of a mobile-hole OPV model with $N_{don}=N_{acc}=3$ (top). The maximally and minimally-separated states for $r_P=1,2,3$ (bottom).}
    \label{fig:deloc_limit}
\end{figure}

In the OPV model system, localized states correspond to a greater range of possible $\langle \hat{R}_{eh}\rangle$ than delocalized states. The dashed lines in the contour plots of Fig. \ref{fig:mech_overview}c-e represent the limits imposed by the geometry of the OPV model system on $\langle \hat{R}_{eh}\rangle$. To find the upper and lower limit on $\langle \hat{R}_{eh}\rangle$ for a given delocalization value $r_P$, we construct states in which both the electron and hole are delocalized evenly over the $r_P$ donor and acceptor molecules furthest from and closest to the boundary, respectively, and calculate the associated $\langle \hat{R}_{eh}\rangle$. In Fig. \ref{fig:deloc_limit}, we visualize this for a simple $N_{acc}=N_{don}=3$ OPV chain, outlining the states associated with maximal and minimal separation in dashed and dotted lines, respectively. In the fixed-hole case, we do the same process, with the electron delocalized evenly over the $r_P$ molecules furthest from and closest to the sole donor molecule.

\section{Convergence Scans}

We present the convergence parameters used in all calculations in the main text in Table \ref{tab:convergence_params}.

\begin{figure}
    \centering
    \includegraphics{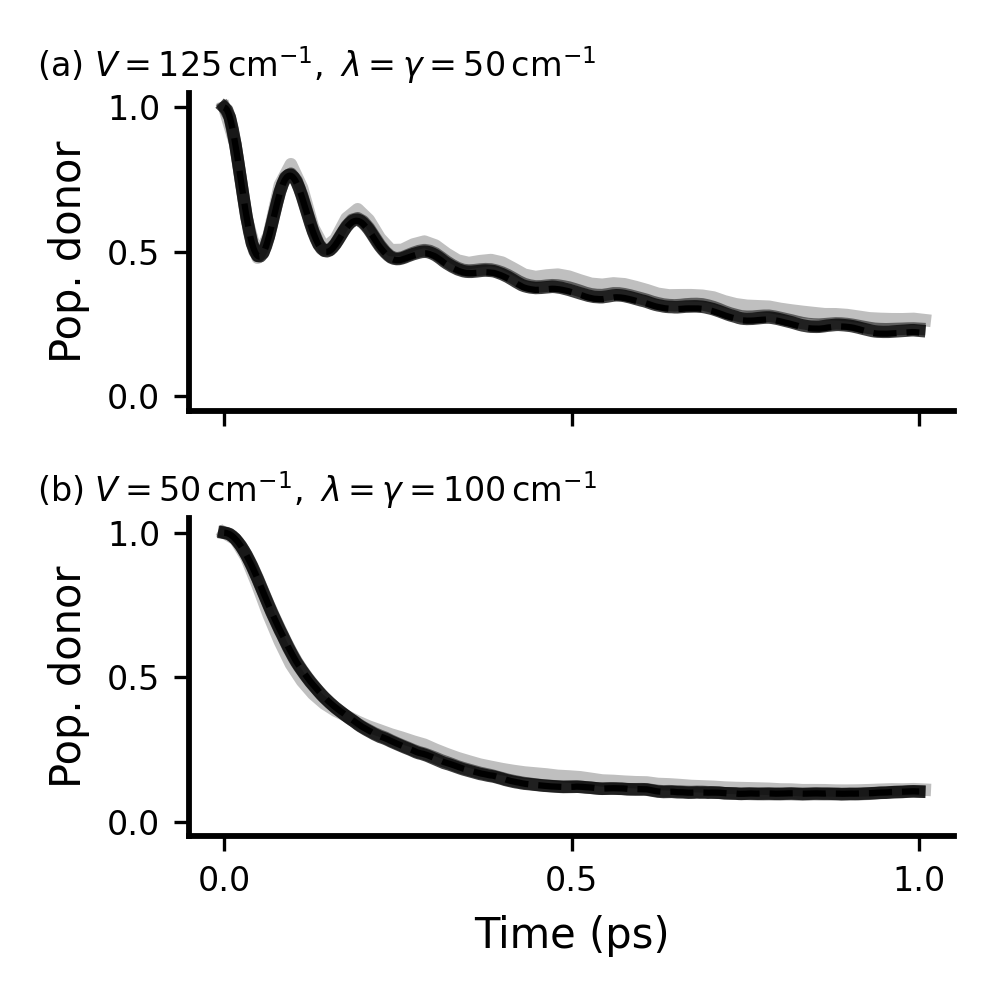}
    \caption{Convergence of dimer calculations. The dimer dynamics as seen in Fig. \ref{fig:LTC_dimer} in terms of the donor population dynamics. Darker lines represent more-converged calculations. In all cases, an effective integration of the noise was used with a noise time step of $\tau=0.05$ fs. Convergence parameters are given in Table \ref{tab:convergence_params_conv}.}
    \label{fig:LTC_dimer_convergence}
\end{figure}

\begin{figure}
    \centering
    \includegraphics{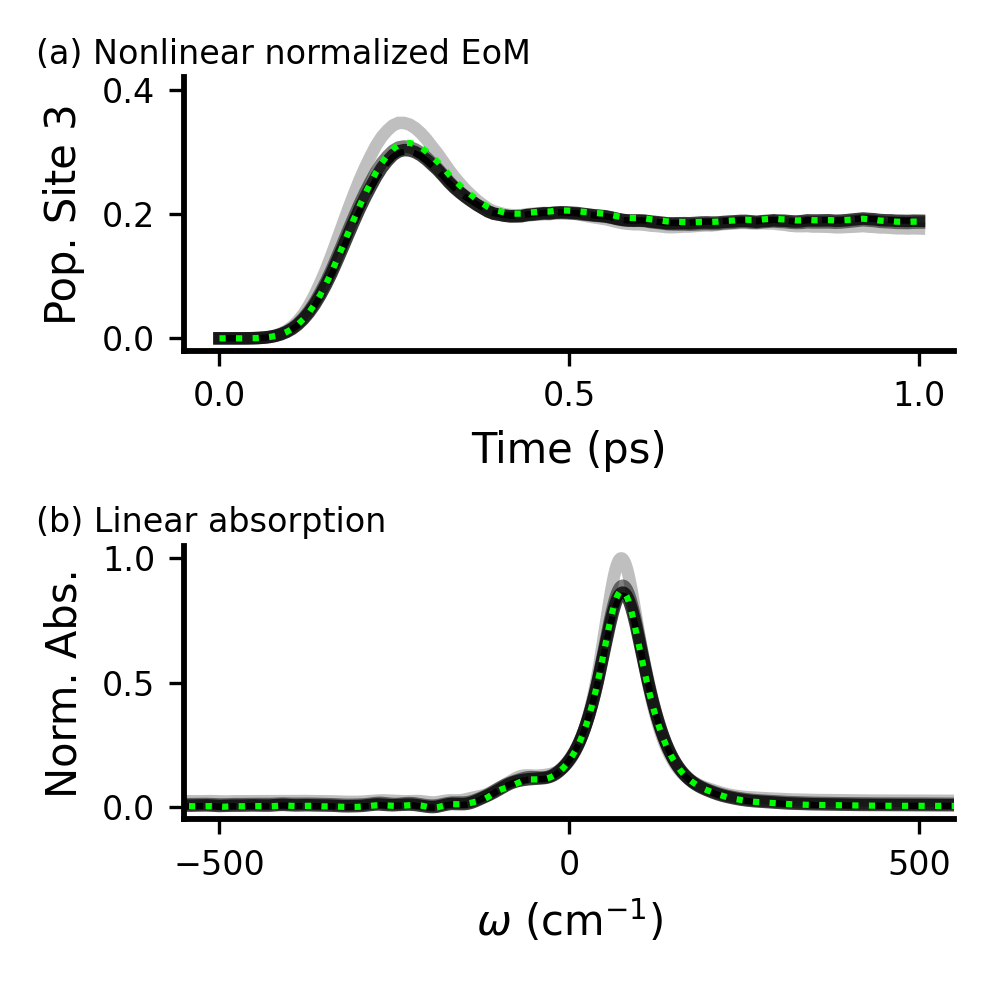}
    \caption{Convergence of four-site linear chain calculations. (a) The linear chain dynamics in terms of the final site population as seen in Fig. \ref{fig:LTC_Linchain}a. (b) The absorption spectrum as seen in Fig. \ref{fig:LTC_Linchain}b. Darker lines represent more-converged calculations. The black lines represent calculations where all Matsubara modes are Markovian-filtered, while the green dotted lines represent cases in which the first Matsubara mode was not filtered. In all four-site linear chain calculations, $f_{dis}=0$ and an effective integration of the noise was used with a noise time step of $\tau=0.5$ fs. Convergence parameters are given in Table \ref{tab:convergence_params_conv}.}
    \label{fig:LTC_Linchain_convergence}
\end{figure}

\begin{figure}
    \centering
    \includegraphics{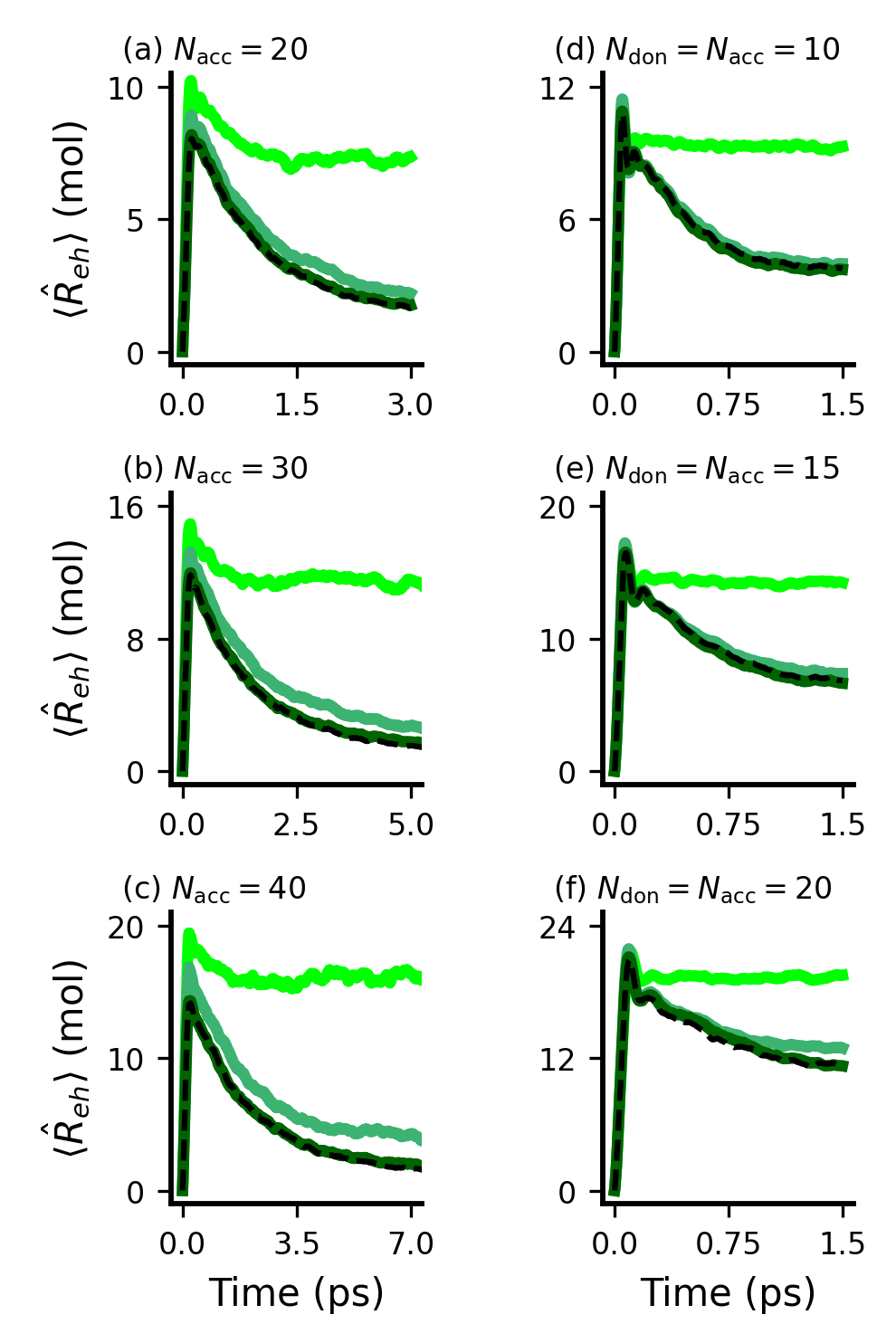}
    \caption{Convergence of OPV calculations. Dynamics of OPV linear chains in terms of $\langle \hat{R}_{eh} \rangle_t$ dynamics at various chain lengths. Darker lines represent more-converged calculations. In all OPV calculations, $f_{dis}=0.9$ and an effective integration of the noise was used (excepting Fig. \ref{fig:OPV_time step}b) with a noise time step of $\tau=0.025$ fs. Convergence parameters are given in Table \ref{tab:convergence_params_conv}.}
    \label{fig:OPV_convergence}
\end{figure}

We illustrate convergence in terms of the expectation dynamics of some characteristic observable for each calculation. In Fig. \ref{fig:LTC_dimer_convergence}, we show convergence of the calculations in Fig. \ref{fig:LTC_dimer} in terms of the donor population. In Fig. \ref{fig:LTC_Linchain_convergence}, we show the convergence of the calculations in Fig. \ref{fig:LTC_Linchain} in terms of the population of the final site in the chain (a) and the absorption spectrum (b). In Fig. \ref{fig:OPV_convergence}, we show convergence of OPV calculations in terms of expectation electron-hole separation distance $\langle \hat{R}_{eh} \rangle_t$. The convergence parameters of all convergence scan simulations are recorded in Table \ref{tab:convergence_params_conv}.

\begin{table*}
\centering
\begin{tabular}{ |c|c|c|c|c|c|c|c| } 
 \hline
 Simulation & $k_{\textrm{max}}$ & $k_{\textrm{Mats}}$ & $dt$ (fs) & $\delta_A$ & $\delta_S$ & $u_t$ (fs) & $N_{traj}$ \\
 \hline
 Fig. \ref{fig:SizeInvariance}a & 15 & 0/0 & 4.0 & 0.0005 & 0.001 & 8.0 & $[3998, 4000]$ \\
 \hline
 Fig. \ref{fig:SizeInvariance}b & 15 & 0/0 & 4.0 & 0.0005 & 0.001 & 4.0 & $4000$ \\
 \hline
 Fig. \ref{fig:SizeInvariance}c & 25 & 0/0 & 2.0 & 0.00025 & 0.0005 & 8.0 & $[999,1000]$ \\
 \hline
 Figs. \ref{fig:mech_overview}, \ref{fig:model_extension} - fixed-hole $(N_{\textrm{sites}} \leq 40)$ & 3 & 5/0 & 0.5 & 0.0075 & 0 & 16 & $5000$ \\
 \hline
 Figs. \ref{fig:mech_overview}, \ref{fig:model_extension} - mobile-hole $(N_{\textrm{sites}} \leq 40)$ & 3 & 5/0 & 0.5 & 0.025 & 0 & 16 & $5000$ \\
 \hline
  Fig. \ref{fig:model_extension} - fixed-hole $(N_{\textrm{sites}} > 40)^\dag$ & 3 & 5/0 & 0.5 & 0.0075 & 0 & 16 & $1000$ \\
 \hline
 Fig. \ref{fig:model_extension} - mobile-hole $(N_{\textrm{sites}} > 40)^\dag$ & 3 & 5/0 & 0.5 & 0.025 & 0 & 16 & $1000$ \\
 \hline
  Fig. \ref{fig:LTC_dimer} & 5 & * & 4.0 & 0 & 0 & - & $1000$ \\
 \hline
 Fig. \ref{fig:LTC_Linchain}a - HTA & 15 & 0/0 & 1.0 & 0.001 & 0 & 16 & $5000$ \\
 \hline
 Fig. \ref{fig:LTC_Linchain}a - LTC & 15 & 2/0 & 1.0 & 0.001 & 0 & 16 & $5000$ \\
 \hline
 Fig. \ref{fig:LTC_Linchain}a - Conv. & 15 & 2/2 & 1.0 & 0.001 & 0 & 16 & $5000$ \\
 \hline
 Fig. \ref{fig:LTC_Linchain}b - HTA & 25 & 0/0 & 2.0 & 0.00025 & 0 & 2.0 & $5000$ \\
 \hline
 Fig. \ref{fig:LTC_Linchain}b - LTC & 25 & 5/0 & 2.0 & 0.00025 & 0 & 2.0 & $5000$ \\
 \hline
 Fig. \ref{fig:LTC_Linchain}b - Conv. & 25 & 5/5 & 2.0 & 0.00025 & 0 & 2.0 & $5000$ \\
 \hline
 Fig. \ref{fig:OPV_time step} & 3 & 5/0 & * & 0.0075 & 0 & 16 & $5000$ \\
 \hline
 Fig. \ref{fig:exciton_correlation} & 3 & 5/0 & 0.5 & 0.025 & 0 & 16 & $5000$ \\
 \hline

 \end{tabular}
 \caption{The convergence parameters of all converged simulations. Note that in $k_{\textrm{Mats}}$, we indicate two values: the first is the number of Matsubara modes used in the noise, and the second is the number of Matsubara modes explicitly included in the hierarchy (i.e., not low-temperature-corrected). A value of * indicates varying parameters, detailed in the figure associated with the calculation. A value of - indicates non-applicability. A superscripted $\dag$ indicates that the simulation's convergence parameters were asserted based on other calculations.}
 \label{tab:convergence_params}
\end{table*}

\begin{table*}
\centering
\begin{tabular}{ |c|c|c|c|c|c|c|c| } 
 \hline
 Simulation & $k_{\textrm{max}}$ & $k_{\textrm{Mats}}$ & $dt$ (fs) & $\delta_A$ & $\delta_S$ & $u_t$ (fs) & $N_{traj}$ \\
 \hline
 Fig. \ref{fig:LTC_dimer_convergence}a,b & 1,\, 3,\, 5,\, 7 & 0/0,\, 3/1,\, 5/2,\, 10/3 & 16,\, 8.0,\, 4.0,\, 2.0 & 0 & 0 & - & $1000$ \\
 \hline
 Fig. \ref{fig:LTC_Linchain_convergence}a (black) & 10,\, 15,\, 20,\, 30 & 0/0,\, 2/2,\, 5/5,\, 10/10 & 8.0,\, 4.0,\, 2.0,\, 1.0 & 0.0025,\, 0.001,\, 0.0005,\, 0.00025 & 0 & 32,\, 16,\, 8.0,\, 4.0 & $1000$ \\
 \hline
 Fig. \ref{fig:LTC_Linchain_convergence}a (green) & 15 & 2/2 & 4.0 & 0.001 & 0 & 16 & $1000$ \\
 \hline
 Fig. \ref{fig:LTC_Linchain_convergence}b (black) & 10,\, 15,\, 20,\, 30 & 0/0,\, 2/2,\, 5/5,\, 10/10 & 8.0,\, 4.0,\, 2.0,\, 1.0 & 0.001,\, 0.0005,\, 0.00025,\, 0.0001 & 0 & 8.0,\, 4.0,\, 2.0,\, 1.0 & $1000$ \\
 \hline
 Fig. \ref{fig:LTC_Linchain_convergence}b (green) & 40 & 10/10 & 1.0 & 0.0001 & 0 & 1.0 & $1000$ \\
 \hline
 Fig. \ref{fig:OPV_convergence}a & 1,\, 2,\, 3,\, 4 & 0/0,\, 2/0,\, 5/0,\, 10/0 & 2.0,\, 1.0,\, 0.5,\, 0.25 & 0.04,\, 0.02,\, 0.0075,\, 0.005 & 0 & 64,\, 32,\, 16,\, 8.0 & $1000$ \\
 \hline
  Fig. \ref{fig:OPV_convergence}b-c & 1,\, 2,\, 3,\, 4 & 0/0,\, 2/0,\, 5/0,\, 10/0 & 2.0,\, 1.0,\, 0.5,\, 0.25 & 0.04,\, 0.02,\, 0.0075,\, 0.0035 & 0 & 64,\, 32,\, 16,\, 8.0 & $1000$ \\
 \hline
 Fig \ref{fig:OPV_convergence}d-e,  & 1,\, 2,\, 3,\, 4 & 0/0,\, 2/0,\, 5/0,\, 10/0 & 2.0,\, 1.0,\, 0.5,\, 0.25 & 0.1,\, 0.05,\, 0.025,\, 0.01 & 0 & 64,\, 32,\, 16,\, 8.0 & $[995,1000]$ \\
 \hline
 Fig. \ref{fig:OPV_convergence}f & 1,\, 2,\, 3,\, 4 & 0/0,\, 2/0,\, 5/0,\, 10/0 & 2.0,\, 1.0,\, 0.5,\, 0.25 & 0.1,\, 0.05,\, 0.025,\, 0.01 & 0 & 64,\, 32,\, 16,\, 8.0 & $1000,\,999,\,1000,\,510$ \\
 \hline
 \end{tabular}
 \caption{The convergence parameters of all convergence scans. Note that in $k_{\textrm{Mats}}$, we indicate two values: the first is the number of Matsubara modes used in the noise, and the second is the number of Matsubara modes explicitly included in the hierarchy (i.e., not low-temperature-corrected). A value of - indicates non-applicability.}
 \label{tab:convergence_params_conv}
\end{table*}

% \nocite{*}
\bibliographystyle{apsrev4-1.bst}
\bibliography{N-particle_bib}% Produces the bibliography via BibTeX.

\end{document}